\documentclass[3p,times,preprint,11pt]{elsarticle}

\usepackage{tabularx}
\usepackage{amssymb}
\usepackage{graphicx}
\usepackage{subfigure}

\usepackage{caption}

\DeclareCaptionLabelFormat{boldfigure}{\textbf{Fig. #2}}
\DeclareCaptionLabelSeparator{boldsep}{\textbf{. }}
\usepackage{amsmath}
\usepackage{bm}
\usepackage{booktabs} %插表格用的宏包调整表格线与上下内容的间隔
\usepackage{diagbox} %插表格用的宏包
\usepackage{multirow} %插多行表格用的宏包
\usepackage[section]{placeins}

\usepackage{algorithm}
\usepackage{algpseudocode}

\journal{Elsevier}

\begin{document}

\begin{frontmatter}

%% Title, authors and addresses

%% use the tnoteref command within \title for footnotes;
%% use the tnotetext command for theassociated footnote;
%% use the fnref command within \author or \address for footnotes;
%% use the fntext command for theassociated footnote;
%% use the corref command within \author for corresponding author footnotes;
%% use the cortext command for theassociated footnote;
%% use the ead command for the email address,
%% and the form \ead[url] for the home page:
%% \title{Title\tnoteref{label1}}
%% \tnotetext[label1]{}
%% \author{Name\corref{cor1}\fnref{label2}}
%% \ead{email address}
%% \ead[url]{home page}
%% \fntext[label2]{}
%% \cortext[cor1]{}
%% \affiliation{organization={},
%%             addressline={},
%%             city={},
%%             postcode={},
%%             state={},
%%             country={}}
%% \fntext[label3]{}

% \title{Physics-informed neural networks for unstructured adaptive mesh refinement}
%\title{Residuals-guided adaptive unstructured mesh refinement using physics-informed neural networks for solving compressible and incompressible flows}
\title{An unstructured adaptive mesh refinement for steady flows based on physics-informed neural networks}
%% use optional labels to link authors explicitly to addresses:
%% \author[label1,label2]{}
%% \affiliation[label1]{organization={},
%%             addressline={},
%%             city={},
%%             postcode={},
%%             state={},
%%             country={}}
%%
%% \affiliation[label2]{organization={},
%%             addressline={},
%%             city={},
%%             postcode={},
%%             state={},
%%             country={}}

\author[inst1]{Yongzheng Zhu}
%\ead{zhuyz@zju.edu.cn}

\author[inst2]{Shiji Zhao}

\author[inst3]{Yuanye Zhou}

\author[inst2]{Hong Liang}

\author[inst1]{Xin Bian\corref{cor1}}
\ead{bianx@zju.edu.cn}

% Department and Organization
\affiliation[inst1]{organization={State Key Laboratory of Fluid Power and Mechatronic Systems, Department of Engineering Mechanics, Zhejiang University},
            city={Hangzhou},
            postcode={310027}, 
            country={China}}
\affiliation[inst2]{organization={Department of Physics, Hangzhou Dianzi University},
            city={Hangzhou},
            postcode={310018}, 
            country={China}}
\affiliation[inst3]{organization={Shanghai Academy of AI for Science},
            city={Shanghai},
            postcode={200030}, 
            country={China}}
\cortext[cor1]{Corresponding author.}

% --------------------------------- Abstract ---------------------------------
\begin{abstract}
Mesh generation is essential for accurate and efficient computational fluid dynamics simulations.
To resolve critical features in the flow,
adaptive mesh refinement (AMR) is routinely employed in certain regions of the computational domain,
where gradients or error estimates of the solution are often considered as the refining criteria.
In many scenarios, however, these indicators can lead to unnecessary refinement over a large region, making the process a matter of trial and error and resulting in slow convergence of the computation. 
To this end, we propose a heuristic strategy that employs the residuals of the governing partial differential equations (PDEs) as a novel criterion to adaptively guide the mesh refining process. 
In particular, we leverage on the physics-informed neural networks (PINNs) to integrate imprecise data obtained on a coarse mesh and the governing PDEs.
Once trained, PINNs are capable of identifying regions of highest residuals of the Navier-Stokes/Euler equations 
and suggesting new potential vertices for the coarse mesh cells. 
Moreover, we put forth two schemes to maintain the quality of the refined mesh through the strategic insertion of vertices and the implementation of Delaunay triangulation. By applying the residuals-guided AMR to address a multitude of typical incompressible/compressible flow problems and comparing the outcomes with those of gradient-based methods, we illustrate that the former effectively attains a favorable balance between the computational accuracy and cost. 

\end{abstract}

%%Graphical abstract
% \begin{graphicalabstract}
% \includegraphics{grabs}
% \end{graphicalabstract}

%%Research highlights
% \begin{highlights}
% \item Research highlight 1
% \item Research highlight 2
% \end{highlights}

\begin{keyword}
%% keywords here, in the form: keyword \sep keyword
unstructured mesh; adaptive mesh refinement; physics-informed neural networks; residuals of PDEs ; computational fluid dynamics; incompressible and compressible flows. 

%% PACS codes here, in the form: \PACS code \sep code
% \PACS 0000 \sep 1111
%% MSC codes here, in the form: \MSC code \sep code
%% or \MSC[2008] code \sep code (2000 is the default)
% \MSC 0000 \sep 1111
\end{keyword}

\end{frontmatter}

%% \linenumbers

%% For citations use: 
%%       \citet{<label>} ==> Jones et al. [21]
%%       \citep{<label>} ==> [21]
%%

%% main text
% -------------------------------- I. INTRODUCTION ---------------------------------
\section{Introduction}
\label{sec:Introduction}
Mesh generation is a pivotal component in the numerical resolution of partial differential equations~(PDEs), facilitating accurate solutions with rapid convergence and a reduction in numerical diffusion.
A variety of PDE solvers including finite volume and finite element methods~\citep{brenner2008mathematical, karniadakis2005spectral, moukalled2016finite}, typically rely on either structured or unstructured meshes. 
Structured meshes are predominantly quadrilateral (2D) or hexahedral (3D) in nature and offer enhanced accuracy and convergence, but they require a considerable amount of time to generate. Unstructured meshes are typically based on triangles, tetrahedra, or arbitrary polyhedra, and they have gained popularity in engineering applications due to their rapid generation process and ability to handle intricate geometric boundaries~\citep{tam2000anisotropic, murayama2008comparison, katz2011mesh, alakashi2014comparison}.
In computational fluid dynamics~(CFD) simulations, a coarse mesh suffices in regions where the solution remains smooth.
However, in regions with complex physical features, such as steep gradients or even discontinuities, a finer mesh becomes necessary~\citep{berger1984adaptive, berger1989local, baker1997mesh}.
In such scenarios, a global strategy of refinement based on a uniform mesh is computationally expensive and also unnecessary.

Adaptive mesh refinement~(AMR) represents a class of well-established techniques proposed to adapt the local mesh resolutions based on the solution~\citep{berger1984adaptive, berger1989local, baker1997mesh}. 
The traditional AMR paradigm follows an iterative procedure, that is, SOLVE$\rightarrow$ESTIMATE$\rightarrow$MARK$\rightarrow$REFINE loop~\citep{morin2000data, braess2007convergence, hoppe2009convergence}. This involves numerically solving the PDEs by a selected solver, estimating the error for each mesh cell according to a chosen criterion, marking some of the mesh cells with potentially relatively larger/smaller errors, and refining or coarsening the marked cells to obtain an updated mesh~\citep{braess2007convergence, daniel2018adaptive}.
Accordingly, an optimal mesh can be identified by a sensitivity analysis based on desired metrics among different resolutions. AMR has been demonstrated to be an effective approach for addressing flow problems, offering the requisite spatial resolution for accurate solutions while minimizing memory demands~\citep{berger1984adaptive, berger1989local, baker1997mesh, venditti2002grid, giuliani2019adaptive, cant2022unstructured, freret2022enhanced}.
One of the primary challenges in employing AMR is the lack of reliable error indicators. In its most basic form, AMR assumes that regions with larger gradients are associated with greater errors. This leads to the common strategy of adapting flow characteristics based on these metrics. However, local refinement based on gradients does not ensure a simultaneous reduction in other global error metrics and may even result in inaccurate outcomes. Furthermore, there is considerable scope to improve the efficiency for division and storage of the mesh.
A number of promising approaches have been proposed to enhance the performance of AMR, including the Kelly error estimator~\citep{kelly1983posteriori}, the adjoint-based error correction~\citep{pierce2000adjoint, venditti2002grid}, graphics processing unit (GPU)-based AMR~\citep{giuliani2019adaptive}, Morton code-based cell division~\citep{cant2022unstructured}, and anisotropic block-based AMR~\citep{freret2022enhanced}, among others. 
The prevailing approaches to AMR, including their enhanced versions, rely on the gradient or curvature of the solution as the refinement criterion
and are predominantly applied to structured meshes. In certain scenarios, these refinement criteria still necessitate extensive mesh refinement over large regions, relying on trial-and-error, and exhibiting slow error convergence~\citep{baker1997mesh, venditti2002grid}. 
These pioneering works encourage us to explore a more versatile and reliable indicator of AMR in broader scenarios.

Indeed, there exists a number of exploratory studies on the learning of optimal meshes exclusively using machine learning techniques.
Bohn and Feischl~\citep{bohn2021recurrent} provided a theoretical demonstration of the feasibility of using recurrent neural networks to learn an optimal mesh refinement strategy. The effectiveness of this approach is evaluated using the Poisson equation, 
while its broader applicability to other PDEs remains to be seen. 
Another intriguing approach is to conceptualize the AMR as an observable Markov decision process, wherein policy networks trained through deep reinforcement learning are derived from numerical simulations~\citep{yang2023reinforcement, foucart2023deep}. 
The concept of employing machine learning techniques to learn the optimal mesh appears to circumvent the conventional loop and eschew the advantages of traditional AMR. Moreover, the efficacy of the optimization is contingent upon human expertise in tuning the hyperparameters for model training~\citep{bohn2021recurrent, yang2023reinforcement, foucart2023deep}.

In light of the extensive research on AMR outlined above, we propose the integration of traditional AMR with emerging machine learning techniques, aiming to preserve the established merits of AMR while improving computational accuracy and/or efficiency through the use of machine learning. 
Given our objective of solving PDEs, it is reasonable to consider the residual of the PDEs as an error estimator, taking into account the errors of all differential operators in the equations.
In contrast, a traditional approach to error estimation typically focuses on a single or a limited set of differential operators, such as first- or second-order derivatives. 
It is suggested that the total residual of the governing PDEs provides a more comprehensive error representation and offers a broader view of the error distribution compared to the gradient or curvature of a single physical quantity.
Moreover, various numerical solvers tend to prioritize the resolution of unknown physical quantities over the provision of associated differential operator values. 
In the case of an open-source solver, some effort is required to modify the code to calculate the various differential operators.
However, in the case of a closed-source solver, the source code is not accessible,
and therefore derivatives and other information can only be calculated by interpolating the solutions on the mesh as an additional post-processing step.  
For high-order PDEs, such as the Navier-Stokes (NS) equations, this can be tedious and time-consuming efforts and is not a particularly practical solution. 

To this end, we employ physics-informed neural networks~(PINNs)~\citep{raissi2019physics, karniadakis2021physics} as an estimator for the residuals of PDEs.
This enables the residuals to be considered not only at the specified mesh nodes, but also at any location within the computational domain.
The benefits of utilizing PINNs in this context are readily apparent.
During the training phase, PINNs require only approximate solution data from CFD simulations on a coarse mesh.
Given the pivotal role played by the governing PDEs during the training process,
the trained PINNs model can be regarded as a reasonable surrogate for the underlying physics.
Once trained, PINNs can utilize automatic differentiation~\cite{rall1981automatic} to efficiently compute first- or higher-order derivatives of any physical quantity,
thereby simplifying the computation of various differential operators and eliminating the need for tedious mesh interpolation. 
Moreover, PINNs receive solution data exclusively at the mesh nodes, yet
remain independent of the underlying PDE solvers. 
This flexibility allows PINNs to function as an external "plug-in" and to 
be compatible with any numerical solver. 
PINNs have already stimulated a broad spectrum of research activities and have been extensively utilized in diverse contexts, including shock waves, heat transfer, Stefan problems, multi-phase flows, moving boundaries, and more~\citep{mao2020physics,  liang2024continuous, Meng2020a, cai2021physics, wang2021deep, jin2021nsfnets, qiu2022physics, zhu2024physics}.
If PINNs are regarded as a numerical solver, they do not require a mesh but rely on randomly sampled collocation points instead. 
Consequently, extending the usage to provide PDEs residuals at any location is a relatively straightforward process.
For instance, a residual-based adaptive refinement of PINNs, which employs the PDEs residuals to enhance the distribution of sampling points throughout the training process, has been shown to be highly effective~\citep{lu2021deepxde, wu2023comprehensive}. 
In certain scenarios, the incorporation of a restricted number of supplementary points can significantly enhance the precision of the solution. 

In this work, we present a PDE residuals-guided adaptive mesh refinement~(AMR) framework for the purpose of refining an unstructured mesh.
In particular, the proposed framework employs PINNs as an external plug-in to identify the locations of the highest residuals and assist in solving steady flow problems in conjunction with a PDEs solver.
Furthermore, we propose two schemes for the preservation of mesh quality through the strategic insertion of vertices and the performance of Delaunay triangulation~\citep{chew1989constrained, shewchuk1996triangle, shewchuk2002delaunay}.
The novel AMR methodology enables the same degree of computational accuracy to be achieved with a reduction of approximately 50\% or more in the number of mesh cells typically employed in the traditional AMR.
The remainder of this paper is organized as follows. In Section~\ref{sec:Residuals-guided AMR framework}, we present the PDE residuals-guided AMR framework, which elucidates the process of solving the flow problems, training the PINNs, predicting the PDEs residuals, marking the mesh cells, and refining the mesh.
The numerical results and discussions are validated by four representative cases,
including incompressible and compressible flows, as presented in Section~\ref{sec:Numerical results and discussions}. The paper concludes in Section~\ref{sec:Conclusions}. 

% ------------------------- 2. Residuals-guided AMR framework ---------------------------
\section{Residuals-guided AMR framework}
\label{sec:Residuals-guided AMR framework}

% --------------------------------- 2.1. Overview ----------------------------------
\subsection{Overview}
\begin{figure}[htb]
    \centering
    \includegraphics[width=0.85\linewidth]{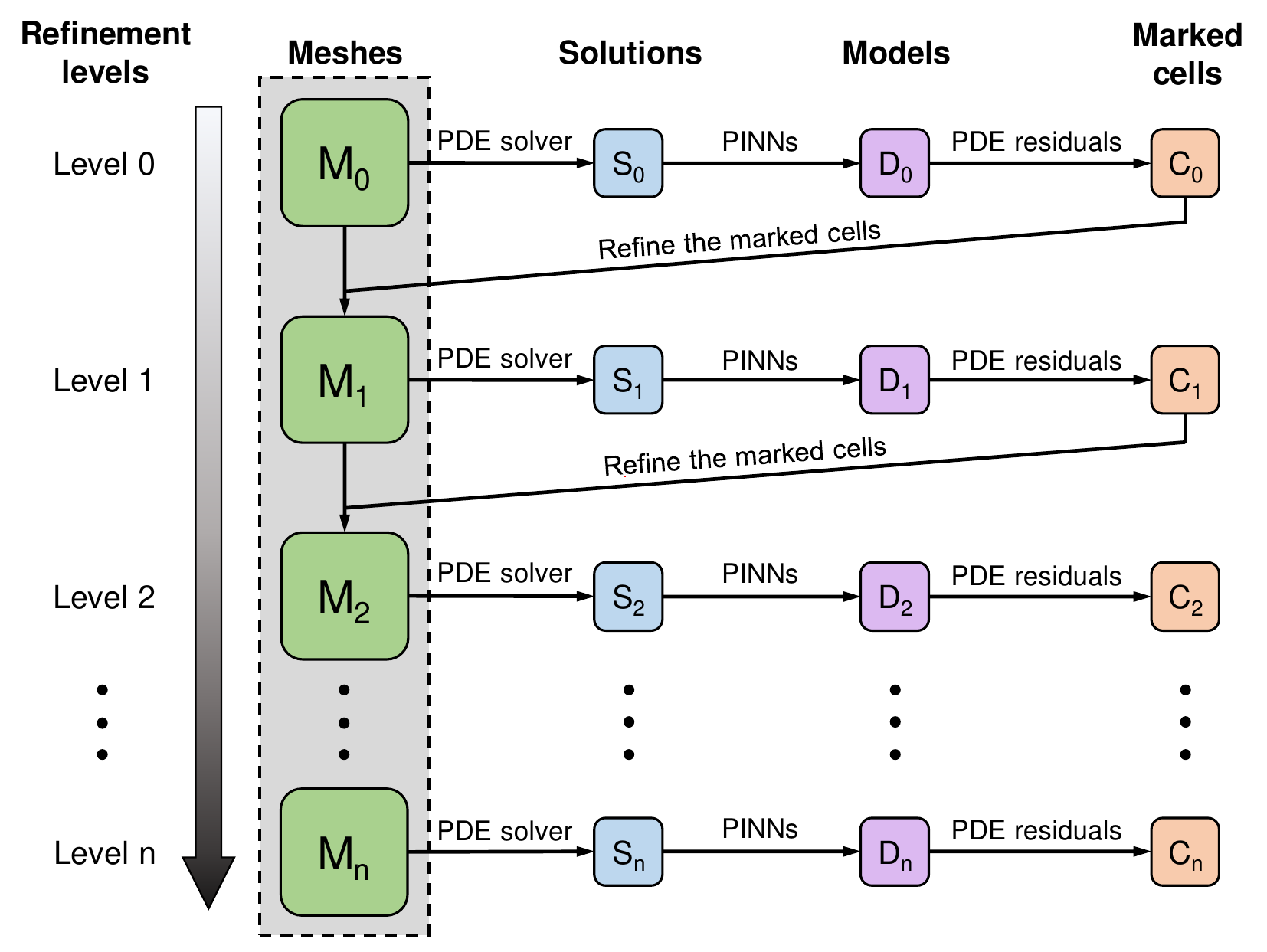}
    \caption{Schematic of the proposed PDE residuals-guided adaptive mesh refinement based on PINNs. It follows the introduced SOLVE$\rightarrow$TRAIN$\rightarrow$ESTIMATE$\rightarrow$MARK$\rightarrow$REFINE loop. Level denotes the degree of mesh refinement, with the initial mesh set at Level 0. $M$, $S$, $D$, and $C$ represent the sets of meshes, solutions, PINNs-trained models, and marked cells, respectively, with their subscripts indicating the refinement level of the mesh.}
    \label{fig:Residuals-guided AMR scheme}
\end{figure}

% 在本章节中，我们简要介绍所提出的残差指导的自适应网格细化技术模拟稳态可压缩与不可压缩流动的过程。所采用的方案遵循AMR的成熟范式，即SOLVE→ESTIMATE→MARK→REFINE循环。利用PINNs集成物理和粗网格的解数据，以PDE残差为准则，来执行非结构网格的自适应细化（见图1流程图）。首先，在SOLVE过程我们对待求解问题生成一个初始的粗网格，并使用传统的CFD求解器对其进行求解，得到粗网格对应的较低精度的流场解。其次，在TRAIN过程中我们利用PINNs整合物理方程和SOLVE过程得到的低精度的解数据，去训练一个满足粗网格对应流场分布的模型，以方便后续进行网格单元上PDE残差的预测和估计。
In this section, we present an overview of the proposed PDEs residuals-based framework for unstructured adaptive mesh refinement in steady flow simulations. 
As we suggested in Section~\ref{sec:Introduction}, the framework leverages physics-informed neural networks~(PINNs) to integrate solution data and physics, identifying mesh cells for refinement according to PDE residuals. 
We integrate PINNs into our AMR scheme prior to error estimation, combining it with the conventional AMR paradigm.  
In addition to the standard SOLVE$\rightarrow$ESTIMATE$\rightarrow$MARK$\rightarrow$REFINE loop~\citep{braess2007convergence, daniel2018adaptive}, we introduce a TRAIN step, forming a modified SOLVE$\rightarrow$TRAIN$\rightarrow$ESTIMATE$\rightarrow$MARK$\rightarrow$REFINE loop. This alteration differentiates our AMR approach from traditional ways.
Fig.~\ref{fig:Residuals-guided AMR scheme} illustrates a schematic of the proposed PDE residuals-guided adaptive mesh refinement based on PINNs. 

Suppose a flow is required to be solved. 
Then, an AMR loop consists of the following main processes.
First, in the SOLVE process, an initial coarse triangular mesh is generated for this flow problem and solved using a classical PDE solver. The resulting flow field solution corresponds to this coarse mesh and may therefore exhibit lower precision compared to solutions on a finer mesh.
During the TRAIN process, we employ PINNs to integrate the governing physical equations and low-resolution solution data obtained from the previous step. Given the pivotal role of the governing PDEs, the trained PINNs model serves as a reliable surrogate for the underlying physics. This allows us to develop a model that reproduces the flow field on the coarse mesh to facilitate the subsequent prediction and estimation of the PDE residuals. 
In the ESTIMATE process, we utilize the trained PINNs model to predict the PDE residuals at the cell centers of the coarse triangular mesh. By analyzing these predicted residuals, we can identify the refinement region where further mesh refinement is necessary. 
In the subsequent MARK stage, we rank the mesh cells according to the PDE residuals at the cell centers. Cells on the coarse mesh with relatively higher PDE residuals are marked for refinement.
For the marked mesh cells, we optionally apply either h-refinement~\citep{guo1986hp, baker1997mesh, venditti2002grid} by edge bisection or Delaunay refinement~\citep{chew1989constrained, shewchuk2002delaunay} by constraining the triangle size and angle, to preserve the overall quality of the refined mesh. 
The iterative AMR loop terminates until the physical quantities of interest converge with the number of mesh cells. A higher level corresponds to a finer mesh. 
The implementation details of each step in the AMR run are presented in the following subsections.

% ------------------- 2.2. Physics equations and finite-volume scheme ------------------
\subsection{Physics equations and finite-volume scheme}
\label{sec:FVM}
The proposed PDE residuals-guided AMR scheme is applied here to solve the Navier-Stokes equations for viscous incompressible flows and the Euler equations for inviscid compressible flows. 
In this study, we focus specifically on steady-state flow problems. For the viscous flow cases, the Reynolds numbers are constrained to ensure laminar flow conditions, eliminating the complexities introduced by turbulence modeling.
For all governing equations, we employ a finite volume discretization scheme and solve them on an unstructured triangular mesh. 
This choice of mesh structure is advantageous for complex geometries, as it allows for flexible cell arrangements and better adaptability to irregular boundaries. 
The finite volume discretization ensures local conservation of mass, momentum, and energy, while the unstructured mesh enables flexible refinement in localized areas of interest, such as shock waves, boundary layers, or regions of high gradients in the flow field.

% ------------------------------ 2.2.1 NS Equations ---------------------------------
\subsubsection{Navier-Stokes equation}
The continuity and momentum equations of the Navier-Stokes equations in two-dimensional Cartesian coordinates ${\bm x}=(x, y)$ are expressed as follows:
\begin{equation}
\label{eq:NS equation}
\begin{split}
    \nabla  \cdot {\bm u}& = 0,
    \\
    \frac{{\partial {\bm u}}}{{\partial t}} + ({\bm u} \cdot \nabla ){\bm u}& =  - \nabla p + \frac{1}{{Re}}{\nabla ^2}{\bm u},
\end{split}
\end{equation}
where $\bm {x}\in \Omega \subset \mathbb{R}^2$, ${\bm u}=\left(u, v\right)$ is the velocity vector of the fluid, and $p$ is the pressure. The Reynolds number, denoted as $Re=UD/v$, is defined by the reference velocity $U$, the characteristic length $D$, and the kinematic viscosity $v$. For the steady-state solution, the time derivative in Eq.~(\ref{eq:NS equation}) is omitted. One or more boundary conditions for velocity and/or pressure are defined as follows:
\begin{eqnarray}
\label{Dirichlet_velocity_for_stationary_boundaries}
{\bm u} = {{\bm u}_\Gamma }({\bm x}),\quad {\bm x} \in {\Gamma _D},
\\
p = {p_\Gamma }({\bm x}),\quad {\bm x} \in {\Gamma _D},
\\
\frac{{\partial {\bm u}}}{{\partial {\bm n}}} = 0,\quad {\bm x} \in {\Gamma _N},
\\
\frac{{\partial p}}{{\partial {\bm n}}} = 0,\quad {\bm x} \in {\Gamma _N},
\end{eqnarray}
where ${\Gamma _D}$ and ${\Gamma _N}$ indicate the Dirichlet and Neumann boundaries, respectively; ${\bm u}_\Gamma$ and $p_\Gamma$ are the corresponding boundary constraints. ${\bm n}$ is the normal vector at the corresponding boundary.

% ------------------------------- 2.2.2 Euler Equations -------------------------------
\subsubsection{Euler equation}
The conservation of mass, momentum, and energy in two-dimensional Euler equations, which govern the compressible flow of inviscid fluids, can be expressed in the following conservative form:
\begin{equation}
\label{eq:Euler}
\begin{split}
    \frac{\partial \bm{U}}{\partial t}+{\nabla} \cdot {\bm{f(U)}}=0,\quad \bm{x}=(x, y)\in \Omega \subset \mathbb{R}^2,
\end{split}
\end{equation}
where $\bm{U}$ is the vector of the conservative quantities and $\bm{f}$ is the solution flux matrix. In two-dimensional Cartesian coordinates, $\bm{U}$ and $\bm{f}$ are given by:
\begin{equation}
    \bm{U}=\left[\begin{array}{c}
    \rho \\
    \rho u \\
    \rho v \\
    \rho E
    \end{array}\right],\quad 
    \bm{f}=(G_1, G_2)
\text {, with } 
    G_1(\bm{U})=\left[\begin{array}{c}
    \rho u \\
    p+\rho u^2 \\
    \rho u v \\
    p u+\rho u E
    \end{array}\right],
    G_2(\bm{U})=\left[\begin{array}{c}
    \rho v \\
    \rho u v \\
    p+\rho v^2 \\
    p v+\rho v E
    \end{array}\right] \text {, }
\end{equation}
where $\left(u, v\right)$ are the velocity components, $p$ is the pressure, $\rho$ is the density, and $E$ is the total specific energy. To close the aforementioned Euler equations, we employ an additional equation of state that characterizes the relationship between pressure $p$ and energy $E$. Specifically, we consider the equation of state for an ideal gas, which is expressed as follows:
\begin{equation}
p=(\gamma-1)\left(\rho E-\frac{1}{2} \rho\|\boldsymbol{u}\|^2\right),
\end{equation}
where $\gamma$ = 1.4 is the specific-heat ratio of air and $\bm{u}= (u,v)$. The Mach number $M=\|\bm{u}\|/a$ is the ratio of the flow velocity to the sound speed $a=\sqrt{\gamma p / \rho}$. 

The steady-state solution is often of interest in aerodynamic scenarios. To obtain this solution, the time derivative in Eq.~(\ref{eq:Euler}) is omitted, reducing the Euler equations to their steady form. The physical properties of air are used in solving the Euler equations via the finite volume scheme.
Flow properties are made dimensionless by the speed of sound $a_0$, pressure $p_0$, and density $\rho_0$ at the stagnation, as follows: 
\begin{equation}
    u^{\prime}=\frac{u}{a_0}, v^{\prime}=\frac{v}{a_0}, p^{\prime}=\frac{p}{p_0}, \rho^{\prime}=\frac{\rho}{\rho_0}.
\end{equation}
These stagnation values satisfy $a_0=\sqrt{\gamma p_0 / \rho_0}$ or $\rho_0={\gamma p_0 / a_0^2}$.
% , which is equivalent to $\rho_0={\gamma p_0 / a_0^2}$.
Furthermore, the steady Euler equations become dimensionless as (we remove all prime symbols $^\prime$ in the equations for convenience):
\begin{equation}
\label{eq:Euler2}
\begin{split}
    {\nabla} \cdot {\bm{f(U)}}=0,\quad \bm{x}=(x, y)\in \Omega \subset \mathbb{R}^2,
\end{split}
\end{equation}
where
\begin{equation}
    \bm{f}=(G_1, G_2)
\text {, with } 
    G_1(\bm{U})=\left[\begin{array}{c}
    \rho u \\
    p+\gamma \rho u^2 \\
    \gamma\rho u v \\
    \rho u E
    \end{array}\right],
    G_2(\bm{U})=\left[\begin{array}{c}
    \rho v \\
    \gamma \rho u v \\
    p+\gamma \rho v^2 \\
    \rho v E
    \end{array}\right] \text {. }
\end{equation}
Note that we use the above dimensionless form of the Euler equations~(Eq.\ref{eq:Euler2}) for both training and prediction in the physics-informed neural networks.

% -------------------------- 2.2.3 finite-volume scheme ----------------------------
\subsubsection{Finite-volume scheme}
The discretization of the governing equations applied herein can be clearly illustrated by considering the steady conservation equation for the transport of a scalar quantity $\varphi$. The governing equation can be expressed in integral form for an arbitrary control volume as follows:
\begin{equation}
\label{eq:integral}
\oint \rho \varphi \vec{V} \cdot d \vec{A}=\oint \Gamma_{\varphi} \nabla \varphi \cdot d \vec{A}+\int_V S_{\varphi} d V
\end{equation}
where $\rho$ is the density, $\vec{V}$ is the velocity vector, $\vec{A}$ is the surface area vector, $\Gamma_{\varphi}$ is the diffusion coefficient for $\varphi$, $\nabla \varphi$ is the gradient of $\varphi$, and $S_{\varphi}$ is the source of per unit volume.
Eq.~\ref{eq:integral} is applied to each control volume, or cell, in the computational domain, such as a two-dimensional triangular cell. Discretizing Eq.~\ref{eq:integral} for a given cell results in:
\begin{equation}
\sum_f^{N_{\text {faces }}} \rho_f \vec{V}_f \varphi_f \cdot \vec{A}_f=\sum_f^{N_{\text {faces }}} \Gamma_{\varphi} \nabla \varphi_f \cdot \vec{A}_f+S_{\varphi} V
\end{equation}
where $N_{\text {faces }}$ denotes the number of faces enclosing the cell, $\varphi_f$ represents the value of $\varphi$ convected through face $f$, $\rho_f \vec{V}_f \cdot \vec{A}_f$ indicates the mass flux through the face, $\vec{A}_f$ is the area vector of face $f$, $\nabla \varphi_f$ is the gradient of $\varphi$ at face $f$, and $V$ is the volume of the cell.

We employ a pressure-based coupling algorithm~\citep{chorin1968numerical} to solve the aforementioned coupled system of equations, which comprises the momentum equation and the pressure-based continuity equation. 
High-order interpolation methods are utilized to calculate the face flux, $\hat{J}_f$, and a pressure gradient correction is applied following a distance-based Rhie-Chow type linear interpolation~\citep{rhie1983numerical}. 
The discrete values of the scalar are stored at the centers of the triangular cells. 
The gradients at the faces, necessary for the equation, are obtained using a least squares cell-based evaluation.
Density, momentum, and energy values at the triangular faces are interpolated from the cell center values using a second-order upwind finite-volume discretization scheme. 
When second-order upwinding is employed, the face value is computed using the expression $\varphi_{f}=\varphi+\nabla \varphi \cdot \vec{r}$, where $\varphi$ and $\nabla \varphi$ are the cell-centered value and its gradient in the upstream cell, and $\vec{r}$ is the displacement vector from the upstream cell centroid to the face centroid.

% ----------------------- 2.3. Physics-informed neural networks ----------------------
\subsection{Physics-informed neural networks}
Here, we introduce physics-informed neural networks~(PINNs) for estimating PDE residuals due to two primary reasons. 
As discussed in Section~\ref{sec:Introduction}, most CFD solvers provide solutions only at the mesh nodes, lacking essential information like first- and second-order derivatives of the solution, which are critical for estimating PDE residuals. 
Typically, deriving these derivatives requires complex and time-consuming interpolation across the mesh. 
By employing PINNs to integrate the solutions data and the physics, the required derivatives can be easily and directly computed by auto-differentiation (AD)~\cite{rall1981automatic}, streamlining the PDE residuals estimation process. This approach leads to a physical and smoother flow field surrogate. 
Consequently, the residuals can be swiftly predicted at any point within the flow field.
In this section, we first provide a brief overview of PINNs for integrating PDEs and data and then present each loss function and technical details. Next, we discuss how to use the PINNs to execute the TRAIN process of the residuals-guided AMR framework to refine the mesh.

% ---------------------- 2.3.1 PINNs for integrating PDEs and data -----------------------
\subsubsection{PINNs for integrating physics and data}
\label{sec:PINNs}
We consider the following parametric PDEs defined on a domain $\Omega$ for the solution $s(\bm{x})$, expressed in a general form:
\begin{equation}
\label{eq:pde}
f\left( {{\bm x};\frac{{\partial s}}{{\partial {x_1}}}, \ldots ,\frac{{\partial s}}{{\partial {x_d}}};\frac{{{\partial ^2}s}}{{\partial {x_1}\partial {x_1}}}, \ldots ,\frac{{{\partial ^2}s}}{{\partial {x_1}\partial {x_d}}}; \ldots ; {\lambda}} \right) = 0, \quad {\bm x} \in \Omega \subset \mathbb{R}^d,
\end{equation}
with the boundary conditions (In PINNs, initial conditions can be treated as boundary conditions):
\begin{eqnarray}
&& {\cal B}(s,{\bm x}) = 0,\quad {\bm x} \in \partial \Omega,
\end{eqnarray}
where $f$ denotes a nonlinear differential operator, ${\bm x}=\left[x_1, x_2, \ldots, x_d\right]$ are independent variables, and $\lambda=\left[\lambda_1, \lambda_2, \ldots\right]$ are the parameters for combining each component.

To solve the Eq.~(\ref{eq:pde}), a fully connected neural network (FCNN) with trainable parameters $\boldsymbol{\theta}$ is generally constructed as a highly nonlinear function $\hat s({\bm x;\bm{\theta}})$ to approximate the solution $s(\bm{x})$ of the PDEs~\citep{raissi2019physics}. The FCNN consists of an input layer, multiple hidden layers, and an output layer. Each layer contains several neurons with associated weights $\bm{w}$, biases $\bm{b}$, and a non-linear activation function $\sigma(\cdot)$. Let ${\bm x}$ represent the input, and ${\bm y}^i$ denote the implicit variable of the $i$th hidden layer. Consequently, a FCNN with $L$ layers can be mathematically expressed as:
\begin{equation}
\left\{
\begin{aligned}
{\bm y}^0& = {\bm x}, \\
{\bm y}^i& = \sigma\left(\bm{w}^i {\bm y}^{i-1}+\bm{b}^i\right), \quad 1 \leq i \leq L-1, \\
{\bm y}^i& = \bm{w}^i {\bm y}^{i-1}+\bm{b}^i, \quad i=L.
\end{aligned}
\right.
\end{equation}

A major advantage of PINNs is that the same formulation of loss functions can be used not only for labeled data but also for known PDEs.
Furthermore, we define a composite loss function as the sum of the residuals of the PDEs, boundary conditions, and labeled data:
\begin{equation}
{\cal L}({\bm{\theta};\cal T}) = {w_f}{{\cal L}_f}\left({\bm{\theta};\cal T}_f \right) + {w_b}{{\cal L}_b}\left({\bm{\theta};\cal T}_b \right) + {w_{d}}{{\cal L}_{d}}\left({\bm{\theta};\cal T}_{d} \right),
\label{eq:totalloss}
\end{equation}
with
\begin{eqnarray}
&&\mathcal{L}_f\left(\bm{\theta} ; \mathcal{T}_f\right)=\frac{1}{\left|\mathcal{T}_f\right|} \sum_{\bm{x} \in \mathcal{T}_f}\left|f\left(\bm{x} ; \frac{\partial \hat{s}}{\partial x_1}, \ldots, \frac{\partial \hat{s}}{\partial x_d} ; \frac{\partial^2 \hat{s}}{\partial x_1 \partial x_1}, \ldots, \frac{\partial^2 \hat{s}}{\partial x_1 \partial x_d} ; \ldots ; \lambda\right)\right|^2, 
\\
&&\mathcal{L}_b\left(\bm{\theta} ; \mathcal{T}_b\right)=\frac{1}{\left|\mathcal{T}_b\right|} \sum_{\bm{x} \in \mathcal{T}_b}|\mathcal{B}(\hat{s}, \bm{x})|^2,
\\
&&\mathcal{L}_{d}\left(\bm{\theta}; \mathcal{T}_{d}\right)=\frac{1}{\left|\mathcal{T}_{d}\right|} \sum_{\mathbf{x} \in \mathcal{T}_{d}}|\hat{s}(\bm{x})-s(\bm{x})|^2
\end{eqnarray}
where ${\cal T}_f$, ${\cal T}_b$ and ${\cal T}_{d}$ are the corresponding sets of training points, respectively. $w_f$, $w_b$ and $w_{d}$ are the weights. The neural network is trained by optimizing the weights and biases with optimizers such as Adam~\citep{kingma2014adam} or L-BFGS~\citep{liu1989limited} to minimize the loss functions. 

% ---------------------- 2.3.2 Technical details for training PINNs -----------------------
\subsubsection{Technical details for training PINNs}
\begin{figure}[htb]
    \centering
    \includegraphics[width=1.0\linewidth]{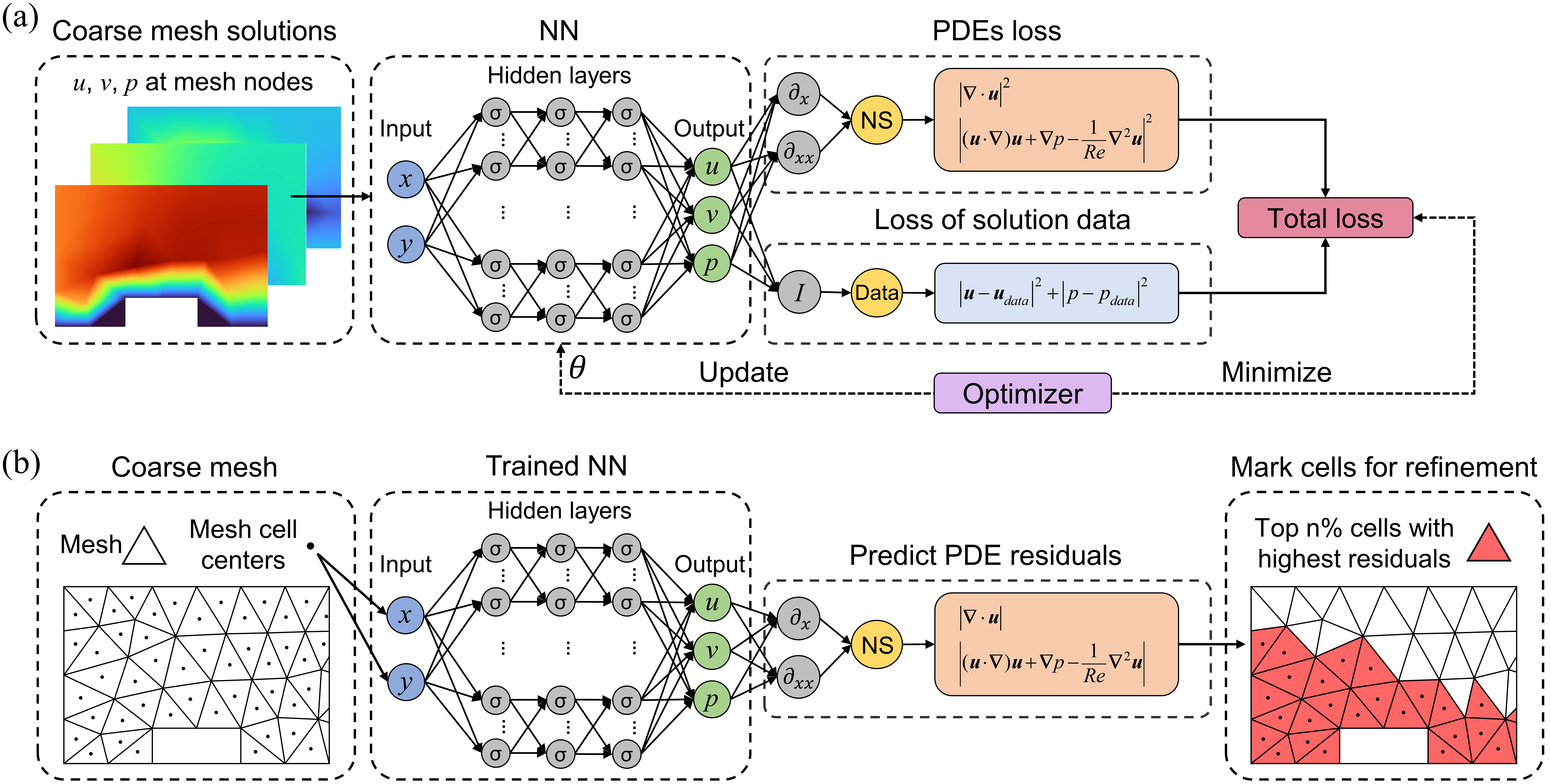}
    \caption{Schematic of the detailed TRAIN$\rightarrow$ESTIMATE$\rightarrow$MARK process of residual-based AMR framework for solving NS equations. (a) TRAIN: Training the PINNs by combining the NS equations with less precise flow field data obtained from coarse mesh. Training points correspond to mesh nodes. The loss functions employed in this approach comprise two components: a physical equation part and a coarse mesh solution data part. (b) ESTIMATE and MARK: The trained model predicts the PDE residuals at the center points of coarse mesh cells. Each mesh cell centroid represents that mesh cell. The mesh cells with larger residuals are marked for a subsequently refinement.}
    \label{fig:Train_estimate_mark}
\end{figure}

We shall introduce the technical details of PINNs' architecture and elaborate each component of the composite loss function. 
Based on the principles outlined in Section~\ref{sec:PINNs}, the neural networks take spatial coordinates $(x,y)$ as inputs and output $(u,v,p)$ to predict the residuals of NS equations, trained with mesh solutions comprising velocity and pressure data. 
Similarly, we take spatial coordinates $(x,y)$ as inputs and $(u,v,p,\rho)$ as outputs of the neural networks for the Euler equations. 
Fig.~\ref{fig:Train_estimate_mark} illustrates a schematic of the detailed TRAIN$\rightarrow$ESTIMATE$\rightarrow$MARK process of the residuals-guided AMR framework for solving NS equations. The derivatives of $u$, $v$, and $p$ with respect to the inputs are calculated using the AD. 
The overall loss function for PINNs is composed of two components: the PDEs loss and the solution data loss, as represented below:
\begin{equation}
{\cal L}({\bf{\theta }}) = {w_f}{{\cal L}_f}({\bf{\theta }}) + {w_{d}}{{\cal L}_{d}}({\bf{\theta }}).
\label{eq:totalloss2}
\end{equation}
The difference from Eq.~(\ref{eq:totalloss}) is that we remove the boundary loss $w_{b}{\cal L}_b$. Boundary conditions are satisfied and included in the solutions, thus we do not consider the extra boundary loss when constructing the loss function of the neural network.
Training with the physical equations allows us to restore the derivatives of the solutions at the mesh nodes, making the fitted flow field more physical and smoother. 
% Moreover, we adopt the Glorot scheme to randomly initialize all weights and biases denoted by $\theta$~\citep{glorot2010understanding}, which are to be optimized.
Taking the NS equation as a demonstration, the definitions for all components in Eq.~(\ref{eq:totalloss2}) are expressed in mean squared errors~(MSE) as:
\begin{eqnarray}
&&{{\cal L}_f} = \frac{1}{{{N_f}}}\sum\limits_{{\bm x} \in \Omega } {\left( {\left| {\nabla \cdot {\bm u}} \right|^2} + {\left| {({\bm u} \cdot \nabla ){\bm u} + \nabla p - \frac{1}{{Re}}{\nabla ^2}{\bm u}} \right|^2} \right)}, 
\\
&&{{\cal L}_{d}} = \frac{1}{{N_{d}}}\sum\limits_{n = 1}^{N_{d}} {\left( {{{\left| {{\bm u}({\bm x}_u^n) - {\bm u}_{d}^n} \right|}^2}} + {{\left| {p({\bm x}_p^n) - p_{d}^n} \right|}^2} \right)},
\end{eqnarray}
where $N_f$ and $N_{d}$ denote the number of training points corresponding to each loss term. The network architecture and each component of the loss function are illustrated in Fig.~\ref{fig:Train_estimate_mark}(a) for a clear overview. Note that we use only mesh nodes as training points for PINNs. Thus, both $N_f$ and $N_{d}$ correspond to the number of mesh nodes.

The number of hidden layers and neurons within each layer of the neural network is typically selected to align with the specific requirements of the problem being addressed.
Following an iterative process of trial-and-error, we universally employ a network architecture consisting of 6 hidden layers, with each layer comprising 40 neurons.
The differentiable and continuous hyperbolic tangent function \textit{tanh} is chosen as the activation function $\sigma(\cdot)$~\citep{lecun2015deep}.
The total loss function is minimized by the Adam optimizer~\citep{kingma2014adam} followed by the L-BFGS optimizer~\citep{liu1989limited}. 
The Adam optimizer works with a decreasing learning rate schedule: $10^{-3}$ for the first 20,000 epochs, $5\times10^{-4}$ for the next 20,000 epochs, and $10^{-4}$ for the final 20,000 epochs. The L-BFGS optimizer follows to further diminish the residuals.
% We obtain approximate solutions when the loss function reaches a sufficiently low level. 
% To evaluate the accuracy of the prediction, we select the relative $L_2$ error as a metric:
% \begin{equation}
% {\varepsilon _V} = \frac{{{{\left\| {V - {V^*}} \right\|}_2}}}{{{{\left\| {{V^*}} \right\|}_2}}},
% \end{equation}
% where $V$ represents the predicted solutions $(u,v,p)$, and $V^*$ denotes the corresponding reference solutions.

% ------------------ 2.4. Residuals estimation and mesh cell marking -----------------
\subsection{Error estimation and mesh cell marking}
\label{sec:Error estimation and mesh cell marking}
The ESTIMATE process calculates the error for each cell of the coarse mesh by predicting the PDE residuals at the cell centers using the trained PINNs model. The total residual ${\cal {R}}_f$ at each cell center is the sum of the residual norms for each equation. 
Taking the NS equations for example, the total residual for each cell can be expressed as:
\begin{eqnarray}
{{\cal {R}}_f} = {\left| {\nabla \cdot {\bm u}} \right|} + {\left| {\frac{{\partial {\bm u}}}{{\partial t}} + ({\bm u} \cdot \nabla ){\bm u} + \nabla p - \frac{1}{{Re}}{\nabla ^2}{\bm u}} \right|}. 
\end{eqnarray}
Note that the absolute values of these residuals are less meaningful; rather, it is the relative magnitudes of these values that indicate which cells require refinement. 
Therefore, we find it unnecessary to scale or normalize the residual values of each equation based on their magnitude or any other metric.

The MARK process selects cells for refinement based on error estimates from the PDE residuals-based indicator. Each mesh cell is represented by its center, and a higher residual at the center indicates a greater error for that cell.
A fixed-fraction strategy~\citep{morin2000data, braess2007convergence} is selected as our cells-refinement rule, which refines a fixed percentage of the total number of cells and controls readily the number of cells in the mesh. Specifically, we rank the cells by their center's PDE residuals and mark the top n\% of the ranked cells with the highest residuals for subsequent refinements. 
The detailed ESTIMATE$\rightarrow$MARK schematic process is presented in Fig.~\ref{fig:Train_estimate_mark}(b) for a clear overview. 
For the test cases in Section~\ref{sec:Numerical results and discussions}, we benchmark the PDE residuals-based indicator against two commonly used AMR error indicators: the gradient indicator and the pressure Hessian indicator. To ensure a fair comparison, we apply a fixed-fraction strategy for AMR execution with both traditional indicators. 
However, if the fixed-fraction strategy fails in some cases, we will consider adopting a fixed-threshold strategy~\citep{morin2000data, braess2007convergence}, where cells with gradient or pressure Hessian values above a specified threshold are marked for refinement.

% ----------------------------- 2.5. Mesh refinement ---------------------------------
\subsection{Mesh refinement}
Delaunay triangulation~\citep{chew1989constrained, shewchuk1996triangle, shewchuk2002delaunay} plays a crucial role in this process, as it reconstructs an optimal triangular mesh by maximizing the minimum angle of the triangles, thereby avoiding poorly shaped, skinny elements. This property ensures improved element quality and numerical stability, making Delaunay triangulation particularly advantageous in finite element mesh generation and refinement. Moreover, Delaunay triangulation can effectively redistribute the mesh connectivity to align with newly inserted nodes, providing a high-quality triangular mesh for further refinement and computation.

\begin{figure}[htb!]
    \centering
    \includegraphics[width=1.0\linewidth]{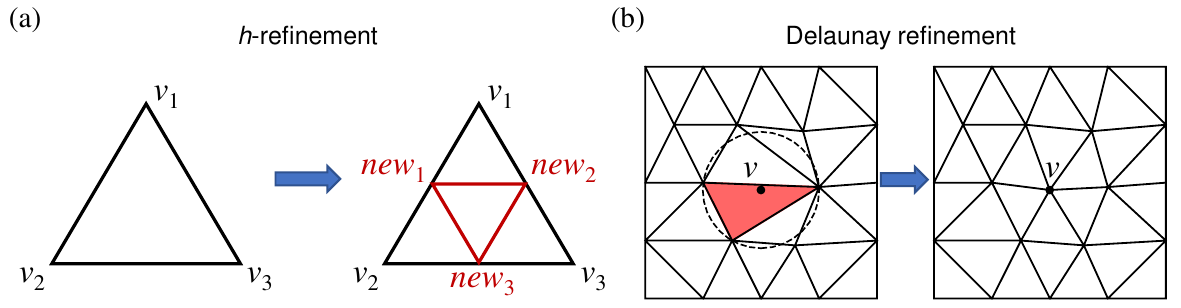}
    \caption{Two refinement algorithms applied to the triangular mesh. (a) h-refinement: It reduces the characteristic length of elements, effectively subdividing each cell into four or more smaller cells without changing their type. we apply edge bisection to a coarse mesh cell and subdivide it into four sub-cells. (b) Delaunay refinement: A triangular cell to be refined is subdivided by inserting a vertex $v$ at its circumcenter. Since no vertices lie inside the circumcircle of the marked triangle before $v$'s insertion, no new edge can be shorter than the circumradius. The insertion of vertex $v$ creates new edges that are solely connected to $v$.}
    \label{fig:mesh_refinements}
\end{figure}

To refine the marked coarse mesh while preserving the quality of the refined mesh, we optionally employ two strategies that incorporate Delaunay triangulation: h-refinement and Delaunay refinement.
The h-refinement method reduces the characteristic length of elements, effectively subdividing each cell into four or more smaller cells without changing their type. As shown in Fig.~\ref{fig:mesh_refinements}(a), we perform h-refinement by bisecting the edges of a coarse mesh cell, subdividing it into four sub-cells.
The refined mesh is then obtained by inserting the midpoints of the edges into the coarse mesh and conducting Delaunay triangulation. 

Delaunay refinement algorithm~\citep{shewchuk2002delaunay} adds Steiner points to the input geometry, ensuring that the resulting Delaunay or constrained Delaunay triangulation satisfies the desired size and quality standards for the mesh. 
A visual representation of the Delaunay refinement process is shown in Fig.~\ref{fig:mesh_refinements}(b). 
The new mesh is further refined by inserting vertices at the circumcenters of marked cells and conducting Delaunay triangulation until the resulting cells satisfy the specified size and minimum angle constraints.
In this study, the size of each sub-cell in the refined mesh is uniformly constrained to no more than one-fourth of the parent cell size, and the minimum angle of the triangles is maintained at no less than 20 degrees.
In certain cases, the number of mesh cells refined by Delaunay refinement may differ slightly from those refined by h-refinement. This discrepancy arises because Delaunay refinement does not strictly reduce the sub-cells to exactly one-fourth of the parent cell size; instead, it ensures they are no more than one-fourth. Additionally, adhering to the minimum angle constraint may result in an increased number of mesh cells.

% ------------------------ 2.6. Numerical implementation -----------------------------
\subsection{Numerical implementation}
Incorporating the key stages described in the previous section, we summarize the complete procedure for the PDE residuals-guided unstructured adaptive mesh refinement framework for steady flows, as outlined in Algorithm~\ref{alg:AMR}.
The training of PINNs is implemented using the DeepXDE library~\citep{lu2021deepxde}, which employs TensorFlow~\citep{abadi2016tensorflow} as its backend. Mesh refinement is performed in Python using the Triangle library~\citep{shewchuk1996triangle, shewchuk2002delaunay}, a two-dimensional quality mesh generator and Delaunay triangulator.
The generated mesh topology dictionary contains information about the nodes, edges, cells, boundaries, and other related components.
Using this essential topological information, the mesh can be formatted and saved as a mesh file compatible with the FVM solver.

\begin{algorithm}
\caption{PDEs residuals-guided unstructured adaptive mesh refinement~(AMR) framework}\label{alg:AMR}
\begin{algorithmic}
\State ${\bm M_0} \gets$ \text{Generate an initial coarse triangular mesh} 
\For{${Level = 0} \text{, ...}$}
    \State ${\bm S_{Level}} \gets$ Solutions~(e.g., $u, v, p$) on the mesh ${\bm M_{Level}}$ obtained from the numerical solver \Comment{SOLVE}
    \State ${\bm D_{Level}} \gets$ Train PINNs with solution data ${\bm S_{Level}}$ and physical equations \Comment{TRAIN}
    \State ${{\cal {R}}_f} \gets$ Predict the PDE residuals at the cell centers of ${\bm M_{Level}}$ using the model ${\bm D_{Level}}$
     \Comment{ESTIMATE}
    \State ${\bm C_{Level}} \gets$ Rank the mesh cells according to the residuals ${{\cal {R}}_f}$ at the cell centers and mark the top $n\%$ of the ranked mesh cells with the highest residuals \Comment{MARK}
    \State ${\bm M_{Level+1}} \gets$ Refine the marked cells ${\bm C_{Level}}$ using either h- or Delaunay refinement \Comment{REFINE}
\EndFor
\State \text{Run AMR loops until the key physical metrics converge}  % After the loop
\end{algorithmic}
\end{algorithm}

% -------------------- 3. Numerical results and discussion ----------------------
\section{Numerical results and discussions}
\label{sec:Numerical results and discussions}
% 本段含义：在本节中，我们采用PDE残差引导的AMR框架来解决四个经典流动问题，这些问题涉及钝体的不可压缩或可压缩绕流。此外，我们将基于PDE残差的指标与两种常见的AMR指标(即基于梯度的指标和基于压力梯度的指标)进行了基准测试。我们利用这些误差指标来指导网格细化，并进行网格灵敏度分析，比较关键物理量的收敛性。最优网格被定义为用最少的单元数达到所需精度的配置。最优网格中的单元数是评估误差指标在AMR中有效性的关键指标。
In this section, we apply the PDE residuals-guided AMR framework to solve four classical steady flow problems, including incompressible and compressible flows over blunt bodies. To evaluate its performance, we benchmark the PDE residuals-based indicator against two commonly used AMR indicators: the gradient indicator and the pressure Hessian indicator. These error indicators are used to guide mesh refinement, and mesh sensitivity analyses are conducted to compare the convergence of key physical quantities. The optimal mesh is defined as the configuration that achieves the desired accuracy with the minimum number of cells, with the cell count of the optimal mesh serving as a critical metric for assessing the effectiveness of the error indicator in AMR.

% ------------------ Case1: laminar flow around an obstacle ---------------------
\subsection{Laminar flow around an obstacle}
\label{subsec:laminar flow around an obstacle}
\begin{figure}[htb!]
	\centering  
	\subfigure[Geometry of the computational domain and boundary conditions]{
		\includegraphics[width=0.9\linewidth]{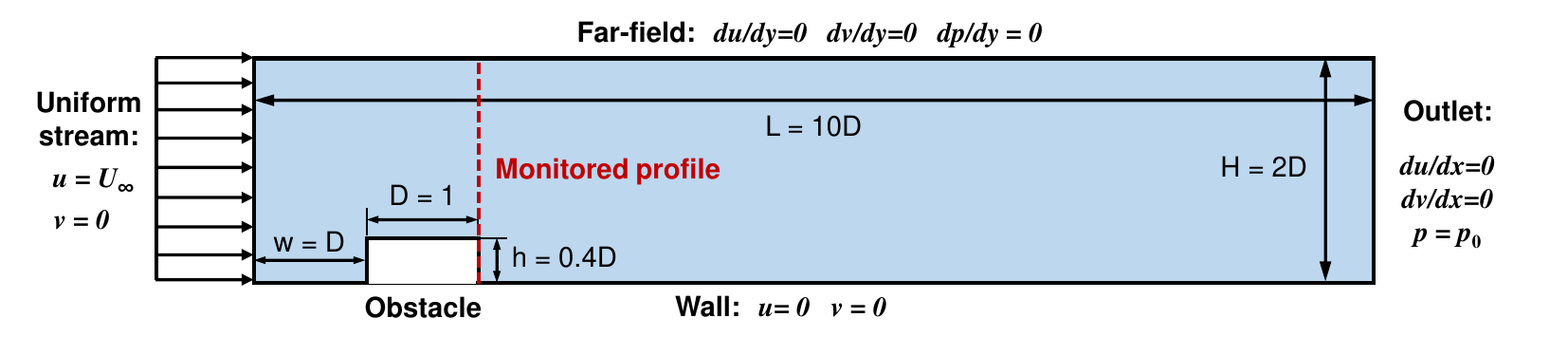}}%
    \\  
	\subfigure[Initial triangular mesh]{
		\includegraphics[width=0.9\linewidth]{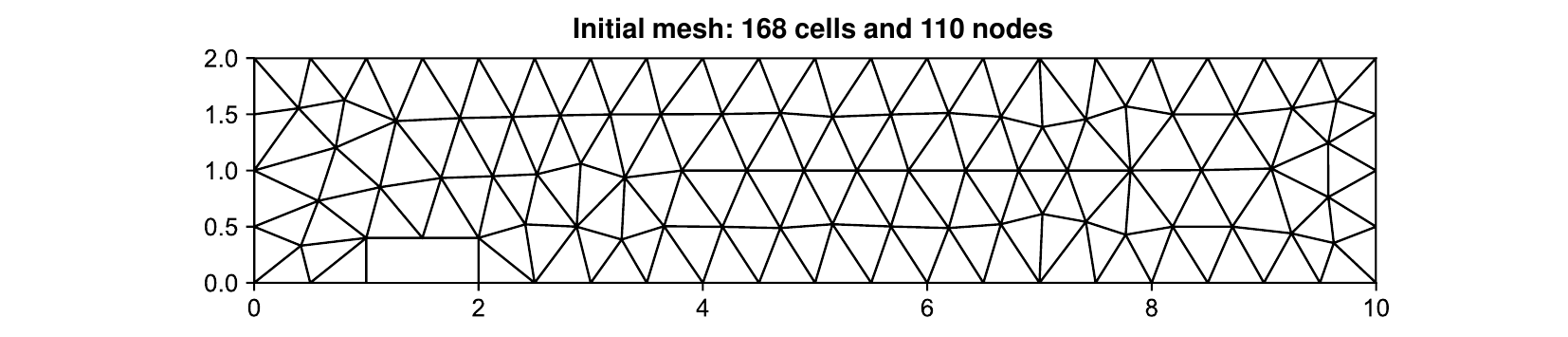}}%
    \\  
	\subfigure[Velocity distribution from the initial mesh]{
		\includegraphics[width=0.9\linewidth]{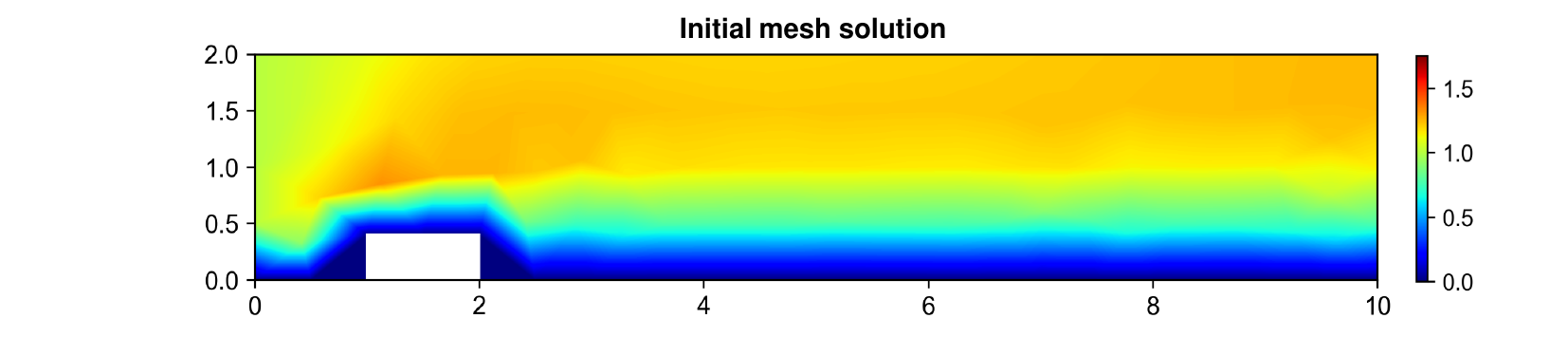}}%
	\caption{Problem setup and initial coarse mesh for laminar flow around an obstacle~(Re = 100). The red dashed line signifies the profile of interest, with its velocity amplitude and distribution serving as the monitoring quantities for our AMR procedure.}
	\label{fig:obstacle_setup}
\end{figure}
% Flow around a rectangular obstacle is examined and provides a multi-faceted evaluation of the proposed AMR framework. 
We first apply the proposed residuals-guided AMR framework to solve the Navier-Stokes equations, examining flow around a rectangular obstacle to provide a comprehensive evaluation of the framework.
For proof-of-concept, we consider the steady laminar flow at a Reynolds number of $Re = {{U_{\infty}}D}/{\nu}=100$, where $U_{\infty}$ denotes the velocity of the uniform stream, $D$ is the length of the rectangular obstacle, and $\nu$ represents the kinematic viscosity of the fluid. 
At the inlet boundary, a uniform horizontal free stream is imposed with conditions $u=U_{\infty}$ and $v=0$. A constant pressure condition and Neumann velocity conditions are applied at the outlet boundary. Neumann conditions for velocity and pressure are also enforced at the far-field boundaries on the top side. No-slip conditions are specified along the obstacle's surface and the bottom wall. 
The geometry and boundary conditions are illustrated in Fig.~\ref{fig:obstacle_setup}(a).
% Velocity distribution and amplitude of a monitored profile behind the obstacle are considered as physical quantities of interest in our AMR procedure, as depicted by the red dashed line in Fig.~\ref{fig:obstacle_setup}(a). 
In the AMR procedure, we monitor the velocity distribution and the amplitude of a profile located downstream of the obstacle, indicated by the red dashed line in Fig.~\ref{fig:obstacle_setup}(a). 
And the initial coarse mesh for this case consists of 168 cells and 110 nodes, as shown in Fig.~\ref{fig:obstacle_setup}(b). 
The velocity distribution obtained from this coarse mesh is presented in Fig.\ref{fig:obstacle_setup}(c).
% We utilize the computational solution obtained from a fine triangular mesh, comprising 46,736 cells and 23,782 nodes, generated from scratch, as the reference.
For reference, we use the computational solution obtained from a fine triangular mesh consisting of 46,736 cells and 23,782 nodes, generated independently. 
We consider the relative $L_2$ error as a metric for assessing the convergence of the velocity distribution in this case.
And the optimal mesh is selected to achieve a relative $L_2$ error of less than 1\% for both the velocity amplitude and its distribution, compared to the reference solution.

Initially, three traditional refinement strategies are implemented for comparison: AMR based on velocity-gradient and pressure-Hessian indicators, as well as global refinement. All these traditional indicators are combined with h-refinement to refine the mesh.
% 这里首先提一下不同百分比的速度梯度准则失败了
We examine the convergence results for the velocity gradient-based AMR indicator in Fig.~\ref{fig:obstacle_velocity_top_n_per}, where different fixed percentages~(i.e., top 40\%, 50\%, and 60\%) of cells marked are applied. However, the number of cells in the optimal mesh obtained using this indicator, which meets the desired accuracy, exceeds that of the globally refined optimal mesh~(43,008 cells). 
% 然后使用了不同固定阈值的策略的速度梯度准则, 但也仅仅>0.01才奏效
Therefore, we consider the velocity gradient indicator combined with the fixed-fraction strategy to be ineffective for this case. We then test it with the fixed-threshold strategy. The convergence results for this strategy are presented in Fig.~\ref{fig:obstacle_velocity_threshold}, where it is observed that only the AMR run with cells marked by a velocity gradient greater than 0.01 leads to a velocity distribution and amplitude that ultimately converge to the desired accuracy. 
The number of cells of the optimal mesh for each corresponding AMR run that meets the desired accuracy is shown near the respective symbol. 
The optimal mesh in this case contains 24,633 cells.
% 这里再提一下不同百分比的压力Hessian准则的收敛结果
Fig.~\ref{fig:obstacle_pressure_top_n_per} shows the convergence results during pressure Hessian-based AMR runs with different fixed percentages~(i.e., top 30\%, 40\%, and 50\%) of cells marked. Among these, the AMR runs with refinement of the top 40\% and 50\% of mesh cells converge to the desired accuracy, with the optimal mesh for the 50\% case containing the fewest cells, totaling 16,914.
% 传统AMR的比较
\begin{figure}[h!]
    \centering
    \subfigure[Velocity amplitude convergence]{
		\includegraphics[width=0.49\linewidth]{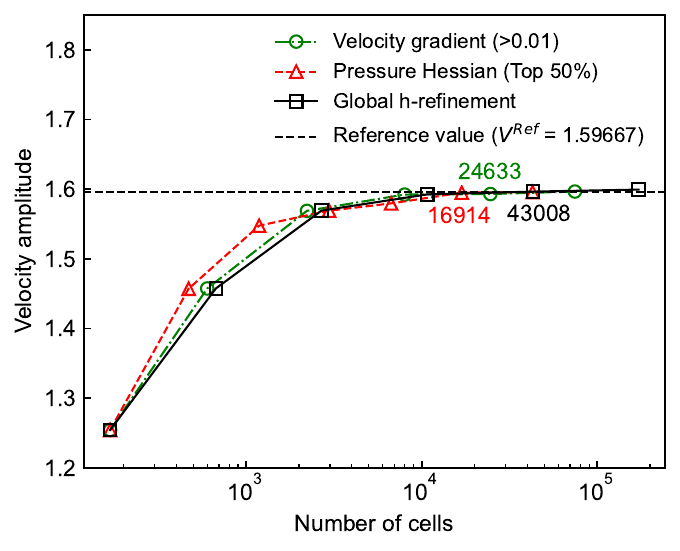}}%  
	\subfigure[Velocity distribution convergence]{
		\includegraphics[width=0.5\linewidth]{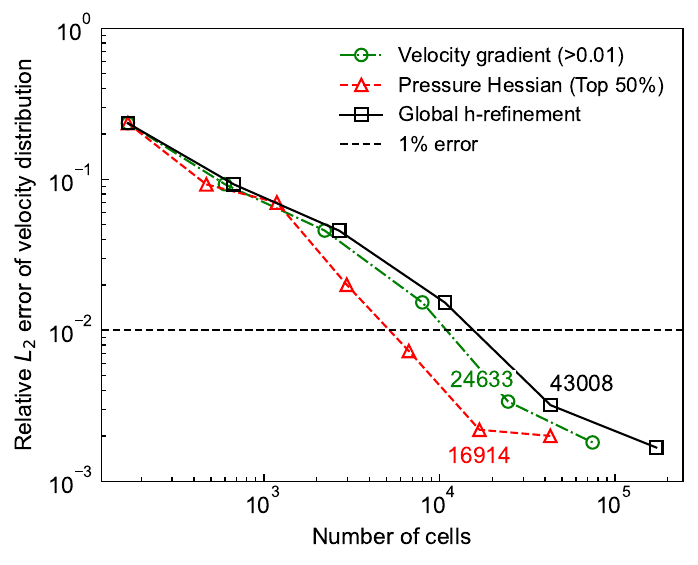}}%
    \caption{Laminar flow around an obstacle. 
    Convergence of the velocity amplitude and distribution along the monitored line during the optimal AMR runs based on two traditional indicators~(Velocity gradient and Pressure Hessian) and the global refinement approach. The relative $L_2$ error is considered as a metric for assessing the convergence of the velocity distribution. The number of cells of the optimal mesh for each corresponding AMR run that meets the desired accuracy is shown near the respective symbol.}
    \label{fig:obstacle_trad_velocity}
\end{figure}
Fig.~\ref{fig:obstacle_trad_velocity} shows a comparison of the convergence results during the optimal AMR runs based on two traditional AMR indicators and the global refinement approach. 
The velocity amplitudes of the monitored line obtained from all three traditional refinement schemes converge to the reference value, and the relative $L_2$ error of the velocity distribution can be reduced to less than 1\%. 
%% 大概总结一下他们的加密等级和最优网格数量
Besides, it is observed that both velocity gradient-based AMR~(\textgreater 0.01) and global refinement converge to a reference solution close to the fine mesh after 4 levels of refinement, while pressure Hessian-based AMR~(Top 50\%) approaches the reference solution after 5 levels of refinement. 
% Despite the relatively fewer cells (12858) of the optimal mesh obtained through the pressure Hessian-based indicator, it does not necessarily imply its superiority, as a higher number of refinement iterations also entails increased computational and time costs.

\begin{figure}[htb!]
    \centering
    \subfigure[Velocity amplitude convergence (Delaunay refinement)]{
		\includegraphics[width=0.49\linewidth]{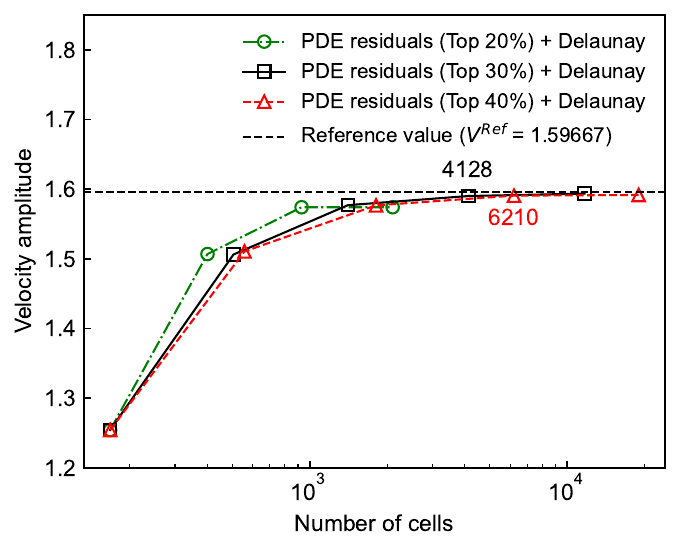}}%  
	\subfigure[Velocity distribution convergence (Delaunay refinement)]{
		\includegraphics[width=0.5\linewidth]{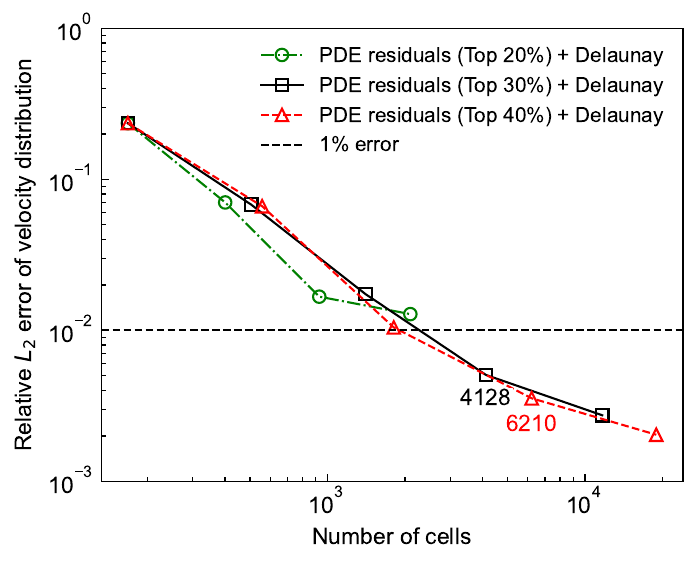}}%
    \\
    \subfigure[Velocity amplitude convergence (h-refinement)]{
		\includegraphics[width=0.49\linewidth]{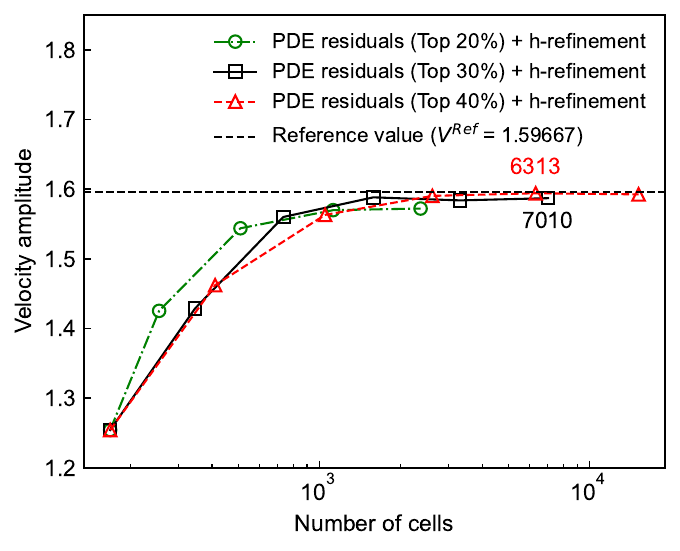}}%  
	\subfigure[Velocity distribution convergence (h-refinement)]{
		\includegraphics[width=0.51\linewidth]{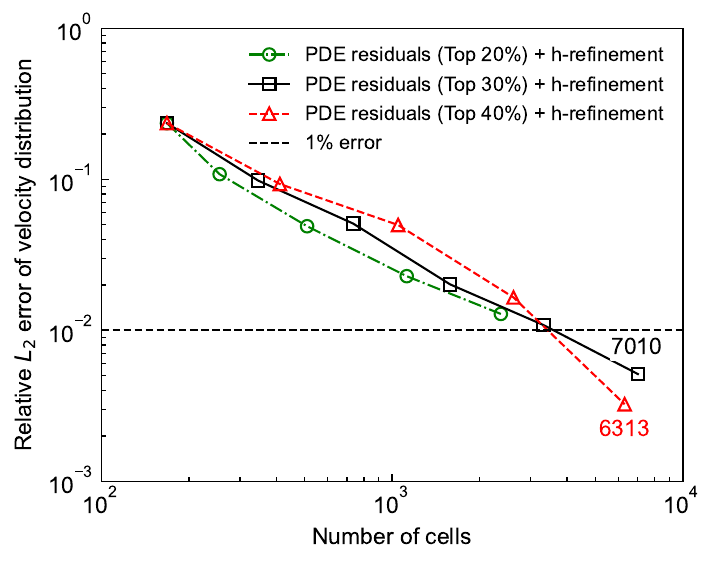}}%
    \caption{Laminar flow around an obstacle. Convergence of the velocity amplitude and distribution along the monitored line during PDE residuals-guided AMR runs with three fixed percentages~(i.e., top 20\%, 30\%, and 40\%) of cells marked and using two types of refinement~(i.e., h-refinement and Delaunay refinement). The number of cells of the optimal mesh for each corresponding AMR run that meets the desired accuracy is shown near the respective symbol.} 
    \label{fig:obstacle_PDE_velocity}
\end{figure}
% 这里提的是不同百分比的PDE残差准则的收敛结果
Then we use the PDE residuals-based indicator in combination with both Delaunay refinement and h-refinement strategies to refine the mesh. 
The purpose of employing these two refinement types is to explore the sensitivity of the PDE residuals-based AMR to the type of mesh refinement.
Additionally, we also investigate the effect of different fixed percentages of marked cells during the PDE residuals-based AMR runs.
Fig.~\ref{fig:obstacle_PDE_velocity} presents the convergence results during PDE residuals-based AMR runs with three fixed percentages~(i.e., top 20\%, 30\%, and 40\%) of cells marked and using two types of refinement~(i.e., h-refinement and Delaunay refinement).
Among these, the AMR runs with refinement of the top 30\% and 40\% of mesh cells converge to the desired accuracy. 
For two types of refinement with the same fixed percentage, AMR combined with Delaunay refinement yields the optimal mesh that satisfies the error convergence criterion (\textless 1\%) after 3 levels of refinement, while h-refinement requires 4-5 levels. 
Delaunay refinement tends to result in a higher number of mesh cells after each level of refinement compared to h-refinement. 
The reason for this is that the quarter parent cell area constraint of Delaunay refinement tends to result in relatively finer sub-cells compared to h-refinement, and produces a larger number of mesh cells after each refinement. 
Thus, the Delaunay refinement method requires fewer refinement levels than h-refinement to attain a mesh resolution of comparable refinement extent.
For the same fixed percentage, the number of the optimal mesh cells obtained by Delaunay refinement is slightly less than that obtained by h-refinement.
For Delaunay refinement, the optimal mesh occurs in the top 30\% case and contains the fewest cells (4,128).
For h-refinement, the optimal mesh occurs in the top 40\% case and contains the fewest cells (6,313).

\begin{figure}[htb!]
    \centering
    \subfigure[Velocity amplitude convergence]{
		\includegraphics[width=0.49\linewidth]{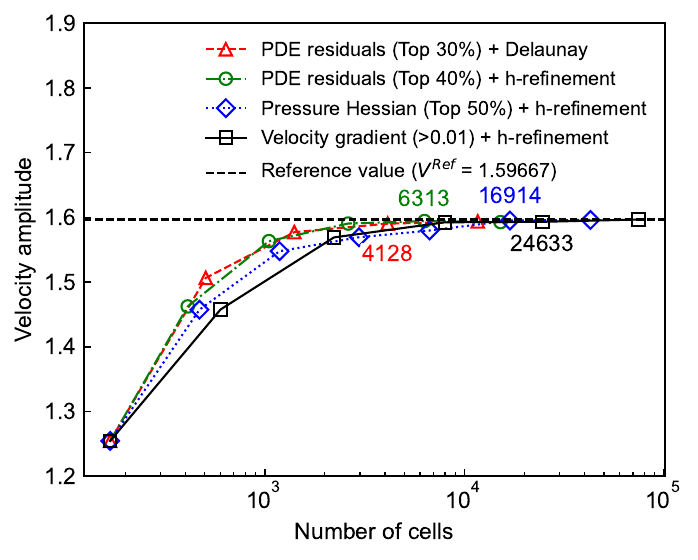}}%  
	\subfigure[Velocity distribution convergence]{
		\includegraphics[width=0.5\linewidth]{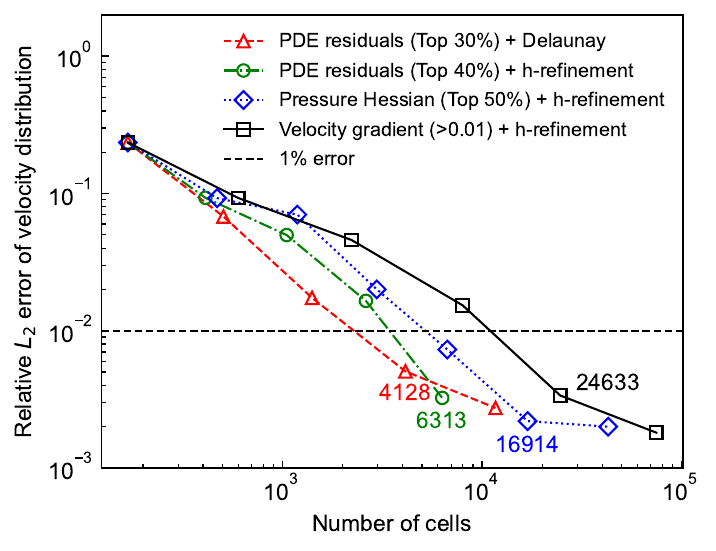}}%
    \caption{Laminar flow around an obstacle. Comparison of the convergence results during the optimal AMR runs based on the proposed PDE residuals-based indicator and two traditional AMR indicators. The number of cells of the optimal mesh for each corresponding AMR run that meets the desired accuracy is shown near the respective symbol.} 
    \label{fig:obstacle_compare_velocity}
\end{figure}
% 这里是PDE残差和传统AMR指示器的收敛结果对比
For clarity, a comparison of the convergence results during the optimal AMR runs based on the proposed PDE residuals-based indicator and two traditional indicators is shown separately in Fig.~\ref{fig:obstacle_compare_velocity}. 
The number of cells in the optimal meshes for each AMR run is also labeled in the figure.
It is evident that the AMR based on the PDE residuals-based indicator converges more quickly than the AMR runs based on the two conventional indicators, irrespective of the refinement type used. It achieves the desired accuracy with fewer refinement levels or a lower growth rate in the number of mesh cells per layer.
This is further reflected in the significantly lower number of mesh cells in the resulting optimal meshes.
Notably, the proposed PDE residuals-based indicator results in a significant reduction in the number of cells, with optimal meshes containing 4 to 6 times fewer cells than those obtained using velocity-gradient or pressure-Hessian indicators. 

\begin{figure}[htb!]
    \centering
    \subfigure[Velocity gradient (\textgreater 0.01) + h-refinement]{
        \includegraphics[width=0.47\linewidth]{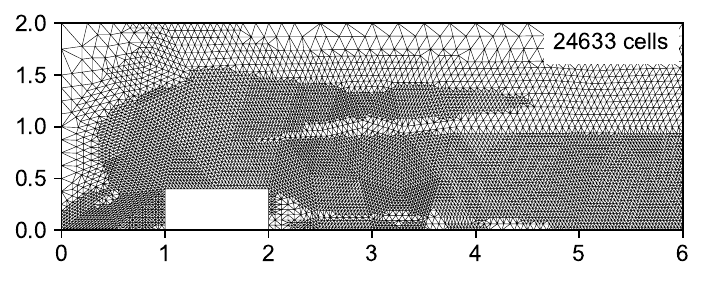}
        \includegraphics[width=0.53\linewidth]{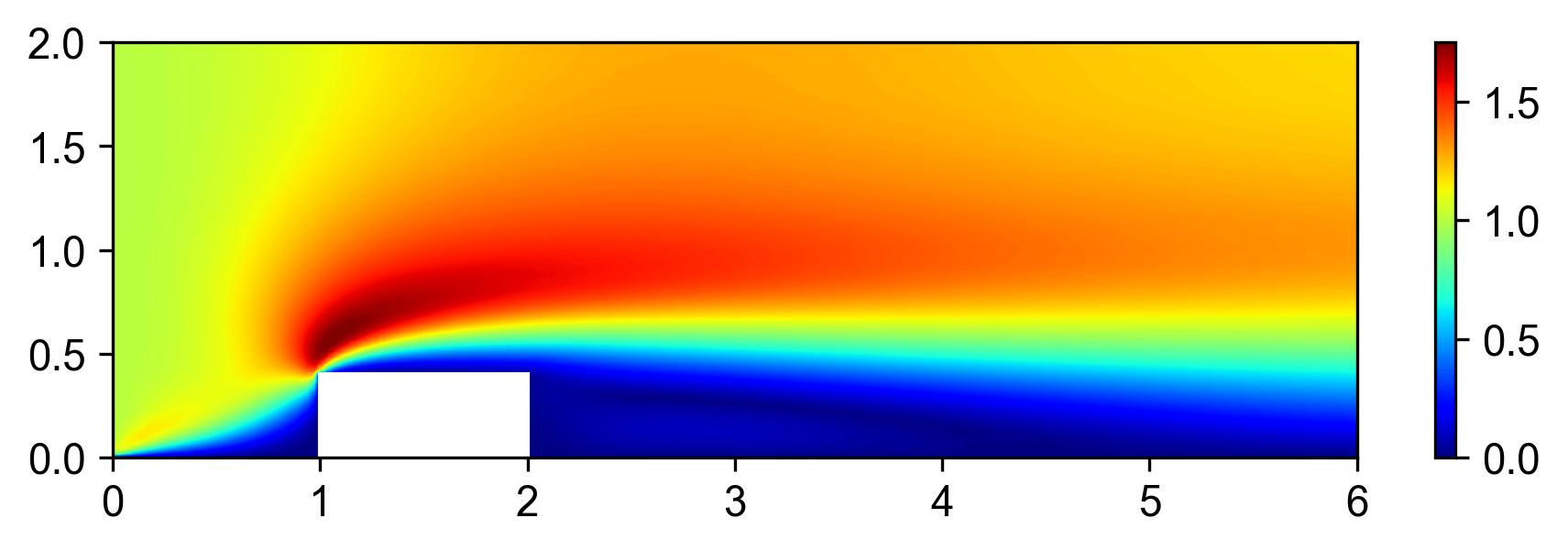}}%  
    \\
    \subfigure[Pressure Hessian (Top 50\%) + h-refinement]{
        \includegraphics[width=0.47\linewidth]{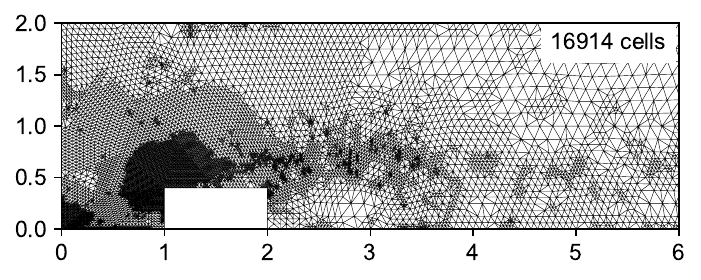}
        \includegraphics[width=0.53\linewidth]{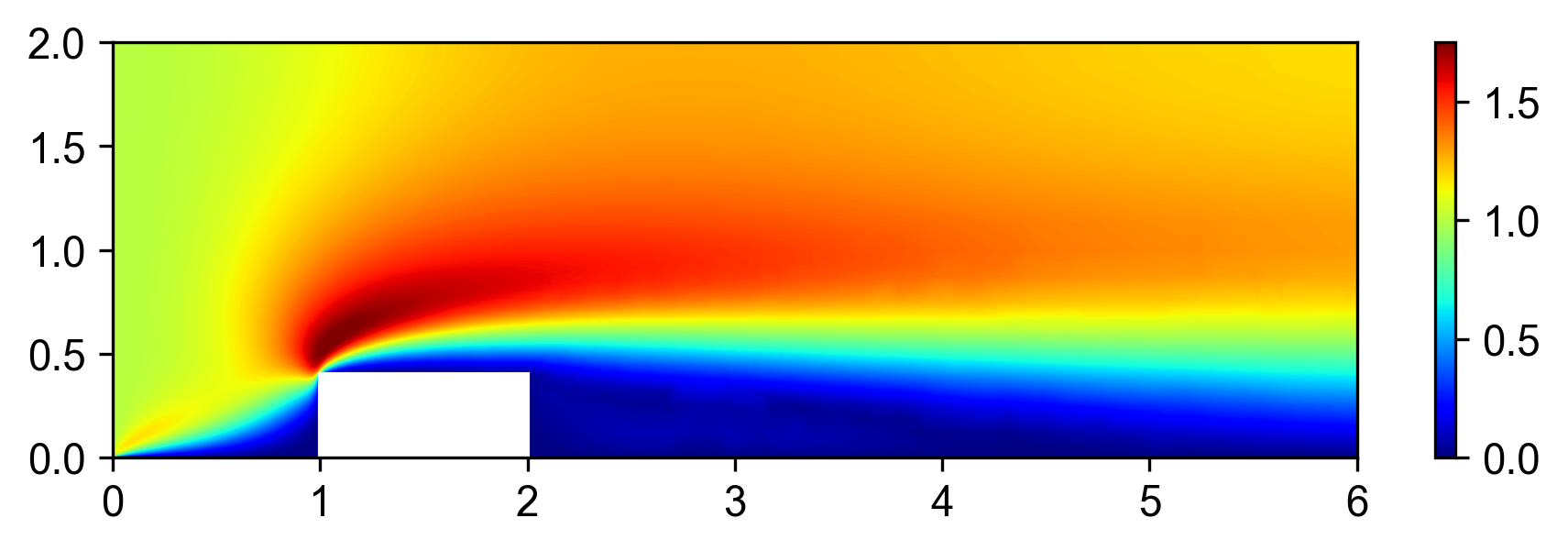}}%
    \\
    \subfigure[PDE residuals (Top 40\%) + h-refinement]{
        \includegraphics[width=0.47\linewidth]{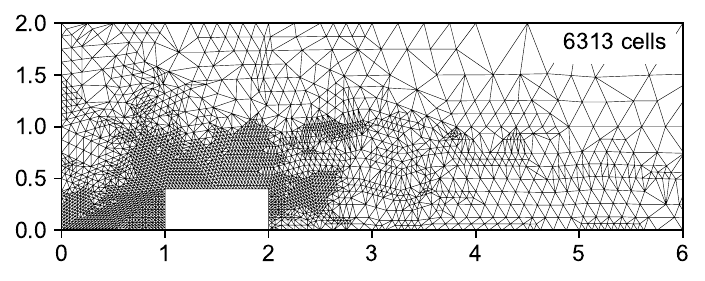}  
        \includegraphics[width=0.53\linewidth]{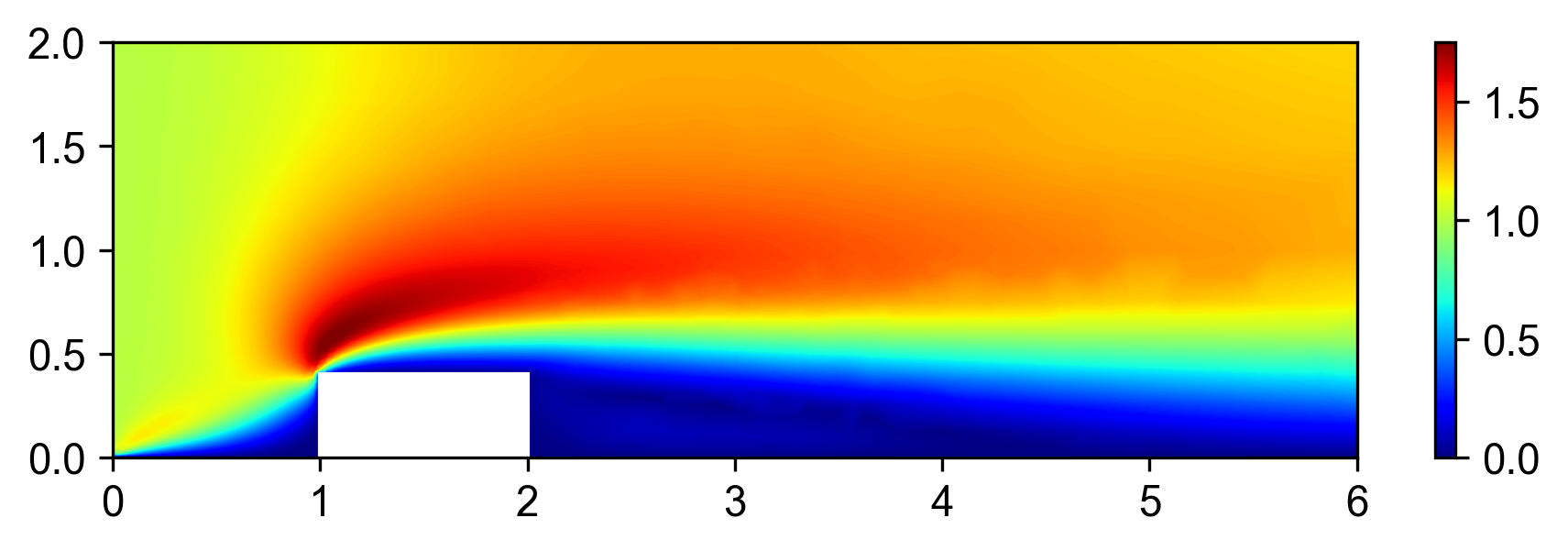}}%
    \\
    \subfigure[PDE residuals (Top 30\%) + Delaunay refinement]{
        \includegraphics[width=0.47\linewidth]{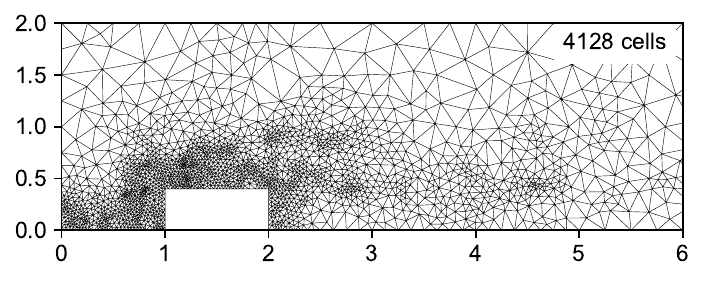}  
        \includegraphics[width=0.53\linewidth]{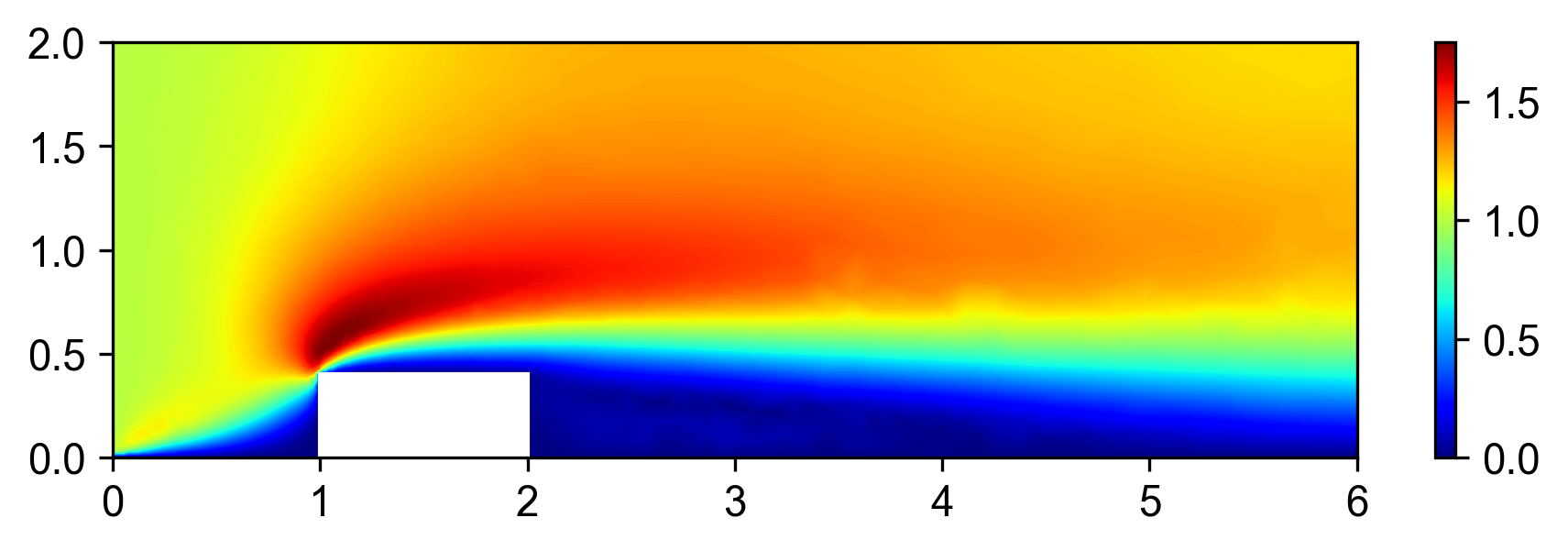}}%
    \caption{Laminar flow around an obstacle. Adapted optimal meshes and the corresponding computed velocity distribution contours from the four AMR schemes. The top right of each mesh shows its corresponding number of cells.} 
    \label{fig:obstacle_mesh_contour}
\end{figure}
% 这里是PDE残差和传统AMR指示器的最优网格分布对比
In Fig.~\ref{fig:obstacle_mesh_contour}, we depict the adapted optimal meshes and the corresponding computed velocity distribution contours from the four AMR schemes~(corresponding to Fig.~\ref{fig:obstacle_compare_velocity}). The top right of each mesh shows its corresponding number of cells.
Examining the velocity distribution contours, all four meshes effectively capture the flow characteristics near the obstacle.
% 基于速度梯度的指示器不仅在障碍物附近，而且在距障碍物较远的上方和后方区域也进行了大量的加密，最终得到满足精度指标的最优网格数目也是最多的。基于压力Hessian的指示器对障碍物的左上角的大片区域着重加密，然而对障碍物背部附近加密相对较少。基于PDE残差的指示器结合两种网格加密方式加密后的网格相似，其着重对障碍物三个面的附近区域进行加密，而距离障碍物较远的其他区域则仍然分布着较粗的网格。这也使得基于PDE残差的指示器指导加密后得到的最优网格在满足精度指标的同时数目最少。
The velocity gradient-based indicator results in significant refinement not only near the obstacle but also in the upper and rear regions farther away from the obstacle, leading to the highest cell number of the optimal mesh that meets the desired accuracy metrics. 
The pressure Hessian-based indicator concentrates on refining a large area in the upper left corner of the obstacle while showing relatively little refinement near the back. 
In contrast, the optimal mesh refined by the PDE residual-based indicator, which combines two types of refinement, focuses on the areas near the left, top, and right surfaces of the obstacle, leaving a coarser mesh in the more distant regions. 
Consequently, this approach yields the least number of meshes while still satisfying the accuracy metrics.

\begin{figure}[htb!]
    \centering
    \subfigure[Velocity distributions at each level]{
		\includegraphics[width=0.5\linewidth]{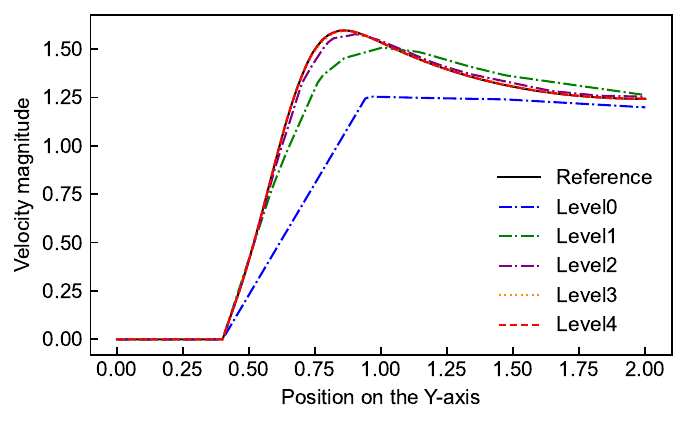}}%
    \subfigure[Velocity distributions at level 3]{
		\includegraphics[width=0.5\linewidth]{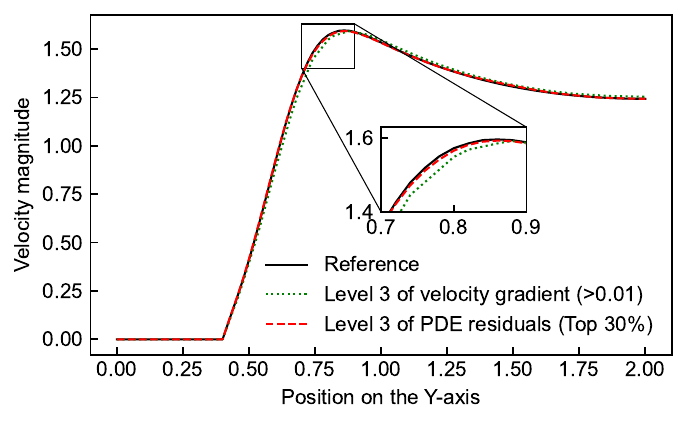}}%
    \caption{Laminar flow around an obstacle. 
    (a) AMR results using PDE residuals-based indicator (Top 30\%) with Delaunay refinement, showing computed velocity distributions along the monitored line for adaptive meshes at each level. (b) Comparison of velocity distributions between velocity gradient-based AMR (\textgreater 0.01) with h-refinement (Level 3: 7995 cells) and PDE residuals-based AMR (Top 30\%) with Delaunay refinement (Level 3: 4128 cells) at the same adaptive level.} 
    \label{fig:back_compare_velocity}
\end{figure}

Fig.~\ref{fig:back_compare_velocity}(a) presents the velocity distributions along the monitored profile, obtained from the PDE residual-based AMR (Top 30\%) with Delaunay refinement of the adaptive mesh refinement at each level. 
Fig.~\ref{fig:back_compare_velocity}(b) shows the comparison of velocity distributions between velocity gradient-based AMR~(\textgreater 0.01) with h-refinement (Level 3: 7995 cells) and PDE residual-based AMR (Top 30\%) with Delaunay refinement (Level 3: 4128 cells) at the same adaptive level.
The velocity gradient-based AMR at level 3 not only utilizes a greater number of meshes than the PDE residuals-based AMR, but its computed velocity distributions also exhibit a slight bias compared to the reference solution.

% ------------------ Case2: Subsonic inviscid flow over an airfoil ------------------
\subsection{Subsonic inviscid flow over an airfoil}
\label{subsec:Subsonic inviscid flow over an airfoil}
Here, we employ the proposed framework to solve the Euler equations as an extension to compressible flows. 
Specifically, we investigate subsonic inviscid flow over the NACA 0012 airfoil with a free-stream Mach number of $M$=0.5, and an angle of attack of $\alpha$=1.25°. 
The resulting flow is subsonic, meaning that the Mach number is consistently less than one throughout the domain, and it is inviscid. Consequently, the flow field exhibits smooth variation without the presence of shocks.
In this scenario, the drag coefficient approaches zero. Therefore, our primary focus shifts to the lift coefficient as the physical quantity of interest. 
We adopt the reference value of lift coefficient, $C_L^{ref}$ = 0.1757, obtained from the numerical simulations conducted by Dolej{\v{s}}{\'\i} and May~\citep{dolejvsi2022anisotropic}.
We take the chord length of the airfoil as the unit, with the far-field boundary situated approximately 25 chord lengths away from the center-placed airfoil.
Referring to the numerical simulations of Vassberg and Jameson~\citep{vassberg2010pursuit}, we employ the following analytic equation
\begin{equation}
y(x)= \pm \frac{0.12}{0.2}\left(0.2969 \sqrt{x}-0.1260 x-0.3516 x^2+0.2843 x^3-0.1015 x^4\right)
\end{equation}
to characterize the geometry of the NACA 0012 airfoil, resulting in a sharp trailing edge.
\begin{figure}[htb!]
	\centering  
	\subfigure[Far-field view of initial mesh]{
		\label{Farfield view of initial mesh}
		\includegraphics[width=0.485\linewidth]{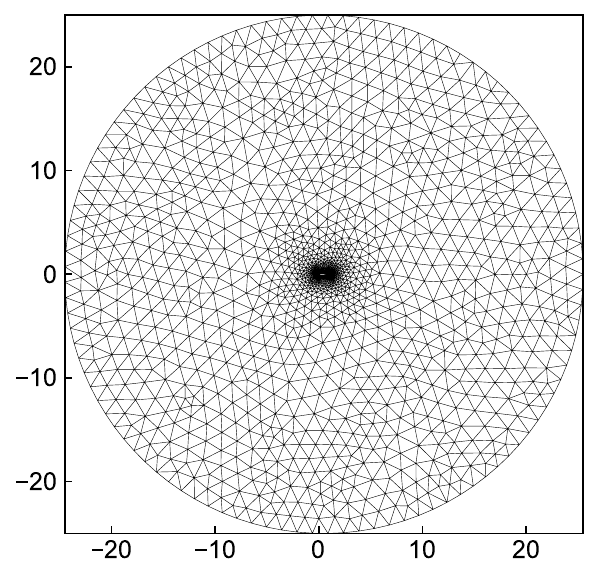}}%
	\subfigure[Closeup view of initial mesh]{
		\label{Closeup view of initial mesh}
		\includegraphics[width=0.515\linewidth]{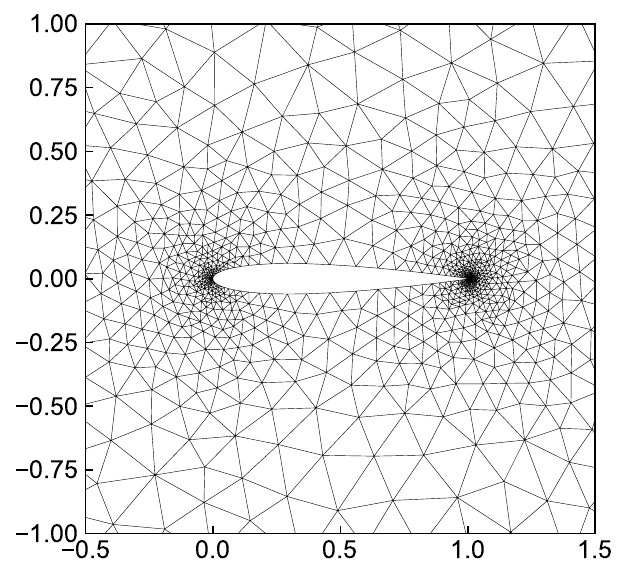}}%
	\caption{NACA 0012: Initial triangular mesh with 4157 cells and 2165 nodes.}
	\label{fig:Initial_mesh_naca0012}
\end{figure}
At the far-field boundary, we impose a steady free stream with a Mach number $M$=0.5 and an angle of attack $\alpha$=1.25° as the Dirichlet boundary condition. Slip boundary conditions are imposed on the airfoil boundary. 
The geometric presentation and generated initial mesh of the domain are provided in Fig.~\ref{fig:Initial_mesh_naca0012}. The initial coarse mesh consists of 4157 cells and 2165 nodes.

\begin{figure}[htb!]
    \centering
    \subfigure[Lift convergence (h-refinement)]{
		\includegraphics[width=0.5\linewidth]{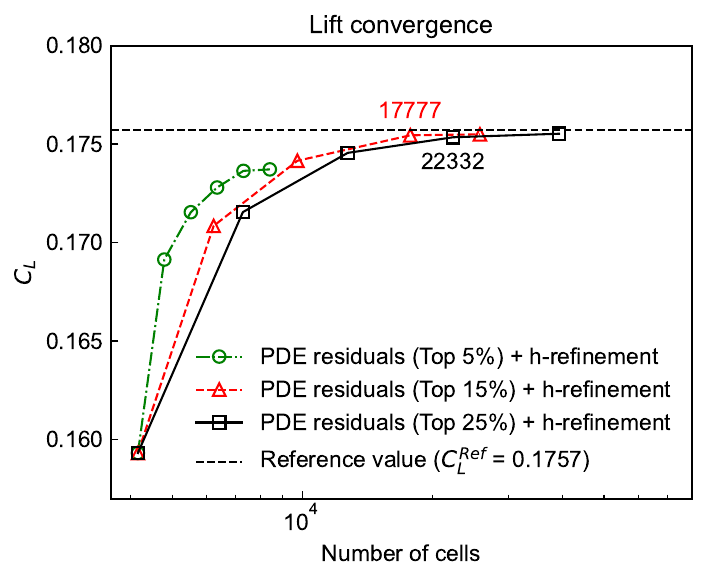}}%  
	\subfigure[Lift convergence (Delaunay refinement)]{
		\includegraphics[width=0.5\linewidth]{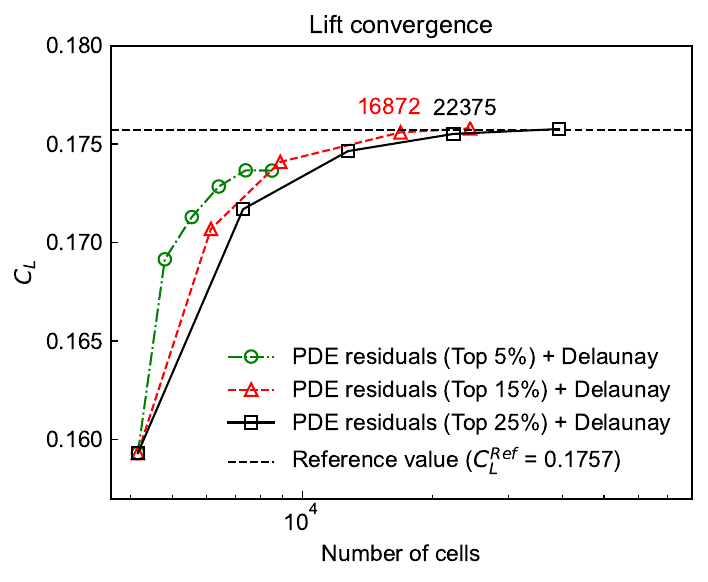}}%
    \caption{Subsonic inviscid flow over the NACA 0012 airfoil~(M=0.5, $\alpha$=1.25°). The lift coefficient convergence during PDE residuals-guided AMR runs with different fixed percentages~(i.e., top 5\%, 15\%, and 25\%) of cells marked. The number of cells of the optimal mesh for each corresponding AMR run that meets the desired accuracy is shown near the respective symbol.} 
    \label{fig:subsonic_PDE_top_n_per_lift}
\end{figure}
% 这里再提一下不同百分比的传统AMR准则的收敛结果
We first examine the convergence results for the density gradient-based and pressure Hessian-based AMR runs in Fig.~\ref{fig:subsonic_traditional_top_n_per_lift}, where different fixed percentages~(i.e., top 15\%, 25\%, 35\%, and 45\%) of cells marked are applied. 
The number of cells of the optimal mesh for each corresponding AMR run that meets the desired accuracy is shown near the respective symbol in the figure.
For the density gradient indicator, the AMR runs with the top 45\% of cells marked converge to the desired accuracy.
For the pressure Hessian indicator, the AMR runs with the top 35\% and 45\% of cells marked converge to the desired accuracy, with the optimal mesh for the 35\% case containing the fewest cells.
% 这里提一下不同百分比的PDE准则的收敛结果
Then, we also investigate the effect of different fixed percentages of marked cells during the PDE residuals-guided AMR runs.
Fig.~\ref{fig:subsonic_PDE_top_n_per_lift} shows the convergence results during PDE residuals-based AMR runs with three fixed percentages~(i.e., top 5\%, 15\%, and 25\%) of cells marked and using two types of refinement~(i.e., h-refinement and Delaunay refinement).
For both types of refinement, the AMR runs with the top 15\% and 25\% of cells marked converge to the desired accuracy, with the optimal mesh for the 15\% case containing the fewest cells. 

\begin{figure}[htb!]
    \centering
    \subfigure[Fixed 15\% of cells marked]{
		\includegraphics[width=0.5\linewidth]{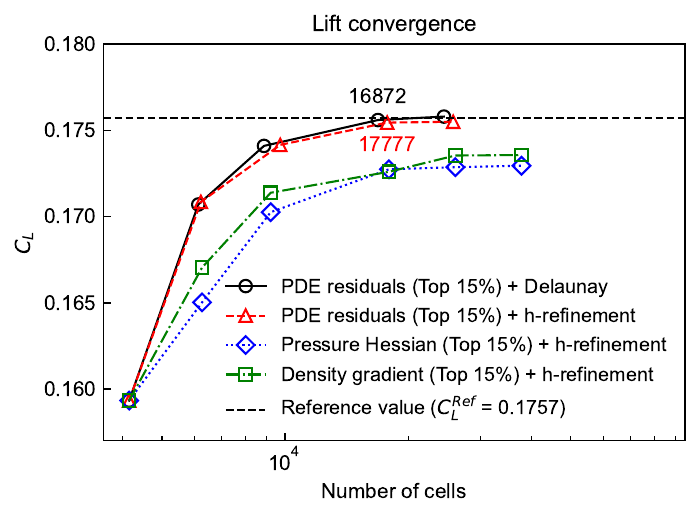}}%  
	\subfigure[Optimal AMR runs]{
		\includegraphics[width=0.5\linewidth]{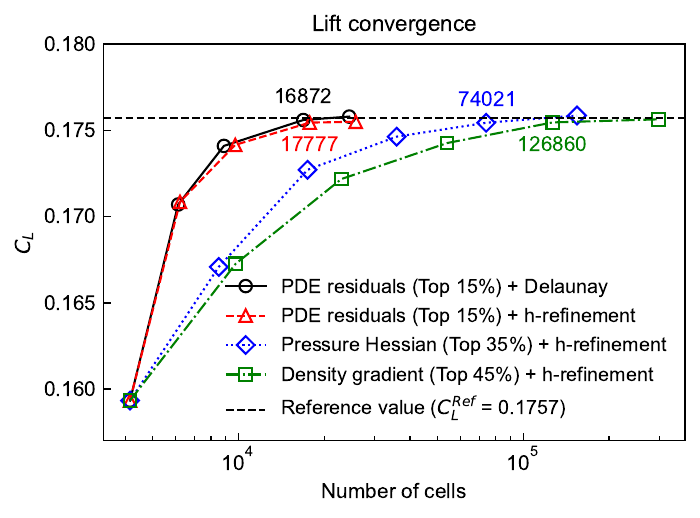}}%
    \caption{The lift coefficient convergence during various types of AMR runs for subsonic flow over the NACA 0012 airfoil~(M=0.5, $\alpha$=1.25°). 
    (a) Comparison of the convergence results during the AMR runs based on the proposed PDE residuals-based indicator and two traditional AMR indicators, using the same fixed-percentage 15\% of cells marked.
    (b) Comparison of the convergence results during the optimal AMR runs based on the proposed PDE residuals-based indicator and two traditional AMR indicators.} 
    \label{fig:Lift_coefficients_sub}
\end{figure}
% 传统AMR准则和PDE准则的比较
We compare the performance of the PDE residuals-based indicator to the density gradient and pressure Hessian indicators, using the same fixed-percentage 15\% of cells marked, as shown in Fig.~\ref{fig:Lift_coefficients_sub}(a). 
At each refinement level, where the number of mesh cells is nearly identical, the lift coefficient obtained using the PDE residuals-guided AMR converges to the reference value. In contrast, the lift coefficients obtained from the two traditional AMR runs exhibit a larger discrepancy from the reference value, even after convergence.
Fig.~\ref{fig:Lift_coefficients_sub}(b) illustrates the convergence of the computed lift coefficient with respect to the number of mesh cells during the optimal AMR runs that converge to the desired accuracy.
In all four adaptive cases, the computed lift values converge to the reference value, and the adaptive process stops after about four or five refinement levels.
The traditional AMR runs converge to a value near the reference after increasing the percentage of marked cells. However, this convergence comes at the cost of a significantly larger number of mesh cells.
In contrast, the PDE residuals-guided AMR runs converge much more quickly than the traditional AMR runs, requiring fewer marked cells and refinement levels. 
This leads to a significantly lower number of optimal mesh cells in comparison to the two traditional AMR runs.
Given that the lift coefficient is a key feature of the exact solution, it is evident that the PDE residuals-guided AMR, combining both refinement strategies, performs particularly well. 

\begin{figure}[htb!]
	\centering  
	\subfigure[Density gradient (Top 45\%) + h-refinement]{
		\includegraphics[width=0.5\linewidth]{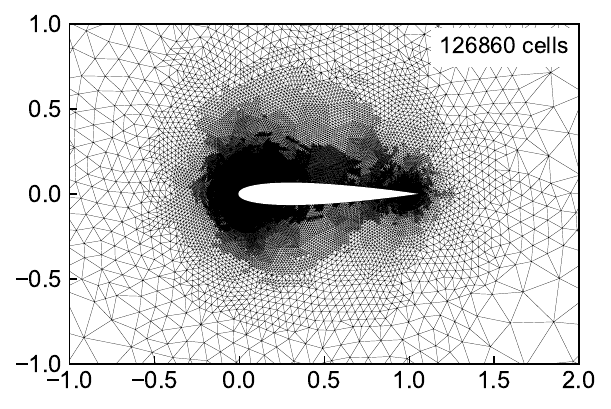}}%
	\subfigure[Pressure Hessian (Top 35\%) + h-refinement]{
		\includegraphics[width=0.5\linewidth]{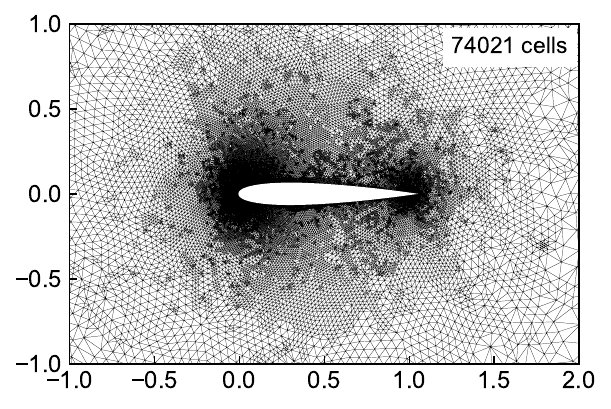}}%
    \\  
    \subfigure[PDE residuals (Top 15\%) + h-refinement]{
		\includegraphics[width=0.5\linewidth]{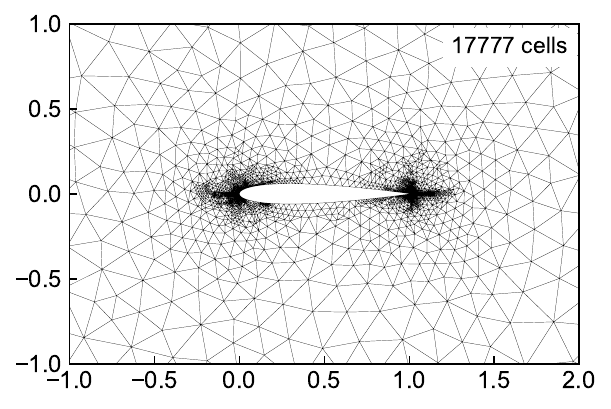}}%
	\subfigure[PDE residuals (Top 15\%) + Delaunay]{
		\includegraphics[width=0.5\linewidth]{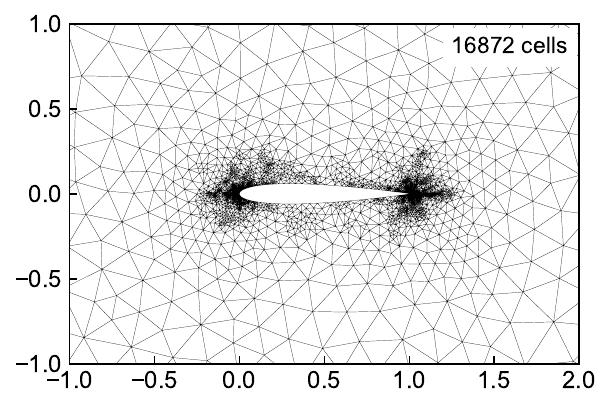}}%
	\caption{Subsonic inviscid flow over the NACA 0012 airfoil~(M=0.5, $\alpha$=1.25°). Adapted optimal meshes corresponding to four AMR schemes. The top right of each mesh shows its corresponding number of cells.}
	\label{fig:subsonic_best_mesh}
\end{figure}
% 传统AMR准则和PDE准则的最优网格对比
In Fig.\ref{fig:subsonic_best_mesh}, we present a closeup view of the adapted optimal meshes for the four AMR schemes~(corresponding to Fig.\ref{fig:Lift_coefficients_sub}(b)), with the number of cells indicated in the top right corner of each mesh.
The PDE residuals-guided AMR, combined with the fixed-fraction strategy, enables a rapid convergence to desired accuracy by selectively refining a limited number of mesh cells in localized regions of the domain.
This is depicted in Fig.~\ref{fig:subsonic_best_mesh}(c) and (d), wherein mesh refinement primarily occurs proximate to the leading edge region where fluid acceleration is most pronounced, with additional refinement observed near the airfoil's trailing edge. 
The pressure Hessian-based AMR exhibits analogous characteristics, though it places more emphasis on refinement at both the leading and trailing edges of the airfoil, while also refining more distant regions.
The density gradient-based AMR approaches tend to perform more comprehensive refinement of the airfoil, resulting in a marked escalation in mesh density. 

\begin{figure}[htb!]
	\centering  
	\subfigure[Density gradient (Top 45\%) + h-refinement]{
		\includegraphics[width=0.5\linewidth]{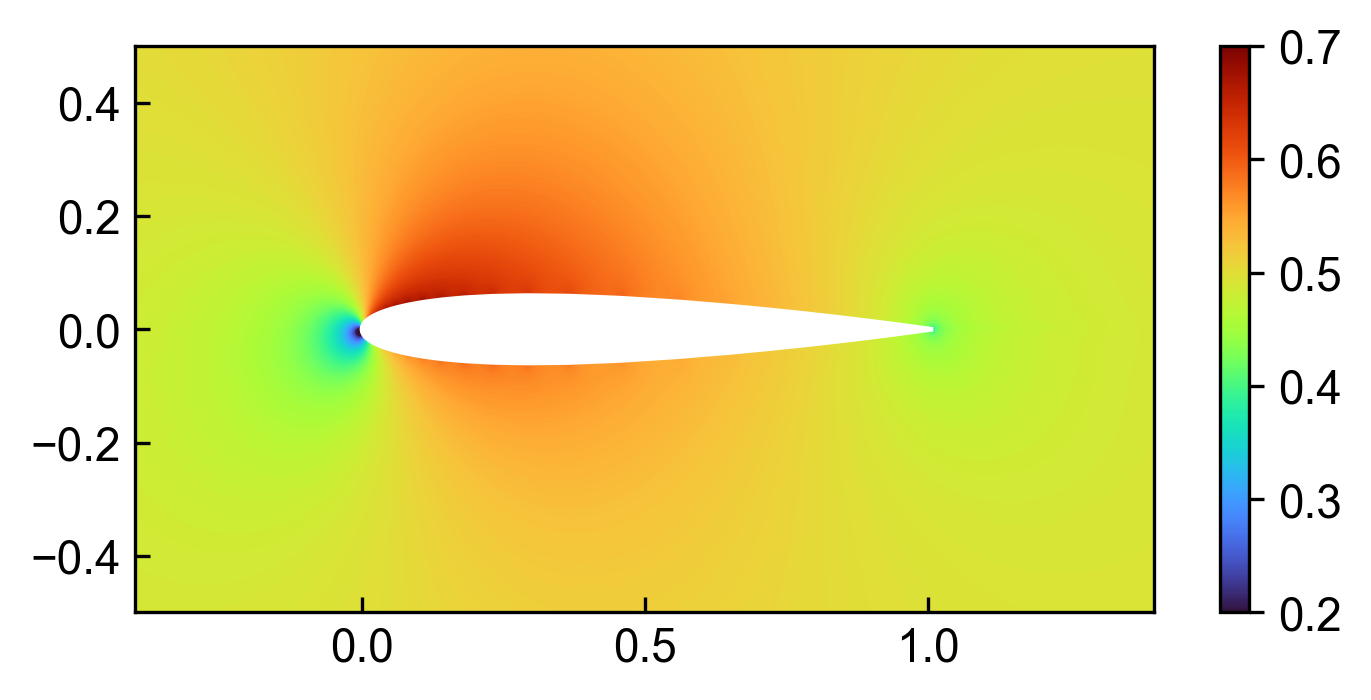}}%
	\subfigure[PDE residuals (Top 15\%) + h-refinement]{
		\includegraphics[width=0.5\linewidth]{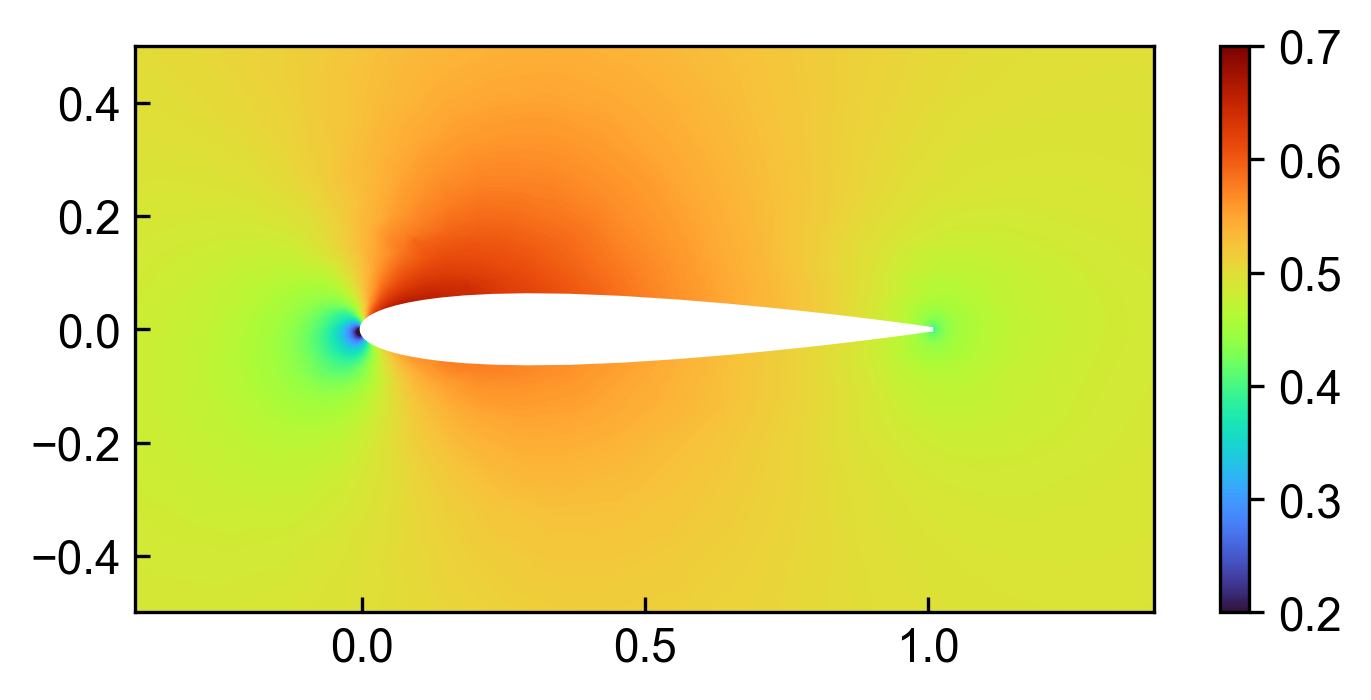}}%
	\caption{Subsonic inviscid flow over the NACA 0012 airfoil~(M=0.5, $\alpha$=1.25°). Mach number distributions computed from the adapted optimal mesh.}
	\label{fig:sub_ma}
\end{figure}

% 传统AMR准则和PDE准则的最优网格马赫数云图对比
Fig.~\ref{fig:sub_ma} depicts plots of the computed Mach number distributions for the adapted mesh solutions obtained using density gradient and PDE residuals AMR indicators. Despite the large difference in the number of cells between the meshes, both can effectively capture the flow characteristics near the airfoil, as seen in the Mach number distribution contours.
The superiority of the PDE residual-based indicator becomes particularly apparent in the final count of optimal mesh cells. Specifically, the utilization of this indicator results in a reduction of optimal mesh count by a factor of 4$\sim$8 compared to meshes obtained through traditional indicators. This substantial decrease in mesh quantity can substantially mitigate the computational expenses.

% ----------------- Case3: Transonic inviscid flow over an airfoil ------------------
\subsection{Transonic inviscid flow over an airfoil}
\label{subsec:Transonic inviscid flow over an airfoil}
We further consider the transonic inviscid flow over the NACA 0012 airfoil with a free-stream Mach number of $M$=0.8, and an angle of attack of $\alpha$=1.25°. 
Due to the presence of the Mach number exceeding one in certain subdomains and the resulting solution discontinuities, this flow demonstrates a strong shock on the upper surface of the airfoil and a weaker shock on the lower surface. 
Both drag and lift coefficients are considered as physical quantities of interest in our analysis.
We refer to the values of lift and drag coefficients in Ref.~\citep{dolejvsi2022anisotropic}, which are $C_L^{ref}$ = 0.333 and $C_D^{ref}$ = 0.02135, respectively.
The computational domain setup and initial mesh remain consistent with those of the subsonic flow case, as illustrated in Fig.~\ref{fig:Initial_mesh_naca0012}. 
We impose a uniform free stream with a Mach number $M$=0.8 and an angle of attack $\alpha$=1.25° on the far-field boundary as the Dirichlet boundary condition. And slip boundary conditions are imposed on the airfoil boundary. 

% 这里再提一下不同百分比的传统AMR准则的收敛结果
The convergence results for the density gradient-based and pressure Hessian-based AMR runs are presented in Fig.~\ref{fig:transonic_density_top_n_per_lift_drag} and Fig.~\ref{fig:transonic_pressure_top_n_per_lift_drag}, with different fixed percentages~(i.e., top 15\%, 25\%, 35\%, and 45\%) of cells marked. 
For the density gradient indicator, the AMR runs with the top 45\% of cells marked converge to the desired accuracy.
For the pressure Hessian indicator, the AMR runs with the top 35\% and 45\% of cells marked converge to the desired accuracy, with the optimal mesh for the 35\% case containing the fewest cells.
% 这里提一下不同百分比的PDE准则的收敛结果
Fig.~\ref{fig:transonic_PDE_top_n_per_lift_drag} shows the convergence results from PDE residuals-based AMR runs with three fixed percentages~(i.e., top 5\%, 15\%, and 25\%) of cells marked and using two types of refinement~(i.e., h-refinement and Delaunay refinement).
For both types of refinement, the AMR runs with the top 15\% and 25\% of cells marked converge to the desired accuracy, with the optimal mesh for the 15\% case containing the fewest cells. 

\begin{figure}[htb!]
    \centering
    \subfigure[Fixed 15\% of cells marked]{
		\includegraphics[width=0.495\linewidth]{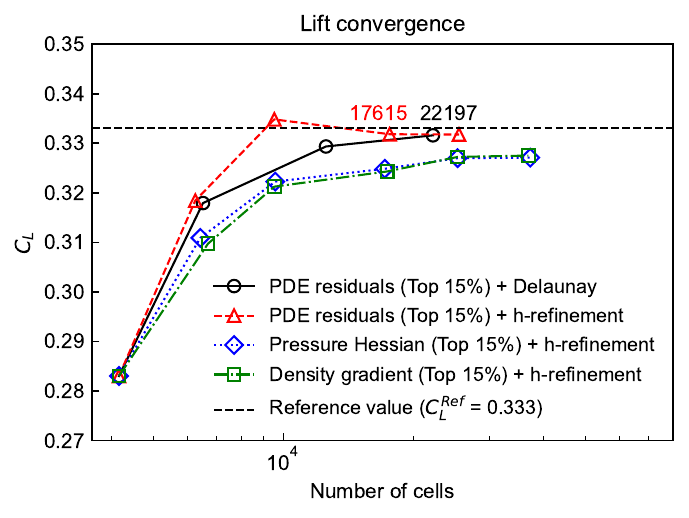}%  
	% \subfigure{
		\includegraphics[width=0.505\linewidth]{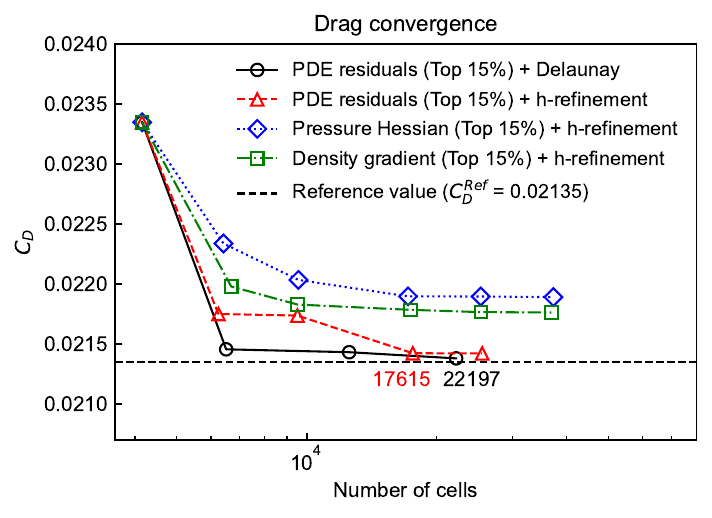}}%
    \\
    \subfigure[Optimal AMR runs]{
		\includegraphics[width=0.495\linewidth]{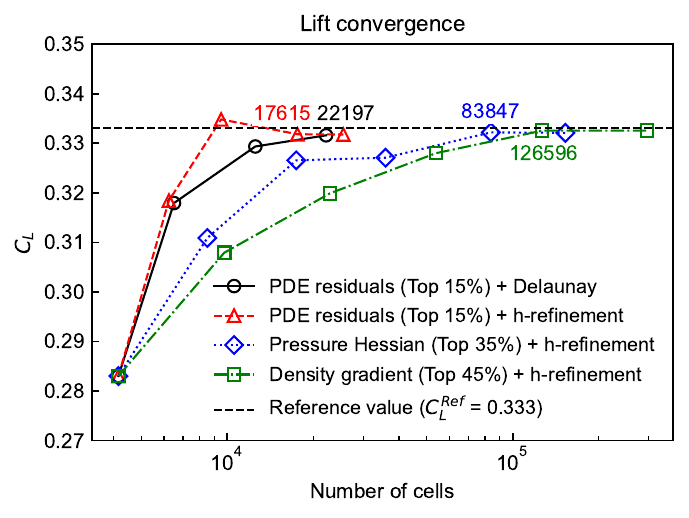}%  
	% \subfigure{
		\includegraphics[width=0.505\linewidth]{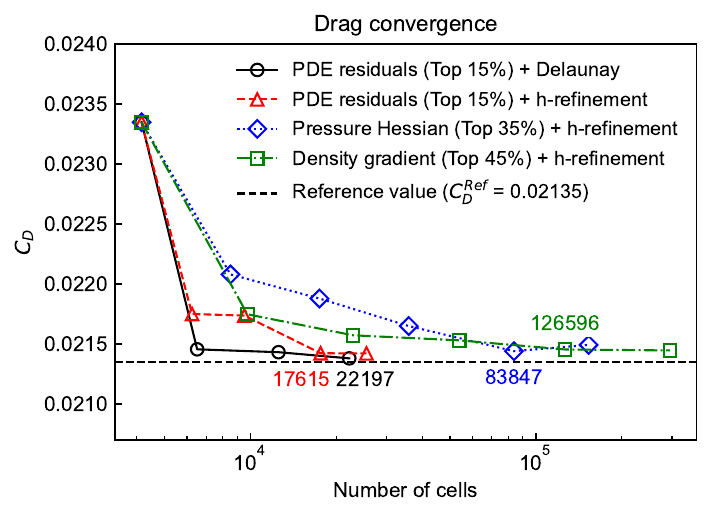}}%    
    \caption{The lift and drag coefficients convergence during various types of AMR runs for transonic flow over the NACA 0012 airfoil~(M=0.8, $\alpha$=1.25°). 
    (a) Comparison of the convergence results during the AMR runs based on the proposed PDE residuals-based indicator and two traditional AMR indicators, using the same fixed-percentage 15\% of cells marked.
    (b) Comparison of the convergence results during the optimal AMR runs based on the proposed PDE residuals-based indicator and two traditional AMR indicators.} 
    \label{fig:Lift_drag_coefficients_trans}
\end{figure}
% 传统AMR准则和PDE准则的比较
We assess the performance of the PDE residuals-based indicator by comparing it with the density gradient-based and pressure Hessian-based error indicators, using the same percentage (15\%) of cells marked, as shown in Fig.~\ref{fig:Lift_drag_coefficients_trans}(a).
Similar to the case of the subsonic flow in Section.~\ref{subsec:Subsonic inviscid flow over an airfoil}, the results for this case show that the lift and drag coefficients computed using the PDE residuals-guided AMR closely match the reference values at each refinement level, where the number of mesh cells remains nearly identical. 
In contrast, the lift and drag coefficients from the traditional AMR runs, based on density gradient and pressure Hessian indicators, exhibit larger discrepancies from the reference values, even after convergence.
Fig.~\ref{fig:Lift_drag_coefficients_trans}(b) illustrates the convergence of the computed lift and drag coefficients with respect to the number of mesh cells during the optimal AMR runs that converge to the desired accuracy.
In all four adaptive cases, the computed lift and drag coefficient values nearly converge to the reference value within approximately three to five levels.
Once more, the PDE residual-guided AMR exhibited efficient performance, showcasing a rapid convergence of computed lift and drag coefficient values towards the reference values. Conversely, the two conventional AMR strategies refined a significantly larger number of mesh cells and demonstrated a relatively slow convergence towards the reference values. 
The traditional AMR runs converge to a value close to the reference as the percentage of marked cells increases, though this comes at the cost of a significantly larger number of mesh cells.
% Fig.~\ref{fig:Lift_drag_coefficients_trans} shows the convergence of error in the computed lift and drag coefficients with respect to the number of mesh cells during various types of AMR runs.

\begin{figure}[htb!]
	\centering  
	\subfigure[Level 0 mesh~(Initial mesh)]{
		\includegraphics[width=0.5\linewidth]{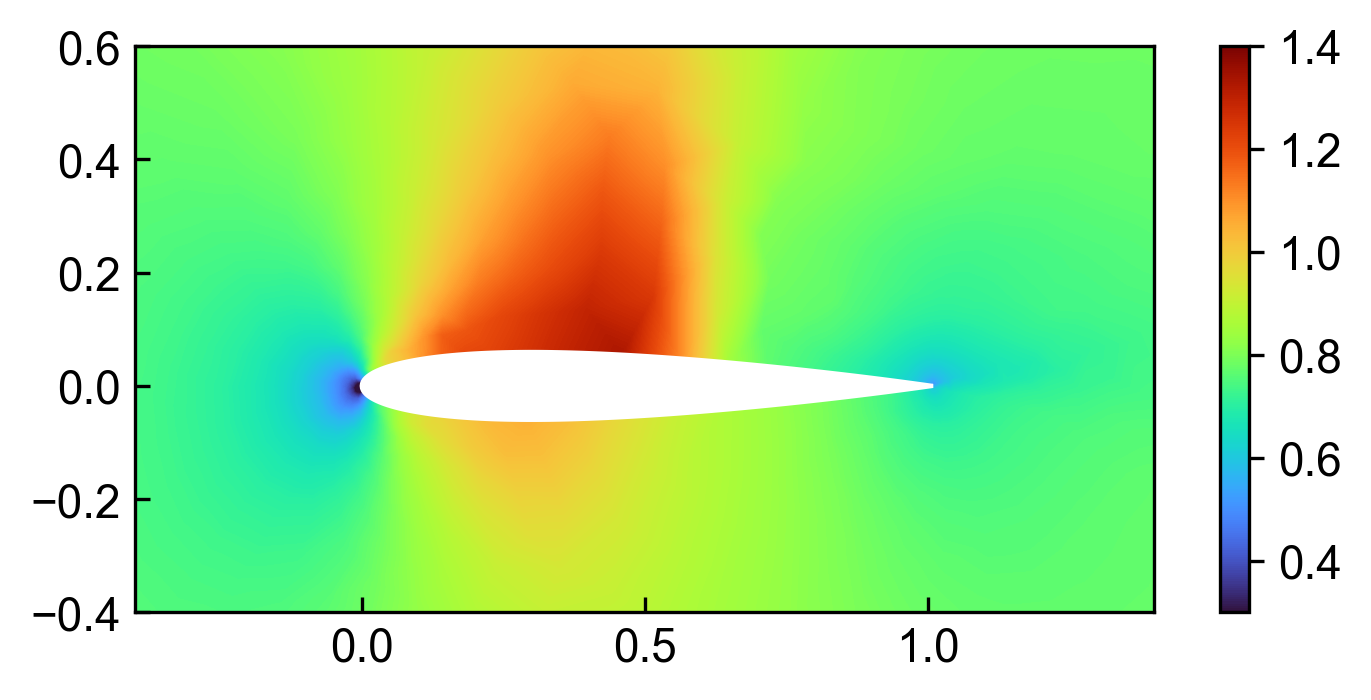}}%
	\subfigure[Level 1 mesh]{
		\includegraphics[width=0.5\linewidth]{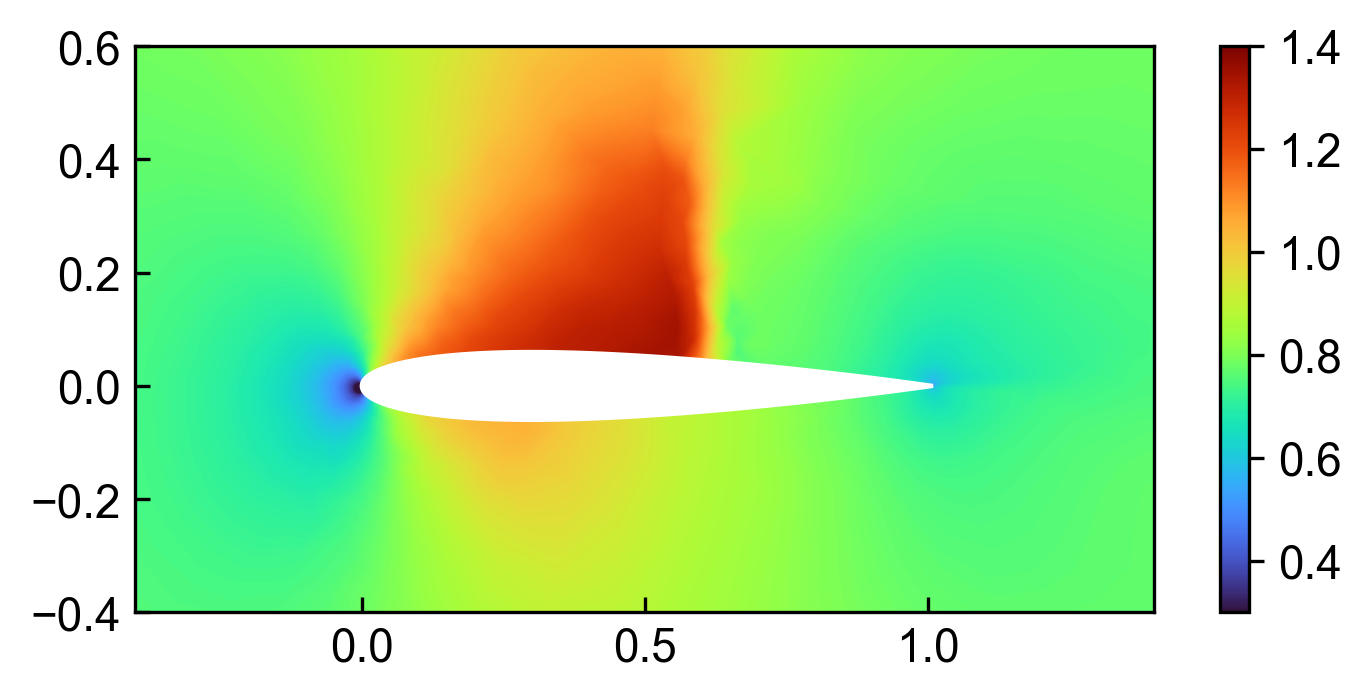}}%
    \\
    \subfigure[Level 2 mesh]{
		\includegraphics[width=0.5\linewidth]{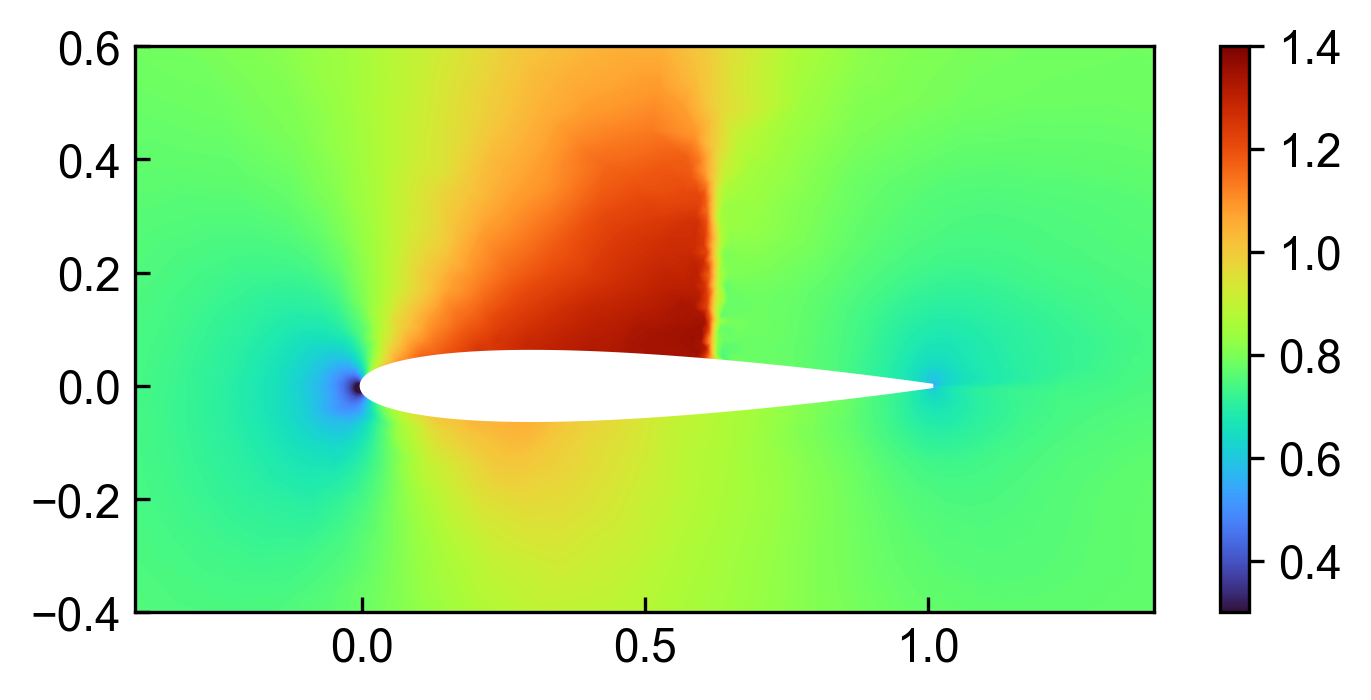}}%
	\subfigure[Level 3 mesh]{
		\includegraphics[width=0.5\linewidth]{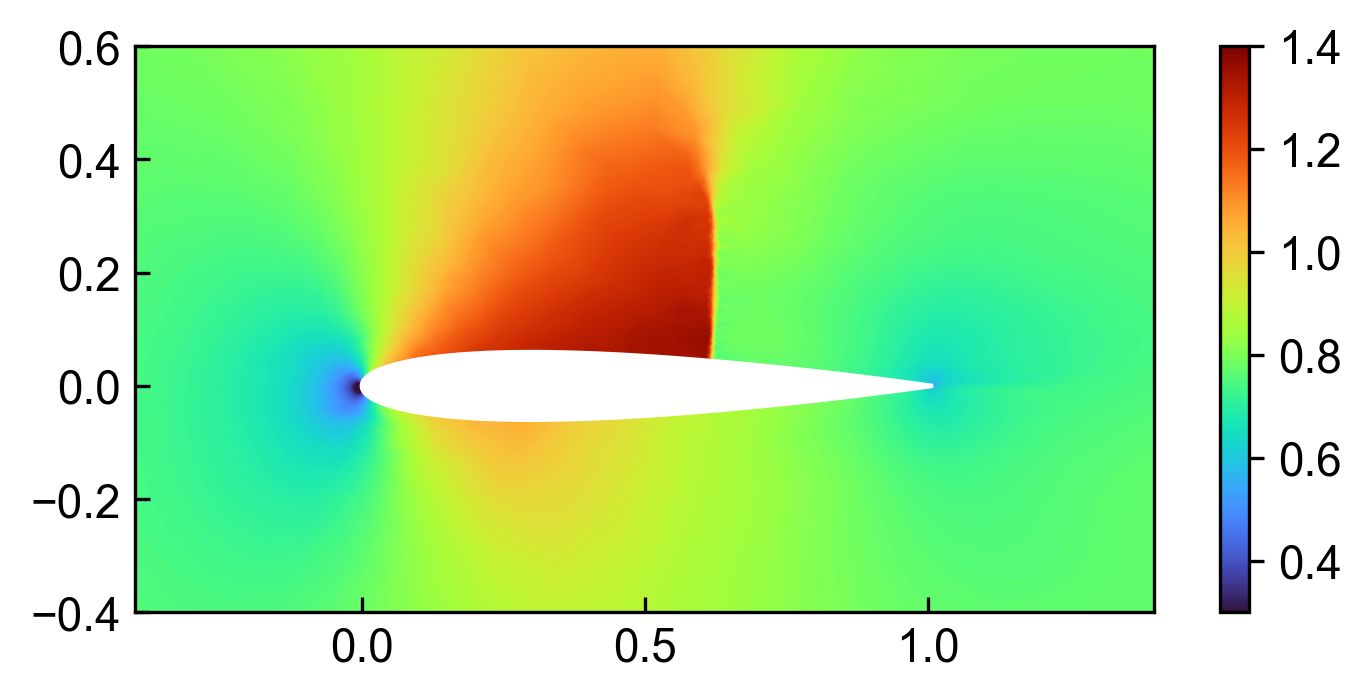}}%
	\caption{Transonic inviscid flow over the NACA 0012 airfoil~(M=0.8, $\alpha$=1.25°). AMR results with PDE residuals-based adaptation~(Top 15\%) and Delaunay refinement. Mach number distributions computed from adaptive meshes at each level.}
	\label{fig:trans_ma}
\end{figure}

Fig.~\ref{fig:trans_ma} depicts plots of the Mach number distributions computed from the adapted meshes at each level, obtained using PDE residuals AMR indicators with Delaunay refinement. A sharp capturing of upper strong waves is clear to see.
Coupled with the rapid convergence observed in Fig.\ref{fig:Lift_drag_coefficients_trans}, these results further confirm that the PDE residuals-based AMR can achieve the desired accuracy with a significantly smaller number of cells.

\begin{figure}[htb!]
	\centering  
	\subfigure[Density gradient (Top 45\%) + h-refinement]{
		\includegraphics[width=0.5\linewidth]{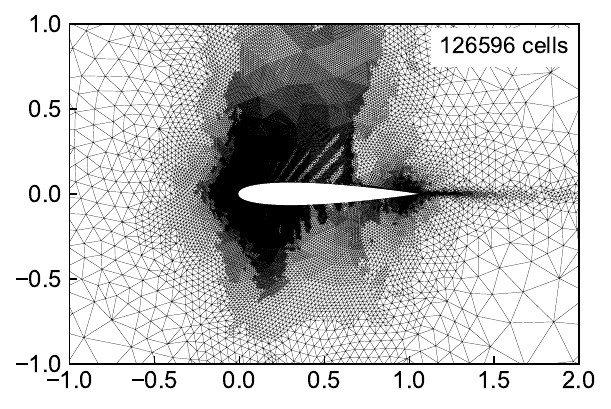}}%
	\subfigure[Pressure Hessian (Top 35\%) + h-refinement]{
		\includegraphics[width=0.5\linewidth]{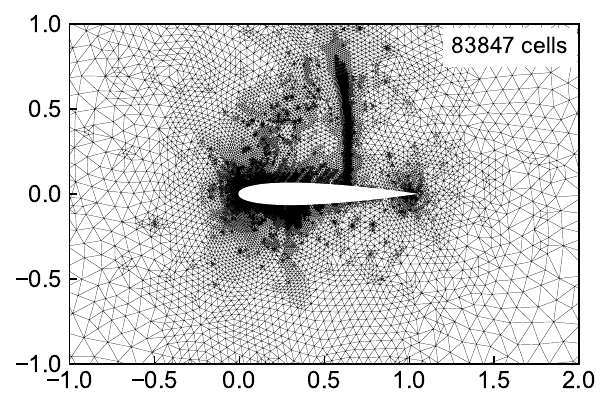}}%
    \\  
    \subfigure[PDE residuals (Top 15\%) + Delaunay]{
		\includegraphics[width=0.5\linewidth]{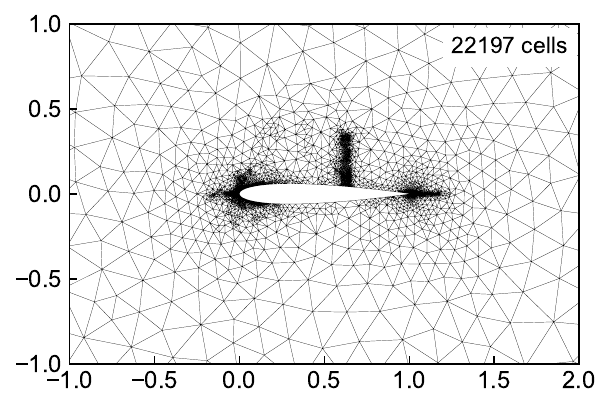}}%
    \subfigure[PDE residuals (Top 15\%) + h-refinement]{
		\includegraphics[width=0.5\linewidth]{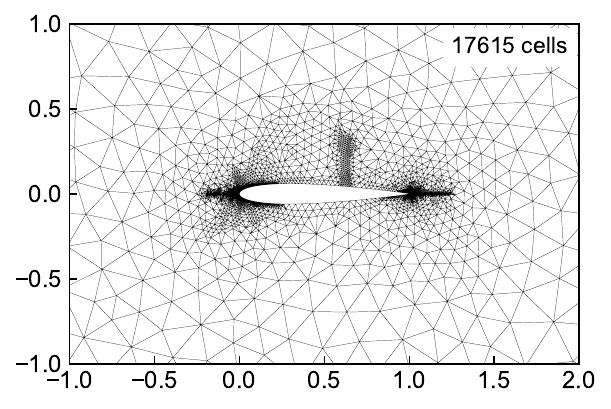}}%
	\caption{Transonic inviscid flow over the NACA 0012 airfoil~(M=0.8, $\alpha$=1.25°). Adapted optimal meshes corresponding to four AMR schemes. The top right of each mesh shows its corresponding number of cells.}
	\label{fig:transonic_best_mesh}
\end{figure}

% 传统AMR准则和PDE准则的最优网格的比较
We present the closeup view of all adapted optimal meshes corresponding to the four AMR schemes in Fig.~\ref{fig:transonic_best_mesh}.
The density gradient indicator not only refines the critical regions but also extends significantly throughout the surrounding areas of the airfoil, as shown in Fig.~\ref{fig:transonic_best_mesh}(a). The extensive refinement reduces its prominence in critical areas, such as the shock.
As shown in Fig.~\ref{fig:transonic_best_mesh}(b), the pressure Hessian indicator prioritizes refinement in critical regions, notably observed by the higher mesh density at the upper strong shock region, while exhibiting relatively sparse and piecemeal refinement in the surroundings compared to the density gradient indicator.

The optimal mesh obtained through the PDE residuals-guided AMR strategy demonstrates a precise and moderate level of refinement in crucial regions, as illustrated in Fig.~\ref{fig:transonic_best_mesh}(c) and (d).
It is apparent from Figs.~\ref{fig:Lift_drag_coefficients_trans} and Figs.~\ref{fig:transonic_best_mesh} that the proposed PDE residuals-guided AMR efficiently provides refinement of the computational mesh in the regions of the leading edge, upper strong shock region, and trailing edge while not introducing unneeded resolution in the region further from the airfoil. As a consequence, the computed lift and drag coefficients appear to rapidly converge to the expected value. 
Notably, the proposed PDE residuals-based AMR leads to a significant reduction in the number of optimal meshes, ranging from 4 to 7 times fewer compared to meshes obtained using conventional indicators.

Although the weak shock region on the lower surface is only slightly refined, the lift and drag coefficients consistently approach a value close to the reference. Clearly, refining the lower weak shock region is not as crucial as refining the leading edge, upper shock region, and trailing edge, which is consistent with the results and conclusions drawn by Venditti and Darmofal~\citep{venditti2002grid} where a similar refinement of weak shock was performed.
As we discussed, the total residual of the governing PDEs provides a more comprehensive error representation and offers a broader view of the error distribution compared to the gradient or curvature of a single physical quantity. 
This might explain why the PDE residual indicator is more effective at capturing key flow features, making it more competitive in certain scenarios compared to traditional indicators such as the velocity gradient or pressure Hessian.

% ---------------- Case4: Supersonic inviscid flow over a wedge-----------------
\subsection{Supersonic inviscid flow over a wedge}
\label{subsec:Supersonic inviscid flow over a wedge}
% 对这个案例的一些说明
In this case, we focus on the ability of the proposed PDE residuals-based indicator to detect and capture large-scale shock waves. For flow features like shock waves, the traditional velocity gradient or Hessian-based indicators can easily detect the sharp gradients across the shock front. However, the residual-based indicator, which is derived from the overall PDE residuals, does not directly correlate with the gradient magnitudes at the shock itself. Thus, the behavior of the PDE residuals-based indicator is more subtle, and it is not guaranteed that the residuals will exhibit a significant peak at the shock.
This raises an important question: Can the large gradient features of the shock be reflected in the overall residuals of the governing PDEs? 
This case provides a perspective to explore whether the PDE residual, as a global error indicator, can effectively and adequately capture such large-scale critical flow features despite not relying on direct gradient information at the shock.

\begin{figure}[htb!]
	\centering  
	\subfigure[Geometry of the computational domain and boundary conditions]{
		\includegraphics[width=0.35\linewidth]{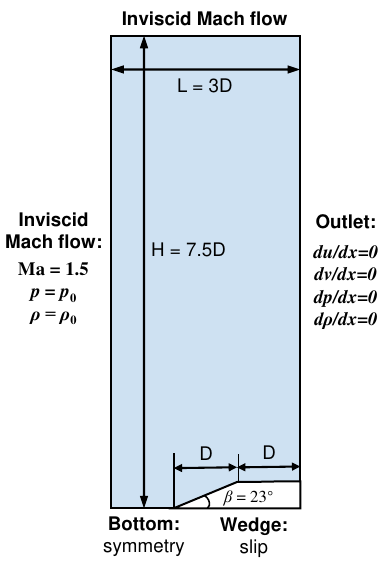}} 
	\subfigure[Initial triangular mesh]{
		\includegraphics[width=0.22\linewidth]{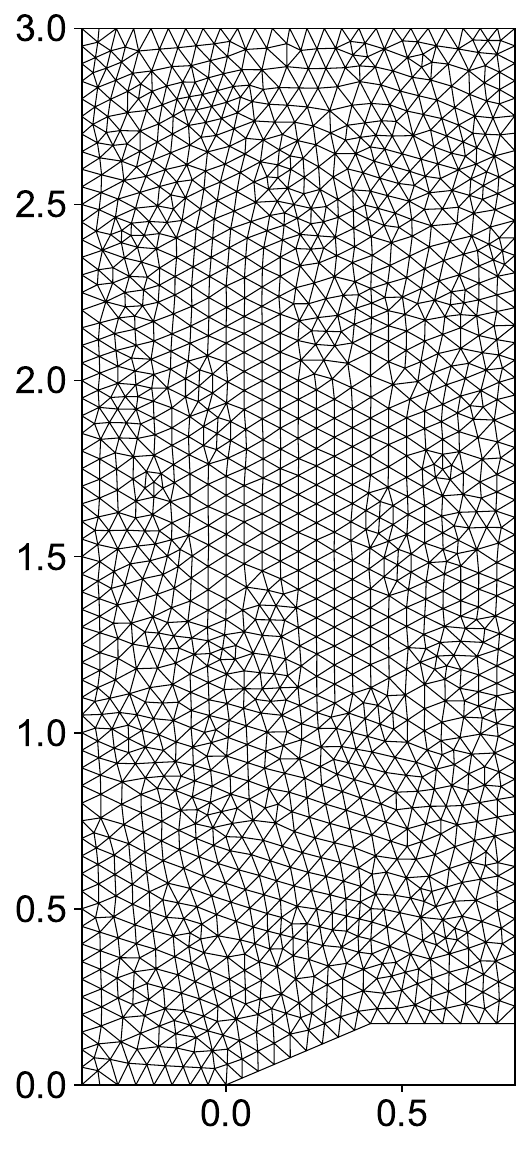}}
    \subfigure[Mach number distribution for the initial mesh]{
		\includegraphics[width=0.296\linewidth]{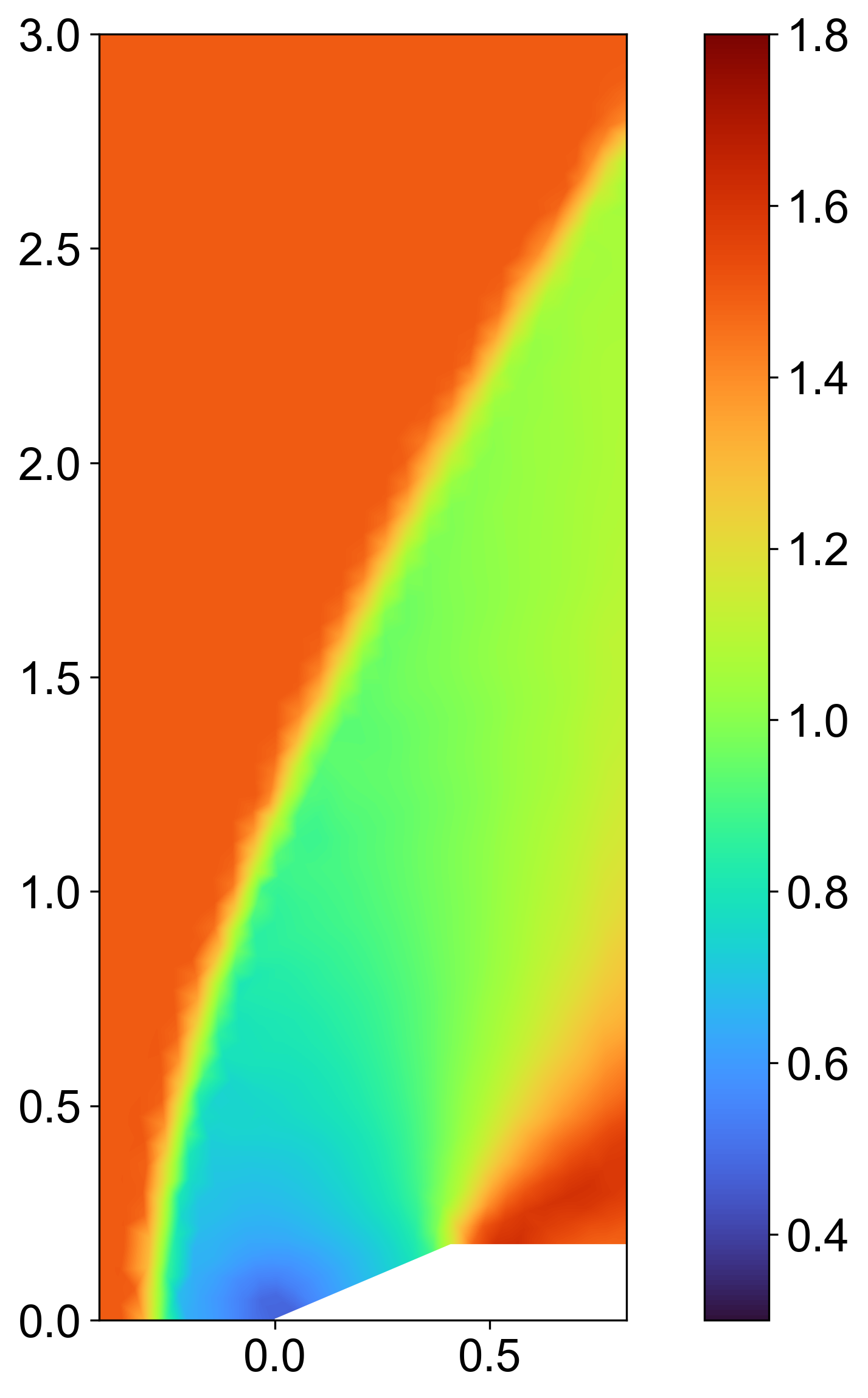}}
	\caption{Problem setup and initial coarse mesh for supersonic inviscid flow over a wedge.}
	\label{fig:wedge_setup}
\end{figure}
We consider supersonic inviscid flow over a 23-degree wedge of finite height. 
In this case, the finite height of the wedge is essential, as the shock wave at the leading edge detaches at a deflection angle of 23 degrees when the incoming Mach number is 1.5, forming a bow shock. Without the finite height, the shock wave standoff distance would become indeterminate.
The geometric design and boundary conditions of this flow are shown in Fig.~\ref{fig:wedge_setup}(a). 
We impose an inviscid uniform stream with a Mach number $M$=1.5 and constant pressure and density at the left inlet boundary and the top side of the far-field boundary, serving as Dirichlet boundary conditions. And slip boundary conditions are imposed on the wedge boundary.
The key to this simulation lies in accurately capturing the bow shock wave in front of the wedge. Since the bow shock extends across the entire computational domain, it is essential to increase the mesh resolution at all locations where the shock wave occurs.
Hence, the resolution of the bow shock wave is considered as physical quantities of interest in our analysis. 
We will measure the post-shock pressure distribution along the centerline at the bottom. The centerline is selected between the inlet and the front of the obstacle.
The initial coarse mesh for this case consists of 3,017 cells and 1,593 nodes, as provided in Fig.~\ref{fig:wedge_setup}(b). Mach number distribution for the initial mesh is depicted in Fig.~\ref{fig:wedge_setup}(c).  

\begin{figure}[htb!]
    \centering
    \includegraphics[width=0.6\linewidth]{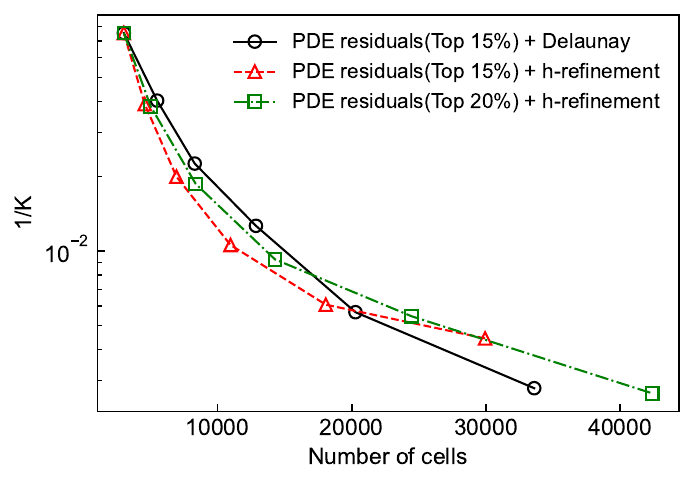}
    \caption{Convergence of the inverse of the slope of the shock wave during a few AMR runs for supersonic inviscid flow over a the wedge.} 
    \label{fig:wedge_inverse_slope}
\end{figure}

We evaluate the performance of the PDE residuals-based indicator using the fixed-fraction strategy combined with two types of refinement.
A key aspect that requires attention is the resolution of the captured shock wave. To assess this, we use the inverse of the shock wave's slope, 1/K, as a physical metric. 
Smaller values of K indicate a sharper capture of the shock, corresponding to higher resolution.
Fig.~\ref{fig:wedge_inverse_slope} shows the convergence of the inverse of the shock wave's slope with respect to the number of mesh cells during a few AMR runs. 
It can be observed that PDE residual-based AMR can attain convergence of the computed inverse slope of the shock wave to values close to zero when coupled with an appropriate mesh cell marking percentage. 
% It can be observed that the convergence of PDE residuals-based indicators is faster than traditional density gradient-based indicators. The refined meshes using the former approach capture sharp shock waves of equal resolution with fewer overall mesh points.
\begin{figure}[htb!]
    \centering
    \includegraphics[width=0.62\linewidth]{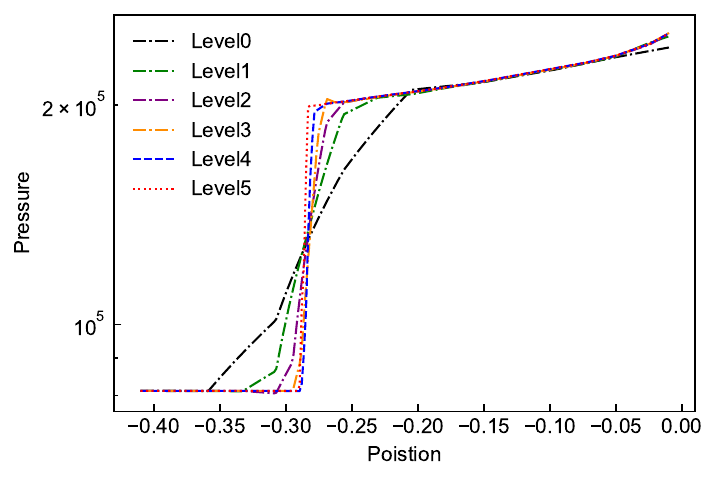}
    \caption{Supersonic inviscid flow over a wedge. AMR results with PDE residuals-based adaptation~(Top 15\%) and Delaunay refinement. Pressure distributions along the bottom centerline computed from adaptive meshes at each refinement level.} 
    \label{fig:wedge_PDED15_pressure}
\end{figure}

\begin{figure}[htb!] 
    \centering 
    \subfigure[PDE residuals~(Top 15\%) + h-refinement]{
		\includegraphics[width=0.24\linewidth]{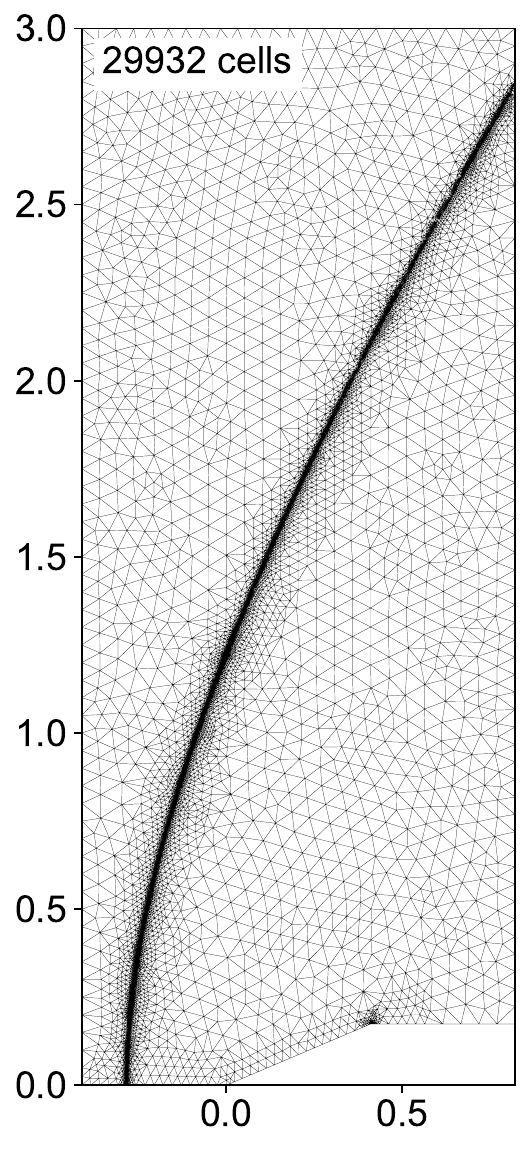}}\quad\quad
    \subfigure[PDE residuals~(Top 15\%) + Delaunay refinement]{
		\includegraphics[width=0.24\linewidth]{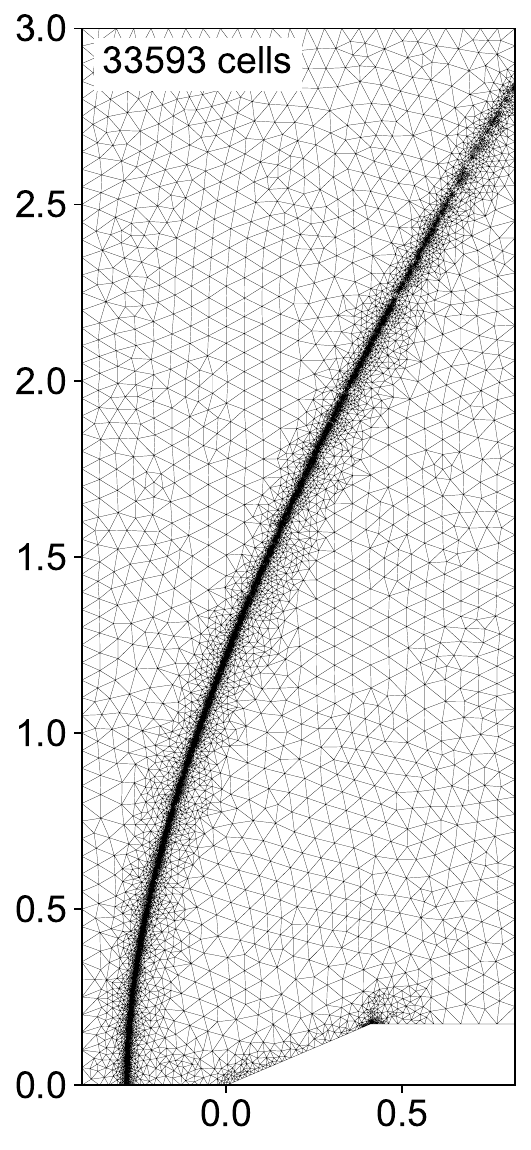}}\quad\quad
    \subfigure[PDE residuals~(Top 20\%) + h-refinement]{
		\includegraphics[width=0.24\linewidth]{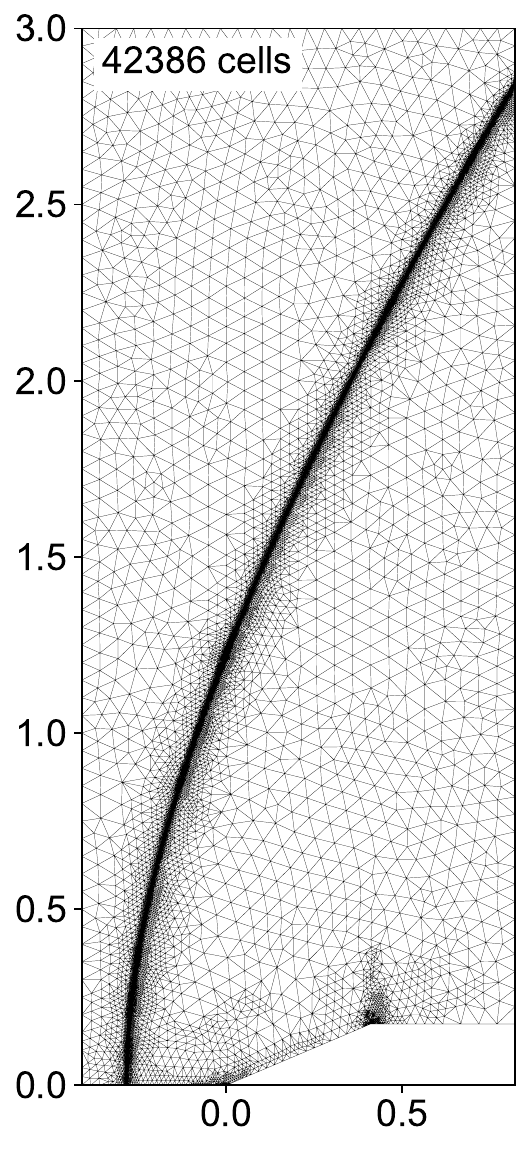}}
	\caption{Supersonic inviscid flow over a wedge. Adapted optimal meshes corresponding to the four adaptive schemes. The top left of each mesh shows its corresponding number of cells.}
	\label{fig:wedge_optimal_mesh}
\end{figure}

In Fig.~\ref{fig:wedge_PDED15_pressure}, we show the pressure distributions along the bottom centerline, computed from adaptive meshes at each refinement level, obtained using the PDE residual-based indicator with Delaunay refinement.
This indicates that the indicator effectively detects the shock wave and progressively refines the mesh at the shock location, resulting in an increasingly sharper capture of the pressure jump across the shock. 
We additionally present, in Fig.~\ref{fig:wedge_optimal_mesh}, the adapted optimal meshes corresponding to the three adaptive schemes. 
% Traditional AMR extensively refines both shock waves and expansion waves. The expanding wave area exhibits significant gradient variations, prompting standard gradient refinement to improve this region.
The PDE residuals-based indicator primarily focuses on refining the entire bow shock wave across the domain, while also performing slight refinement in the expansion wave region and along the surface of the wedge. 
In supersonic flow scenarios, the shock wave is indeed the main flow feature that requires prioritization.
% Traditional AMR dedicates considerable refinement to the regions of both shock waves and expansion waves. Given that the expanding wave area displays notable gradient variations, standard gradient refinement typically enhances this region as well. Conversely, residual-based AMR predominantly focuses on refining discontinuities in the bow shock wave, leading to minimal refinement in the rear swell wave region. In supersonic flow situations, the shock wave is indeed the crucial flow feature to concentrate on.

\begin{figure}[htb!]
	\centering  
    \subfigure[Level 0]{
		\includegraphics[width=0.21\linewidth]{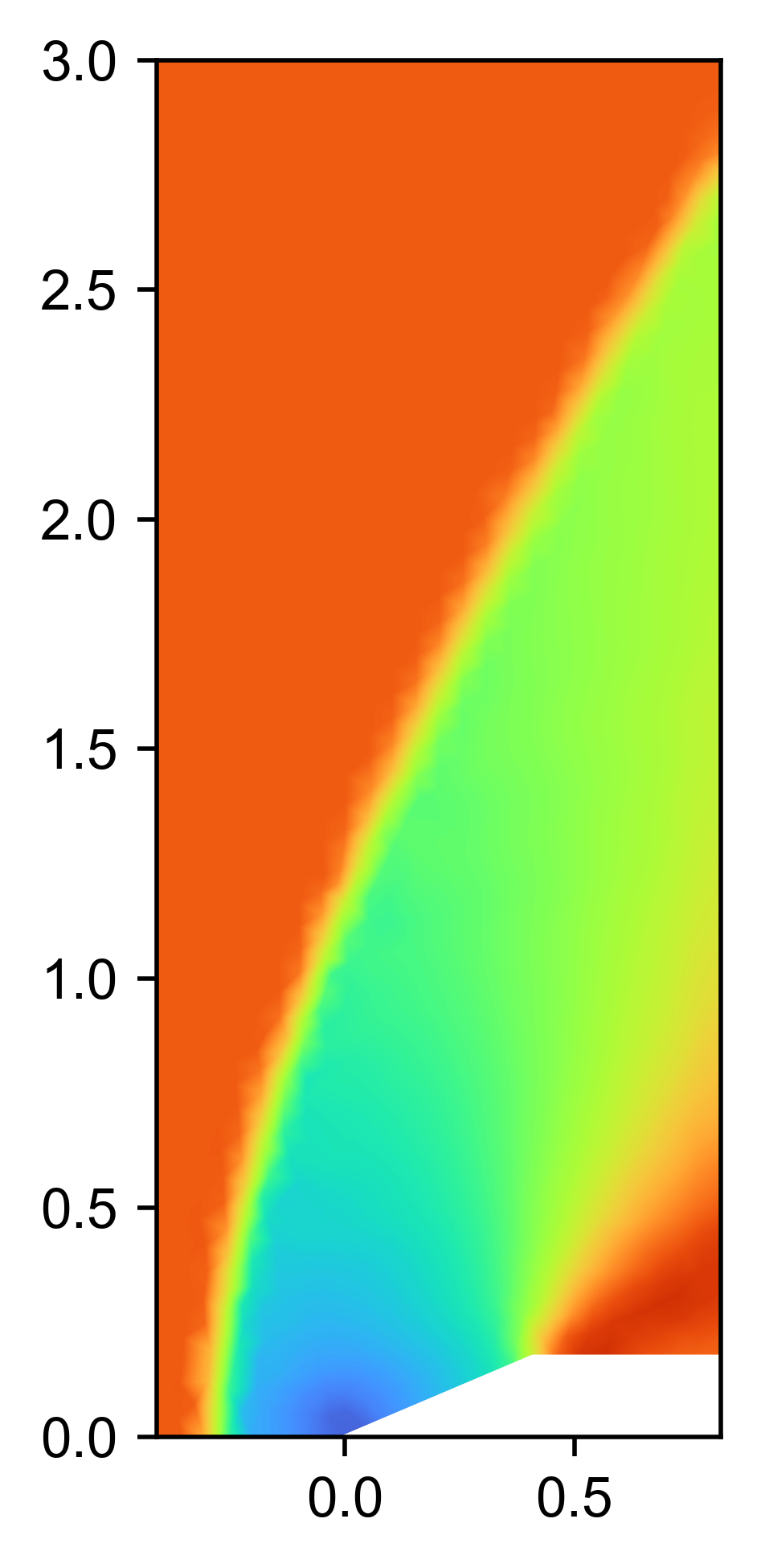}}\quad\quad
	\subfigure[Level 1]{
		\includegraphics[width=0.21\linewidth]{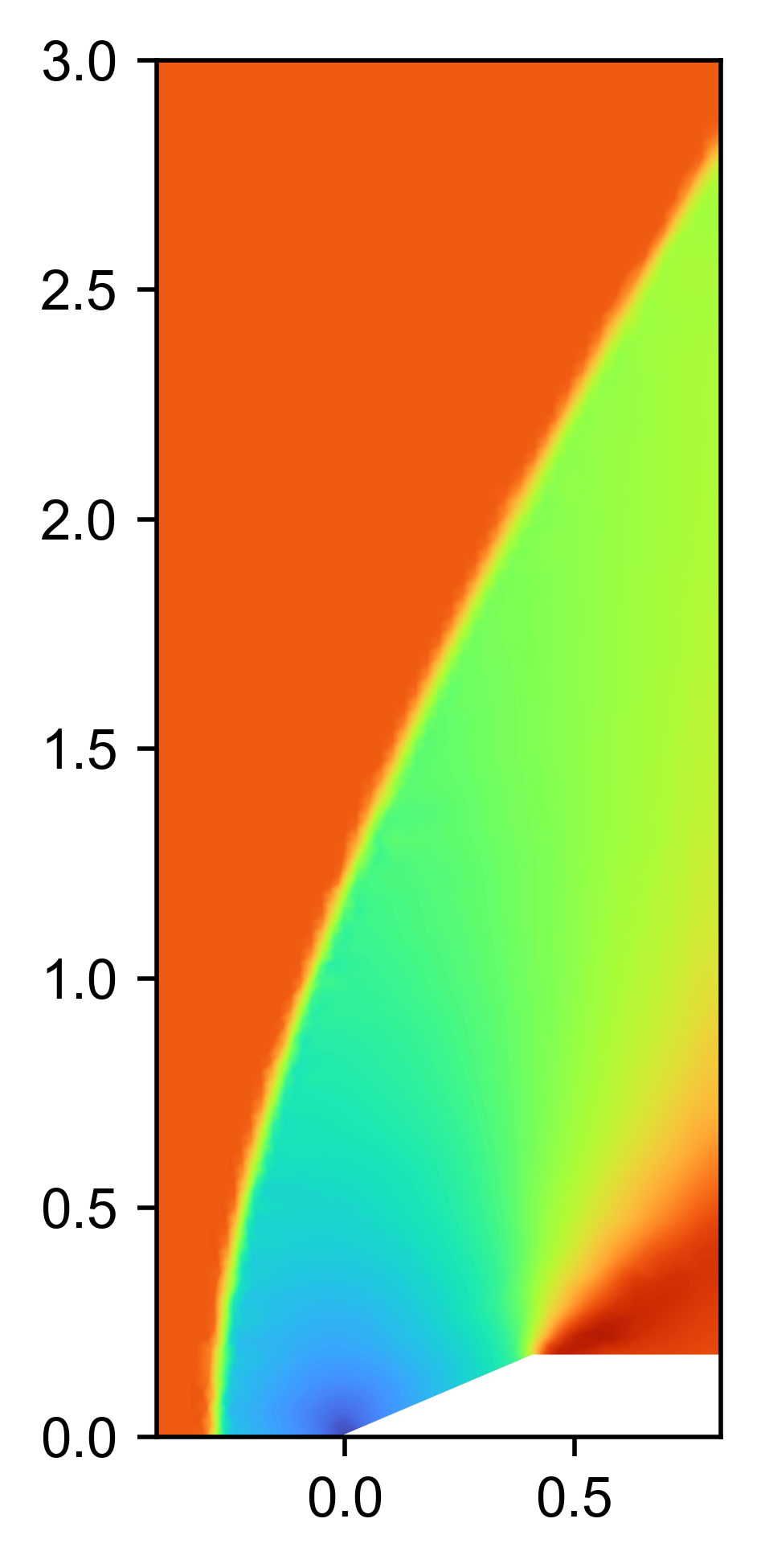}}\quad\quad
    \subfigure[Level 2]{
		\includegraphics[width=0.28\linewidth]{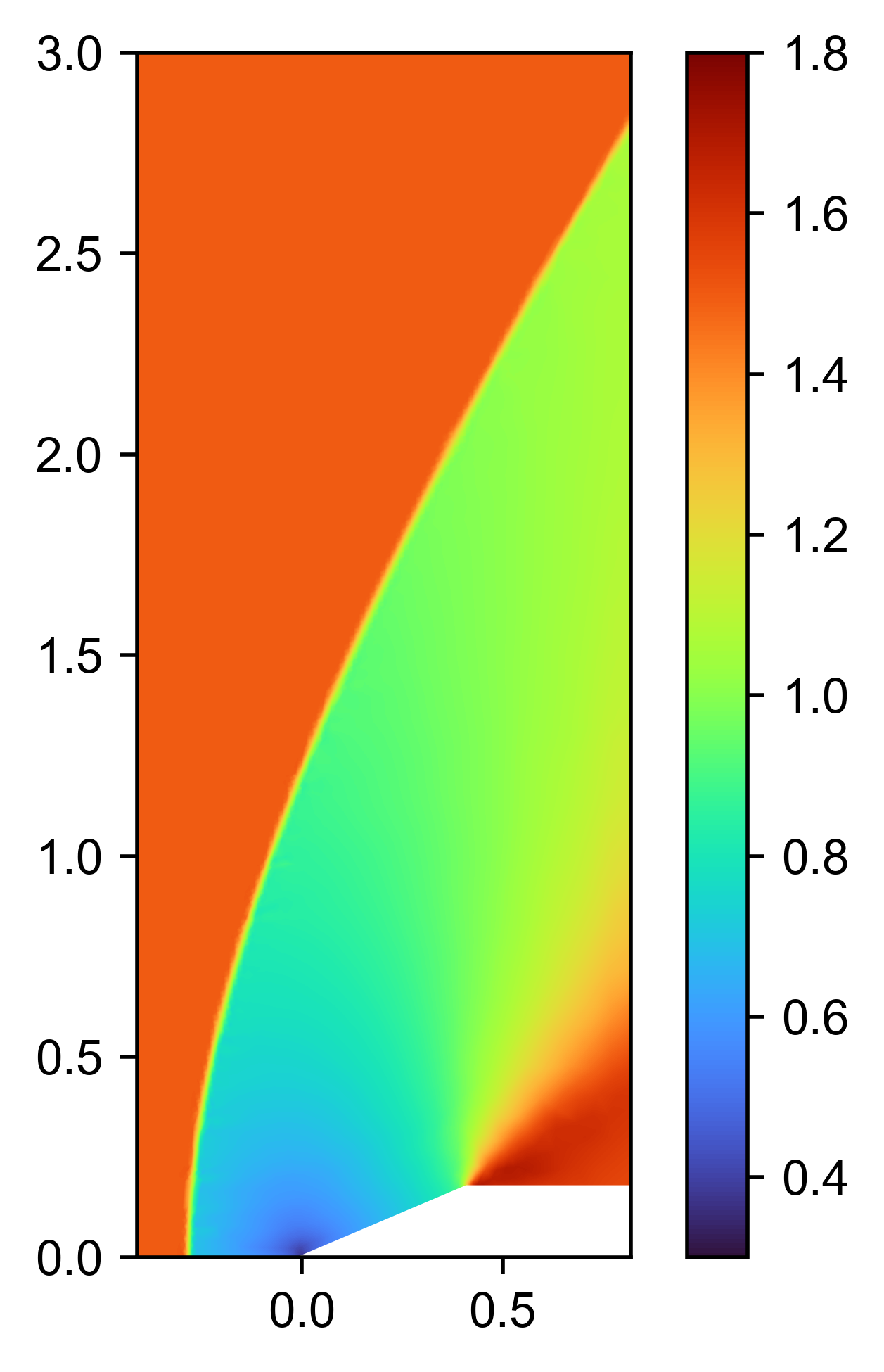}}%
    \\
	\subfigure[Level 3]{
		\includegraphics[width=0.21\linewidth]{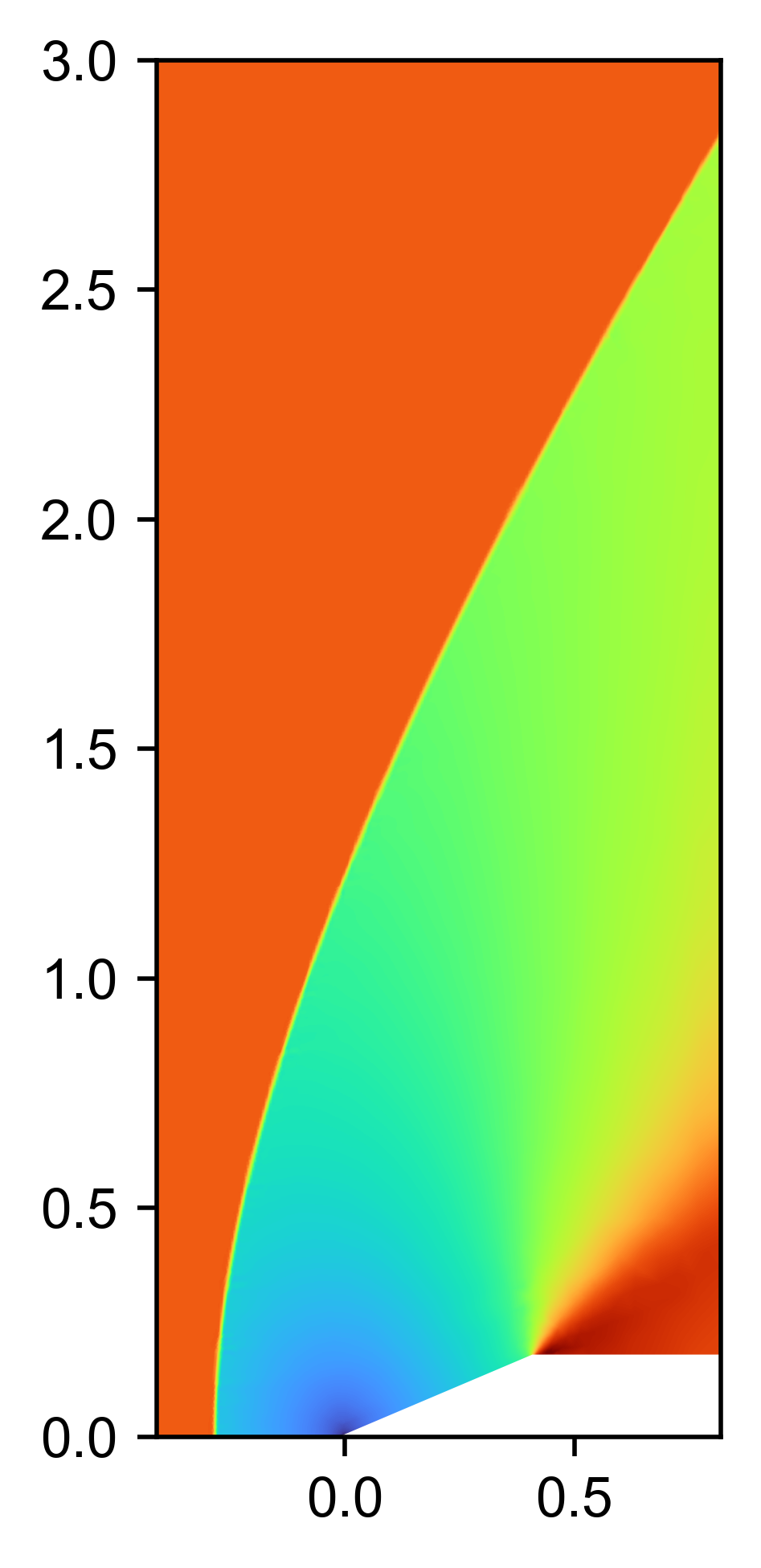}}\quad\quad
    \subfigure[Level 4]{
		\includegraphics[width=0.21\linewidth]{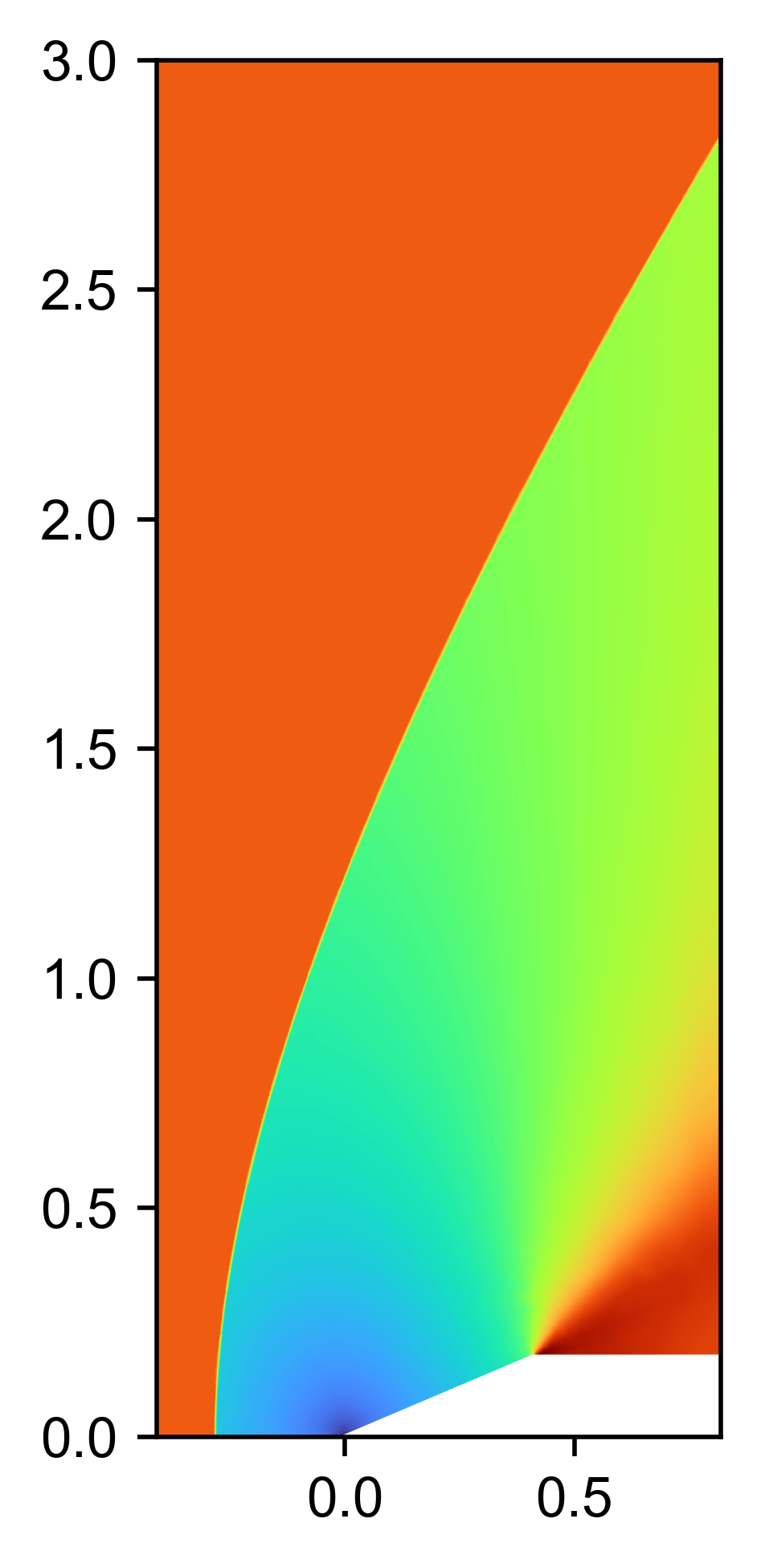}}\quad\quad
	\subfigure[Level 5]{
		\includegraphics[width=0.28\linewidth]{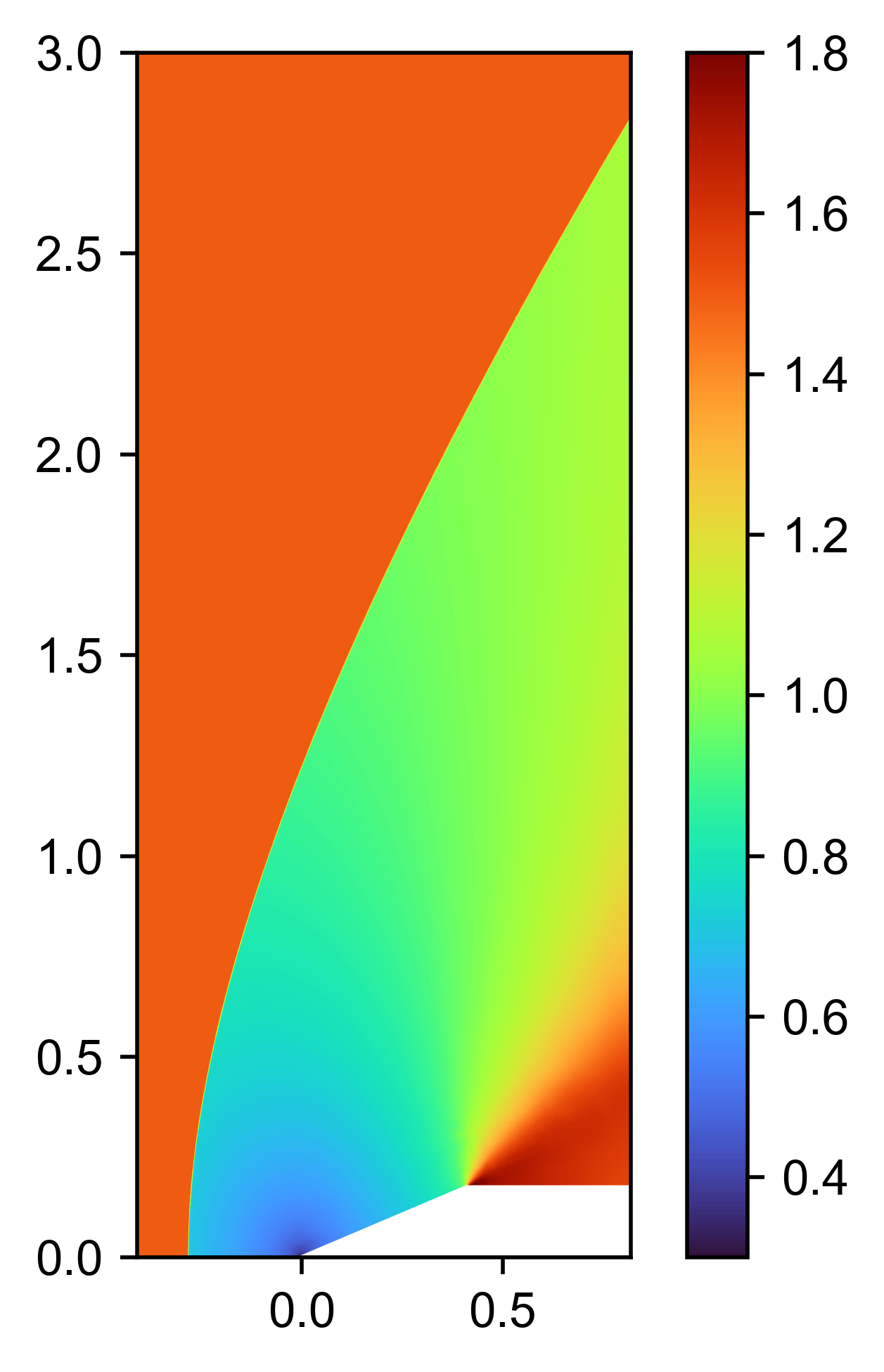}}%
	\caption{Supersonic inviscid flow over a wedge. AMR results with PDE residuals-based adaptation~(Top 20\%) and h-refinement. Mach number distributions computed from adaptive meshes at each refinement level.}
	\label{fig:wedge_ma}
\end{figure}

Fig.~\ref{fig:wedge_ma} depicts plots of the Mach number distributions computed from adaptive meshes at each refinement level, obtained using PDE residuals-based indicator~(Top 20\%) with h-refinement. 
The PDE residuals-based AMR approach, while not directly targeting the gradients, can still recognize and refine areas where such sharp features occur. 
By adapting the mesh at locations with significant overall residuals, the indicator effectively enhances resolution around the shock, where the physical flow features change rapidly. 
As the mesh refinement levels increase, a clear and sharp capturing of the entire bow shock wave becomes evident. 
This observation confirms that the large gradient features of the shock are indeed reflected in the overall PDE residuals.
Therefore, the PDE residuals-based indicator is suitable for completely capturing large-scale shocks with high resolution in supersonic flows.
This ability to capture the large-scale shock front through the PDE residuals rather than relying on local gradients provides a unique and potentially more robust method for refining the mesh in shock-dominated flows.

% Thus, this case demonstrates the potential advantage of using the PDE residuals-based AMR approach in flows with large-scale shock waves, as it offers a more global view of the error distribution and adapts the mesh more efficiently across the entire shock front.

% -------------------------- 6. Conclusions -----------------------------
\section{Conclusions}
\label{sec:Conclusions}
We have developed a PDE residual-based adaptive mesh refinement~(AMR) framework for the refinement of unstructured meshes and the solution of the steady incompressible and compressible flows. 
In particular, we employ physics-informed neural networks~(PINNs) to integrate physical equations with data of limited precision on a coarse mesh. The solution data on this mesh constitutes the data loss component for the PINNs, 
which are combined with the known physical equations to form the overall loss function.
The imposition of physical constraints ensures that the fitted flow field is aligned with the governing physics, while also ensuring that the solution data at the mesh nodes is matched. 
Once trained, PINNs are capable of identifying regions with the highest residuals of the Navier-Stokes/Euler equations at the centers of coarse mesh cells, thereby indicating which cells may require refinement.
In contrast to the conventional SOLVE$\rightarrow$ESTIMATE$\rightarrow$MARK$\rightarrow$REFINE AMR procedure, 
we employed PINNs on the supplementary TRAIN stage to integrate the physical equations and coarse solutions, subsequently predicting the PDE residuals at the mesh cell centers on the ESTIMATE stage with minimal effort.
Furthermore, two types of mesh refinement were introduced: h-refinement and Delaunay refinement. Both refinement methods can guarantee the quality of the mesh after refinement, thus avoiding the potential issue of triangular mesh cells with acute angles.
In contrast to the traditional gradient-based and curvature~(Hessian)-based error indicators for CFD solver, the global PDEs residuals were employed as the criterion for adaptive refinement.
This approach enables the accurate capture of solution features and efficient refinement of unstructured meshes.

The proposed residual-based AMR framework has been validated on a variety of classical steady flow problems. 
For instance, the NS equations were solved for a laminar flow around an obstacle.
Furthermore, the framework was validated by solving Euler equations on three classical inviscid flows. 
We begin with a two-dimensional, steady, incompressible laminar flow problem in Section.~\ref{subsec:laminar flow around an obstacle} as an illustrative example to demonstrate the feasibility of the approach and to provide a detailed examination of the characteristics of the residual-based AMR.
In the context of flow around an obstacle, AMR guided by PDE residuals tends to refine the areas in proximity to the three surfaces of the obstacle. 
Appropriate mesh refinement ratios enable residual-based AMR to achieve convergence of velocity magnitudes and overall distributions along monitoring lines in close proximity to reference values, while simultaneously reducing the mesh cell count by approximately fourfold in comparison to traditional AMR.
In subsonic flow around airfoils with less pronounced flow characteristics, the PDE residuals-guided AMR prioritizes mesh refinement at critical locations, such as the leading and trailing edges of the airfoil. When combined with an appropriate mesh selection ratio, this residual-based AMR can converge the computed lift to values close to the reference value while utilizing approximately 4 to 8 times fewer mesh elements than traditional AMR methods.
In transonic flow around airfoils exhibiting discontinuous shock waves, the residual-based AMR concentrates refinement on critical locations, such as the upper edge shock wave and the leading and trailing edges, while limiting refinement in other areas. The residual-based AMR can attain convergence of the computed lift and drag to values proximate to the reference values, while reducing the number of mesh elements by approximately a factor of 4 to 7 in comparison to traditional AMR methods. 
In the case of supersonic flow over a wedge with a discontinuous bow shock wave, the residual-based AMR concentrates refinement on the entirety of the shock wave, while limiting refinement in other regions. A clear and sharp capturing of the entire bow shock wave becomes evident.
This ability to capture the large-scale shock front through the PDE residuals rather than relying on local gradients provides a unique and potentially more robust method for refining the mesh in shock-dominated flows.
Overall, our proposed PDE residuals-guided AMR framework is both flexible and effective, providing efficient and high-quality solutions for a wide range of flow problems.

The present study yields encouraging outcomes for steady-state problems encompassing both incompressible and compressible flows. In the context of transient problems, the proposed refinement criteria based on PDE residuals remain applicable, enabling PINNs to identify regions of the mesh that require refinement, even when the initial mesh is coarse and the data has low precision. It should be noted, however, that transient problems require a solution-specific mesh at each time step, as well as interpolation of mesh information from one time step to the next. This process entails the coarsening of the mesh and the utilization of dynamic meshing techniques, which will be addressed in future research.
This study primarily focuses on computational fluid dynamics, with an emphasis on the Navier-Stokes equations and the Euler equations. However, the underlying principles of the PDE residual-based AMR framework are broadly applicable. In addition to the two equations mentioned, it can be extended to solve other types of PDEs.
The framework proposed in this study provides valuable insights for the development of more advanced AMR indicators and possesses significant potential, making it a promising field for exploration in the future.

\section*{Declaration of competing interest}
The authors declare that they have no known competing financial interests or personal relationships that could have appeared to influence the work reported in this paper.

% ------------------------------------ ACKNOWLEDGMENTS ------------------------------------
\section*{Acknowledgments}
Support from the grant of the National Key R\&D Program of China under contract number 2022YFA1203200 and the National Natural Science Foundation of China (No. 12172330) is gratefully acknowledged.

% --------------------------------------- Appendix --------------------------------------
%% The Appendices part is started with the command \appendix;
%% appendix sections are then done as normal sections
\appendix
\section{Additional AMR convergence results}
\label{Appendix}
% This appendix presents
This appendix presents additional convergence results and sensitivity analyses related to the AMR approach discussed in Section.~\ref{sec:Numerical results and discussions}. Specifically, we provide a comparison of the AMR convergence behavior under varying refinement criteria, including the PDE residual-based indicator with different percentages of cells marked, as well as sensitivity studies involving different threshold values for other error indicators like the gradient-based indicator.
Furthermore, we include similar convergence analyses for other traditional AMR indicators such as the pressure Hessian-based and gradient-based indicators, with sensitivity to various percentages of cells marked. These results aim to highlight the influence of different refinement strategies on solution accuracy and computational efficiency.

\begin{figure}[h!]
    \centering
    \subfigure[Velocity amplitude convergence]{
		\includegraphics[width=0.488\linewidth]{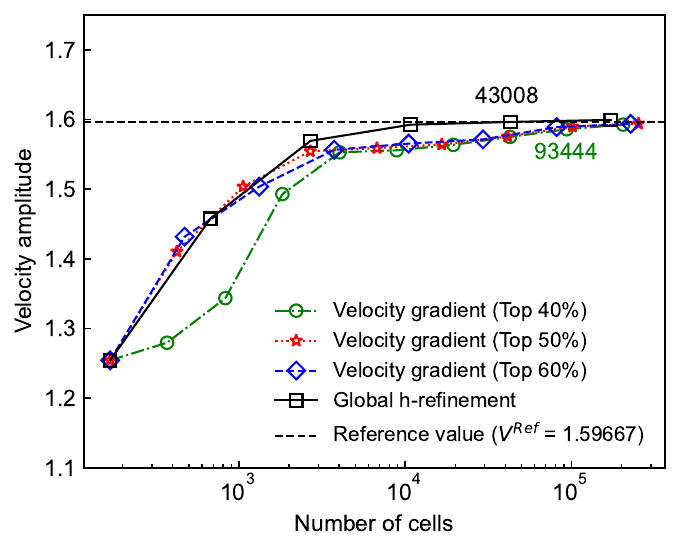}}%  
	\subfigure[Velocity distribution convergence]{
		\includegraphics[width=0.5\linewidth]{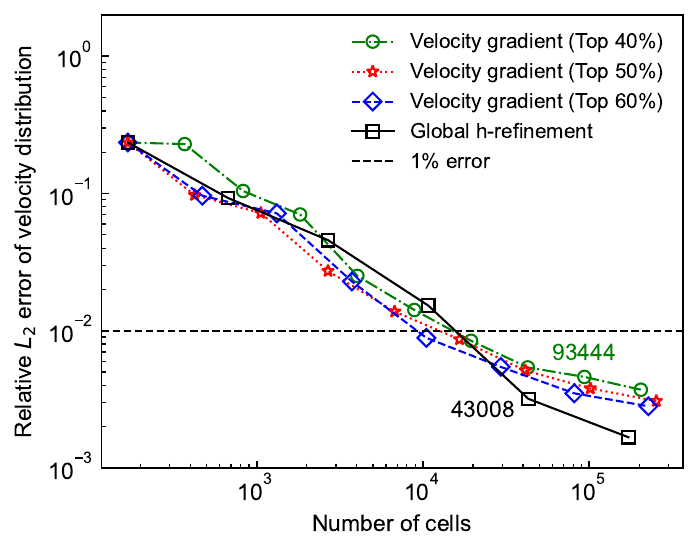}}%
    \caption{Laminar flow around an obstacle. Convergence of the velocity amplitude and distribution along the monitored line during velocity gradient-based AMR runs with different fixed percentages of cells marked. The number of cells in the optimal mesh for each corresponding AMR scheme is shown near the respective symbol.} 
    \label{fig:obstacle_velocity_top_n_per}
\end{figure}

\begin{figure}[h!]
    \centering
    \subfigure[Velocity amplitude convergence]{
		\includegraphics[width=0.488\linewidth]{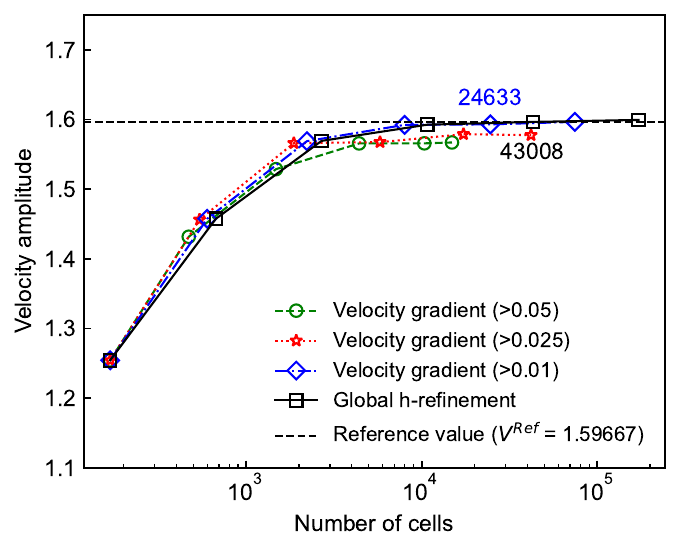}}%  
	\subfigure[Velocity distribution convergence]{
		\includegraphics[width=0.5\linewidth]{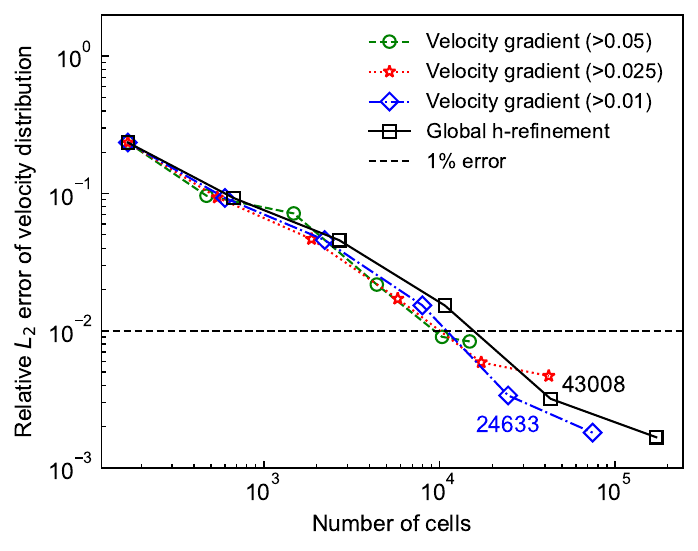}}%
    \caption{Laminar flow around an obstacle. Convergence of the velocity amplitude and distribution along the monitored line during velocity gradient-based AMR runs with different fixed velocity gradient thresholds for cell marking. The number of cells in the optimal mesh for each corresponding AMR scheme is shown near the respective symbol.} 
    \label{fig:obstacle_velocity_threshold}
\end{figure}

\begin{figure}[h!]
    \centering
    \subfigure[Velocity amplitude convergence]{
		\includegraphics[width=0.488\linewidth]{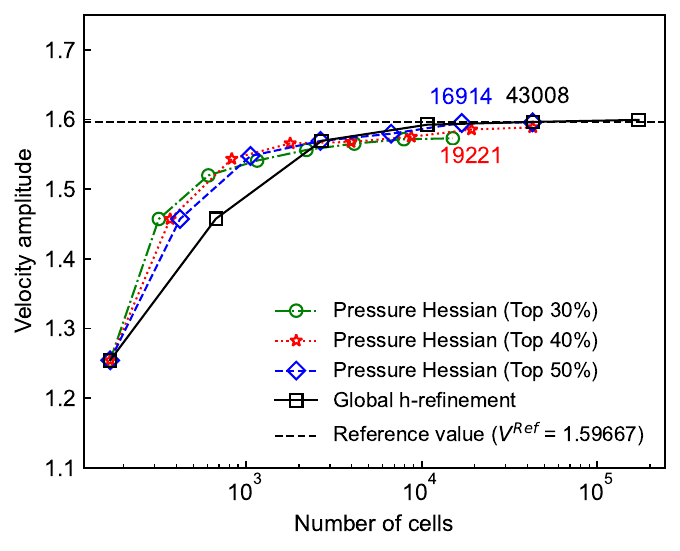}}%  
	\subfigure[Velocity distribution convergence]{
		\includegraphics[width=0.5\linewidth]{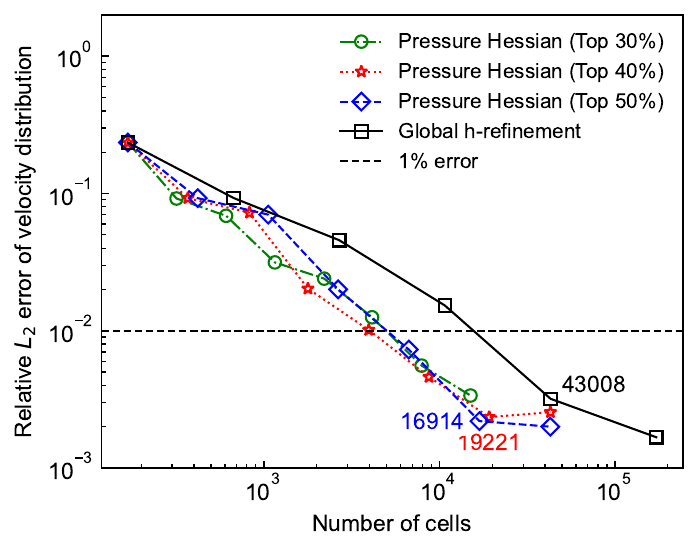}}%
    \caption{Laminar flow around an obstacle. Convergence of the velocity amplitude and distribution along the monitored line during pressure Hessian-based AMR runs with different fixed percentages of cells marked. The number of cells in the optimal mesh for each corresponding AMR scheme is shown near the respective symbol.} 
    \label{fig:obstacle_pressure_top_n_per}
\end{figure}

% ------------------ Case1: laminar flow around an obstacle ---------------------
For the case of laminar flow around an obstacle in Section.~\ref{subsec:laminar flow around an obstacle}, the convergence results for the velocity gradient-based AMR runs are shown in Fig.~\ref{fig:obstacle_velocity_top_n_per}, where different fixed percentages~(i.e., top 40\%, 50\%, and 60\%) of cells marked are applied. 
The convergence results for the velocity gradient-based AMR runs are shown in Fig.~\ref{fig:obstacle_velocity_threshold}, where different fixed velocity gradient thresholds~(i.e., \textgreater 0.05, \textgreater 0.025, and \textgreater 0.01) of cells marked are applied. 
We also examine the convergence results for the pressure Hessian-based AMR runs in Fig.~\ref{fig:obstacle_pressure_top_n_per}, where different fixed percentages~(i.e., top 30\%, 40\%, and 50\%) of cells marked are applied. 

\begin{figure}[h!]
    \centering
    \subfigure[Density gradient-based AMR runs]{
		\includegraphics[width=0.5\linewidth]{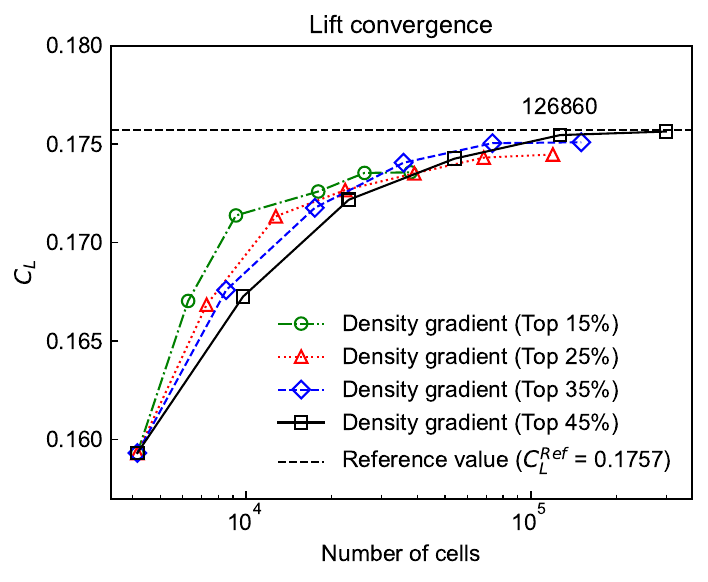}}%  
	\subfigure[Pressure Hessian-based AMR runs]{
		\includegraphics[width=0.5\linewidth]{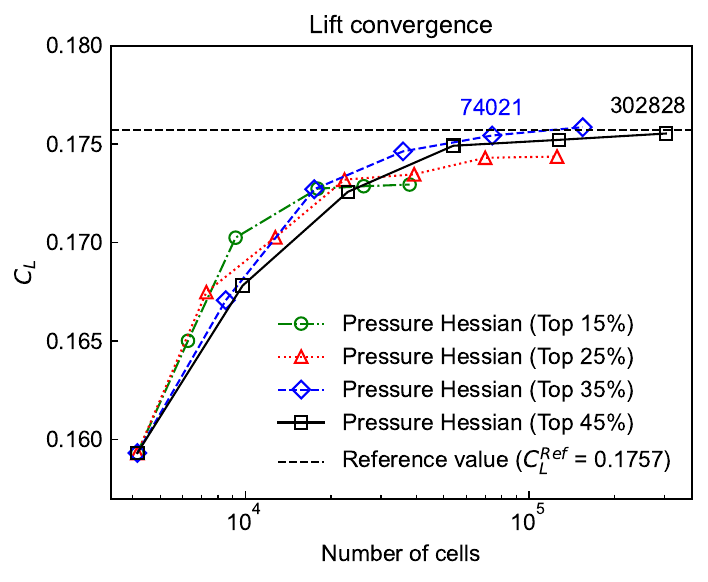}}%
    \caption{Subsonic inviscid flow over the NACA 0012 airfoil~(M=0.5, $\alpha$=1.25°). The lift coefficient convergence during density gradient-based and pressure Hessian-based AMR runs with different fixed percentages of cells marked. The number of cells in the optimal mesh for each corresponding AMR scheme is shown near the respective symbol.} 
    \label{fig:subsonic_traditional_top_n_per_lift}
\end{figure}

% ------------------ Case2: Subsonic inviscid flow over an airfoil ------------------
For the case of subsonic inviscid flow over an airfoil in Section.~\ref{subsec:Subsonic inviscid flow over an airfoil}, we examine the convergence results for the density gradient-based and pressure Hessian-based AMR runs in Fig.~\ref{fig:subsonic_traditional_top_n_per_lift}, where different fixed percentages~(i.e., top 15\%, 25\%, 35\%, and 45\%) of cells marked are applied. 

\begin{figure}[h!]
    \centering
    \subfigure{
		\includegraphics[width=0.486\linewidth]{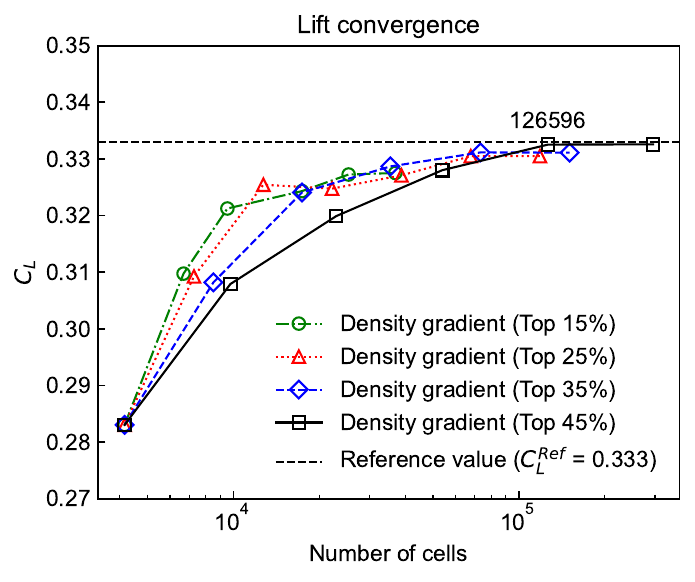}}%  
	\subfigure{
		\includegraphics[width=0.5\linewidth]{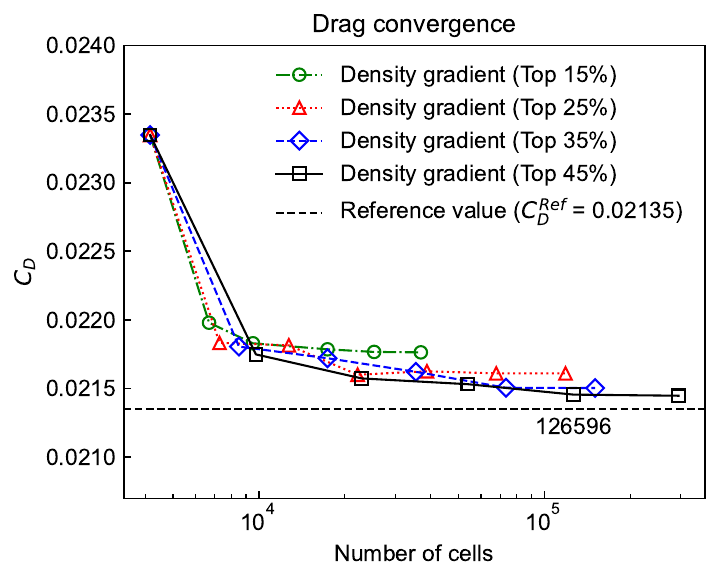}}%
    \caption{Transonic inviscid flow over the NACA 0012 airfoil~(M=0.8, $\alpha$=1.25°). The lift and drag coefficients convergence during density gradient-based AMR runs with different fixed percentages of cells marked. The number of cells in the optimal mesh for each corresponding AMR scheme is shown near the respective symbol.} 
    \label{fig:transonic_density_top_n_per_lift_drag}
\end{figure}

\begin{figure}[h!]
    \centering
    \subfigure{
		\includegraphics[width=0.486\linewidth]{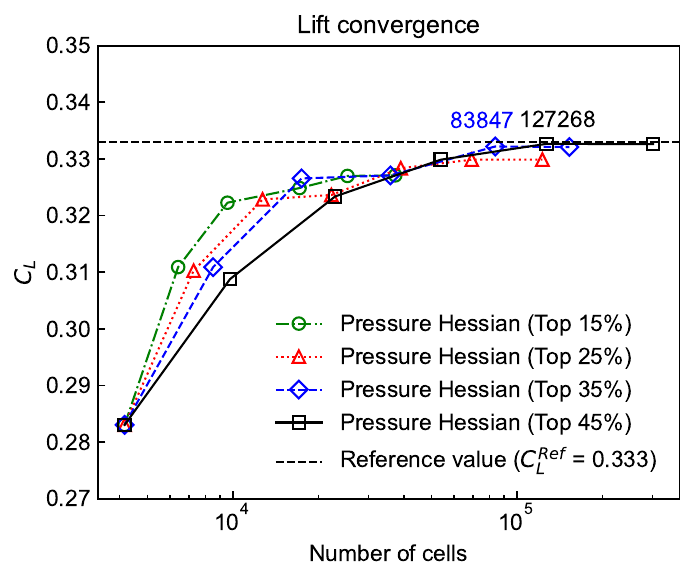}}%  
	\subfigure{
		\includegraphics[width=0.5\linewidth]{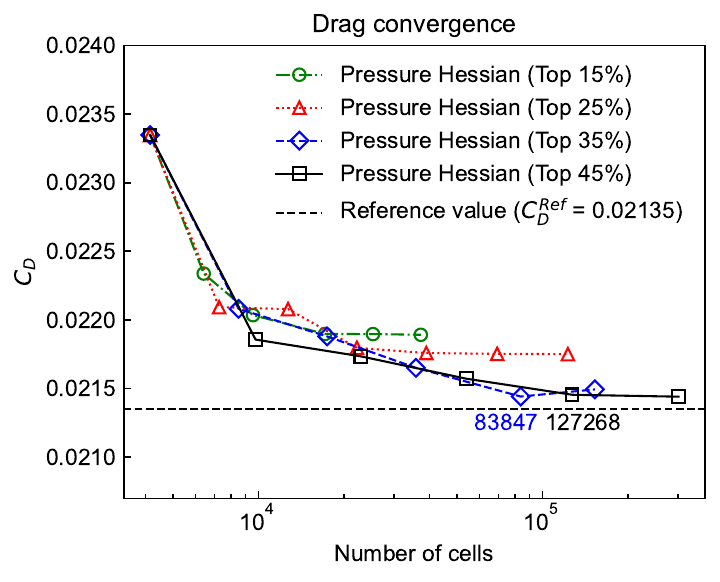}}%
    \caption{Transonic inviscid flow over the NACA 0012 airfoil~(M=0.8, $\alpha$=1.25°). The lift and drag coefficients convergence during pressure Hessian-based AMR runs with different fixed percentages of cells marked. The number of cells in the optimal mesh for each corresponding AMR scheme is shown near the respective symbol.} 
    \label{fig:transonic_pressure_top_n_per_lift_drag}
\end{figure}

\begin{figure}[htb!]
    \centering
    \subfigure[Delaunay refinement]{
		\includegraphics[width=0.49\linewidth]{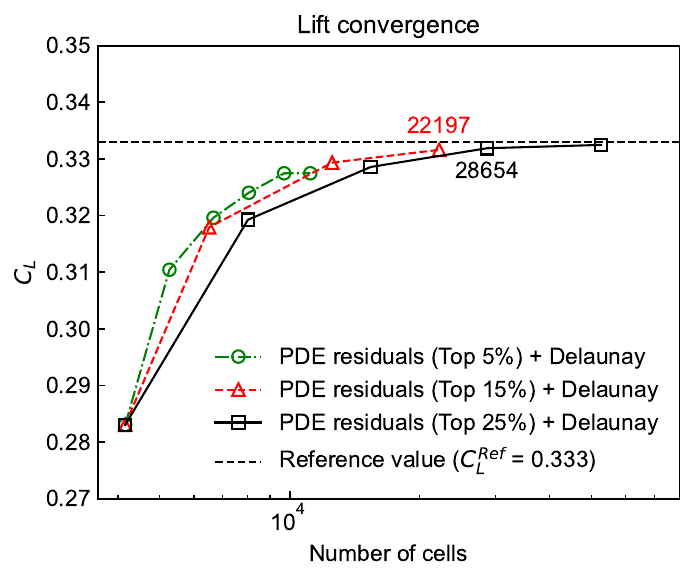}%  
		\includegraphics[width=0.51\linewidth]{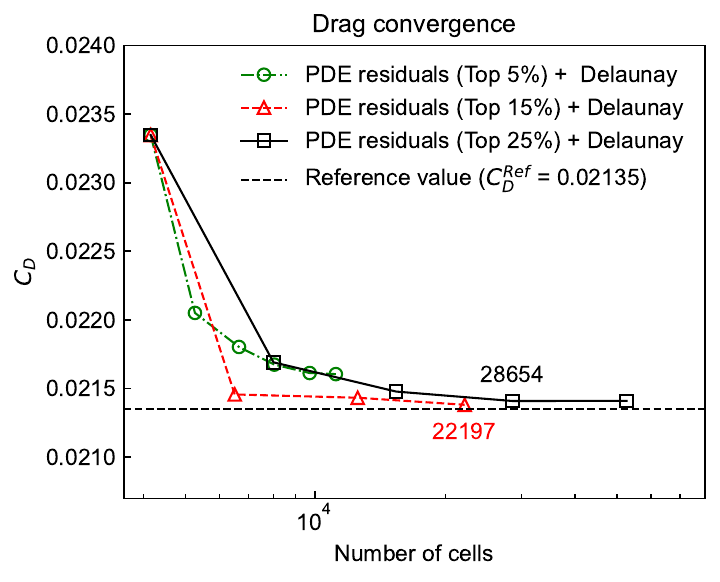}}%
    \\
    \subfigure[h-refinement]{
		\includegraphics[width=0.49\linewidth]{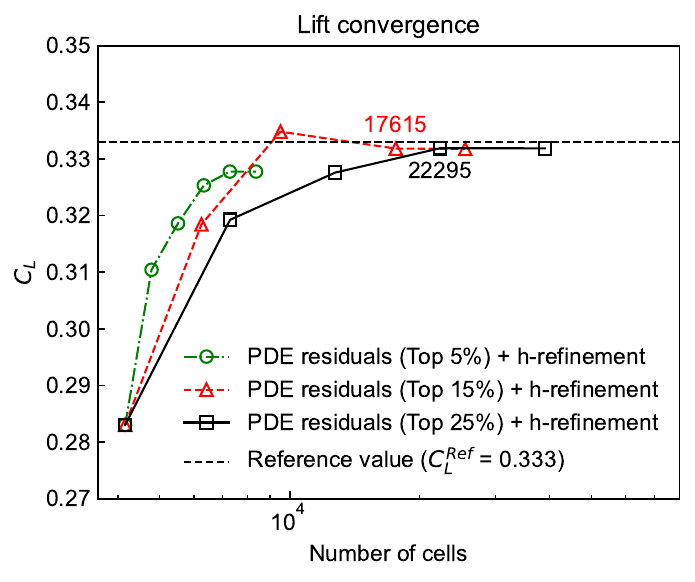}%  
		\includegraphics[width=0.51\linewidth]{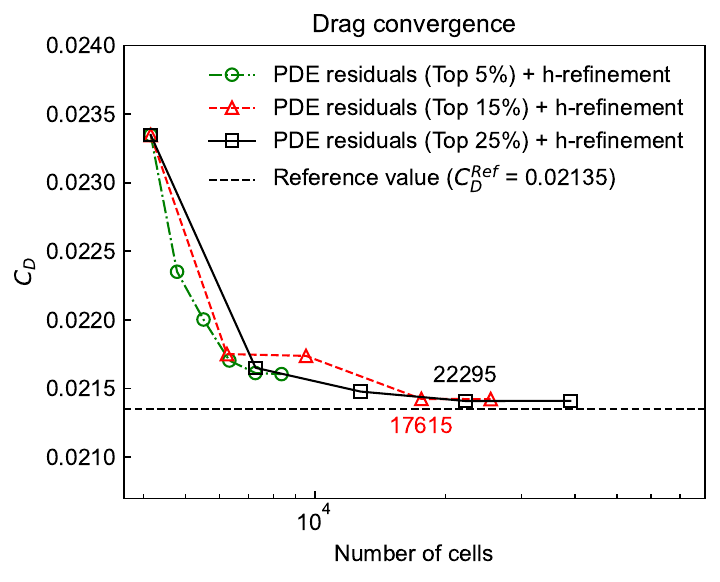}}%
    \caption{Transonic inviscid flow over the NACA 0012 airfoil~(M=0.8, $\alpha$=1.25°). The lift and drag coefficients convergence during PDE residuals-based AMR runs with different fixed percentages of cells marked. The number of cells in the optimal mesh for each corresponding AMR scheme is shown near the respective symbol.} 
    \label{fig:transonic_PDE_top_n_per_lift_drag}
\end{figure}

% ----------------- Case3: Transonic inviscid flow over an airfoil ------------------
For the case of transonic inviscid flow over an airfoil in Section.~\ref{subsec:Transonic inviscid flow over an airfoil}, the convergence results for the density gradient-based AMR runs are shown in Fig.~\ref{fig:transonic_density_top_n_per_lift_drag}, where different fixed percentages~(i.e., top 15\%, 25\%, 35\%, and 45\%) of cells marked are applied. 
And the convergence results for the pressure Hessian-based AMR runs are shown in Fig.~\ref{fig:transonic_pressure_top_n_per_lift_drag}, where different fixed percentages~(i.e., top 15\%, 25\%, 35\%, and 45\%) of cells marked are applied. 
Fig.~\ref{fig:transonic_PDE_top_n_per_lift_drag} shows the convergence results from PDE residuals-based AMR runs with three fixed percentages~(i.e., top 5\%, 15\%, and 25\%) of cells marked and using two types of refinement~(i.e., h-refinement and Delaunay refinement).
% --------------------------------------- References ---------------------------------------

%% If you have bibdatabase file and want bibtex to generate the
%% bibitems, please use
%%
\bibliographystyle{elsarticle-num-names} 
\bibliography{main}

\begin{thebibliography}{48}
\expandafter\ifx\csname natexlab\endcsname\relax\def\natexlab#1{#1}\fi
\providecommand{\url}[1]{\texttt{#1}}
\providecommand{\href}[2]{#2}
\providecommand{\path}[1]{#1}
\providecommand{\DOIprefix}{doi:}
\providecommand{\ArXivprefix}{arXiv:}
\providecommand{\URLprefix}{URL: }
\providecommand{\Pubmedprefix}{pmid:}
\providecommand{\doi}[1]{\href{http://dx.doi.org/#1}{\path{#1}}}
\providecommand{\Pubmed}[1]{\href{pmid:#1}{\path{#1}}}
\providecommand{\bibinfo}[2]{#2}
\ifx\xfnm\relax \def\xfnm[#1]{\unskip,\space#1}\fi
%Type = Book
\bibitem[{Brenner(2008)}]{brenner2008mathematical}
\bibinfo{author}{S.~C. Brenner}, \bibinfo{title}{The mathematical theory of finite element methods}, \bibinfo{publisher}{Springer}, \bibinfo{year}{2008}.
%Type = Book
\bibitem[{Karniadakis and Sherwin(2005)}]{karniadakis2005spectral}
\bibinfo{author}{G.~Karniadakis}, \bibinfo{author}{S.~J. Sherwin}, \bibinfo{title}{{Spectral/hp element methods for computational fluid dynamics}}, \bibinfo{publisher}{Oxford University Press, USA}, \bibinfo{year}{2005}.
%Type = Book
\bibitem[{Moukalled et~al.(2016)Moukalled, Mangani, Darwish, Moukalled, Mangani, and Darwish}]{moukalled2016finite}
\bibinfo{author}{F.~Moukalled}, \bibinfo{author}{L.~Mangani}, \bibinfo{author}{M.~Darwish}, \bibinfo{author}{F.~Moukalled}, \bibinfo{author}{L.~Mangani}, \bibinfo{author}{M.~Darwish}, \bibinfo{title}{The finite volume method}, \bibinfo{publisher}{Springer}, \bibinfo{year}{2016}.
%Type = Article
\bibitem[{Tam et~al.(2000)Tam, Ait-Ali-Yahia, Robichaud, Moore, Kozel, and Habashi}]{tam2000anisotropic}
\bibinfo{author}{A.~Tam}, \bibinfo{author}{D.~Ait-Ali-Yahia}, \bibinfo{author}{M.~Robichaud}, \bibinfo{author}{M.~Moore}, \bibinfo{author}{V.~Kozel}, \bibinfo{author}{W.~Habashi},
\newblock \bibinfo{title}{{Anisotropic mesh adaptation for 3D flows on structured and unstructured grids}},
\newblock \bibinfo{journal}{Comput. Methods Appl. Mech. Eng.} \bibinfo{volume}{189} (\bibinfo{year}{2000}) \bibinfo{pages}{1205--1230}.
%Type = Article
\bibitem[{Murayama and Yamamoto(2008)}]{murayama2008comparison}
\bibinfo{author}{M.~Murayama}, \bibinfo{author}{K.~Yamamoto},
\newblock \bibinfo{title}{Comparison study of drag prediction by structured and unstructured mesh method},
\newblock \bibinfo{journal}{J. Aircr.} \bibinfo{volume}{45} (\bibinfo{year}{2008}) \bibinfo{pages}{799--822}.
%Type = Article
\bibitem[{Katz and Sankaran(2011)}]{katz2011mesh}
\bibinfo{author}{A.~Katz}, \bibinfo{author}{V.~Sankaran},
\newblock \bibinfo{title}{{Mesh quality effects on the accuracy of CFD solutions on unstructured meshes}},
\newblock \bibinfo{journal}{J. Comput. Phys.} \bibinfo{volume}{230} (\bibinfo{year}{2011}) \bibinfo{pages}{7670--7686}.
%Type = Article
\bibitem[{Alakashi and Basuno(2014)}]{alakashi2014comparison}
\bibinfo{author}{A.~M. Alakashi}, \bibinfo{author}{I.~B. Basuno},
\newblock \bibinfo{title}{{Comparison between structured and unstructured grid generation on two dimensional flows based on Finite Volume Method (FVM)}},
\newblock \bibinfo{journal}{Int. J. Min., Metall. Mech. Eng} \bibinfo{volume}{2} (\bibinfo{year}{2014}) \bibinfo{pages}{97--103}.
%Type = Article
\bibitem[{Berger and Oliger(1984)}]{berger1984adaptive}
\bibinfo{author}{M.~J. Berger}, \bibinfo{author}{J.~Oliger},
\newblock \bibinfo{title}{Adaptive mesh refinement for hyperbolic partial differential equations},
\newblock \bibinfo{journal}{J. Comput. Phys.} \bibinfo{volume}{53} (\bibinfo{year}{1984}) \bibinfo{pages}{484--512}.
%Type = Article
\bibitem[{Berger and Colella(1989)}]{berger1989local}
\bibinfo{author}{M.~J. Berger}, \bibinfo{author}{P.~Colella},
\newblock \bibinfo{title}{Local adaptive mesh refinement for shock hydrodynamics},
\newblock \bibinfo{journal}{J. Comput. Phys.} \bibinfo{volume}{82} (\bibinfo{year}{1989}) \bibinfo{pages}{64--84}.
%Type = Article
\bibitem[{Baker(1997)}]{baker1997mesh}
\bibinfo{author}{T.~J. Baker},
\newblock \bibinfo{title}{Mesh adaptation strategies for problems in fluid dynamics},
\newblock \bibinfo{journal}{Finite Elem. Anal. Des.} \bibinfo{volume}{25} (\bibinfo{year}{1997}) \bibinfo{pages}{243--273}.
%Type = Article
\bibitem[{Morin et~al.(2000)Morin, Nochetto, and Siebert}]{morin2000data}
\bibinfo{author}{P.~Morin}, \bibinfo{author}{R.~H. Nochetto}, \bibinfo{author}{K.~G. Siebert},
\newblock \bibinfo{title}{{Data oscillation and convergence of adaptive FEM}},
\newblock \bibinfo{journal}{SIAM J. Numer. Anal.} \bibinfo{volume}{38} (\bibinfo{year}{2000}) \bibinfo{pages}{466--488}.
%Type = Article
\bibitem[{Braess et~al.(2007)Braess, Carstensen, and Hoppe}]{braess2007convergence}
\bibinfo{author}{D.~Braess}, \bibinfo{author}{C.~Carstensen}, \bibinfo{author}{R.~H. Hoppe},
\newblock \bibinfo{title}{Convergence analysis of a conforming adaptive finite element method for an obstacle problem},
\newblock \bibinfo{journal}{Numer. Math.} \bibinfo{volume}{107} (\bibinfo{year}{2007}) \bibinfo{pages}{455--471}.
%Type = Article
\bibitem[{Hoppe et~al.(2009)Hoppe, Kanschat, and Warburton}]{hoppe2009convergence}
\bibinfo{author}{R.~H. Hoppe}, \bibinfo{author}{G.~Kanschat}, \bibinfo{author}{T.~Warburton},
\newblock \bibinfo{title}{{Convergence analysis of an adaptive interior penalty discontinuous Galerkin method}},
\newblock \bibinfo{journal}{SIAM J. Numer. Anal.} \bibinfo{volume}{47} (\bibinfo{year}{2009}) \bibinfo{pages}{534--550}.
%Type = Article
\bibitem[{Daniel et~al.(2018)Daniel, Ern, Smears, and Vohral{\'\i}k}]{daniel2018adaptive}
\bibinfo{author}{P.~Daniel}, \bibinfo{author}{A.~Ern}, \bibinfo{author}{I.~Smears}, \bibinfo{author}{M.~Vohral{\'\i}k},
\newblock \bibinfo{title}{An adaptive hp-refinement strategy with computable guaranteed bound on the error reduction factor},
\newblock \bibinfo{journal}{Comput. Math. Appl.} \bibinfo{volume}{76} (\bibinfo{year}{2018}) \bibinfo{pages}{967--983}.
%Type = Article
\bibitem[{Venditti and Darmofal(2002)}]{venditti2002grid}
\bibinfo{author}{D.~A. Venditti}, \bibinfo{author}{D.~L. Darmofal},
\newblock \bibinfo{title}{Grid adaptation for functional outputs: application to two-dimensional inviscid flows},
\newblock \bibinfo{journal}{J. Comput. Phys.} \bibinfo{volume}{176} (\bibinfo{year}{2002}) \bibinfo{pages}{40--69}.
%Type = Article
\bibitem[{Giuliani and Krivodonova(2019)}]{giuliani2019adaptive}
\bibinfo{author}{A.~Giuliani}, \bibinfo{author}{L.~Krivodonova},
\newblock \bibinfo{title}{Adaptive mesh refinement on graphics processing units for applications in gas dynamics},
\newblock \bibinfo{journal}{J. Comput. Phys.} \bibinfo{volume}{381} (\bibinfo{year}{2019}) \bibinfo{pages}{67--90}.
%Type = Article
\bibitem[{Cant et~al.(2022)Cant, Ahmed, Fang, Chakarborty, Nivarti, Moulinec, and Emerson}]{cant2022unstructured}
\bibinfo{author}{R.~Cant}, \bibinfo{author}{U.~Ahmed}, \bibinfo{author}{J.~Fang}, \bibinfo{author}{N.~Chakarborty}, \bibinfo{author}{G.~Nivarti}, \bibinfo{author}{C.~Moulinec}, \bibinfo{author}{D.~Emerson},
\newblock \bibinfo{title}{An unstructured adaptive mesh refinement approach for computational fluid dynamics of reacting flows},
\newblock \bibinfo{journal}{J. Comput. Phys.} \bibinfo{volume}{468} (\bibinfo{year}{2022}) \bibinfo{pages}{111480}.
%Type = Article
\bibitem[{Freret et~al.(2022)Freret, Williamschen, and Groth}]{freret2022enhanced}
\bibinfo{author}{L.~Freret}, \bibinfo{author}{M.~Williamschen}, \bibinfo{author}{C.~P. Groth},
\newblock \bibinfo{title}{Enhanced anisotropic block-based adaptive mesh refinement for three-dimensional inviscid and viscous compressible flows},
\newblock \bibinfo{journal}{J. Comput. Phys.} \bibinfo{volume}{458} (\bibinfo{year}{2022}) \bibinfo{pages}{111092}.
%Type = Article
\bibitem[{Kelly et~al.(1983)Kelly, De~SR~Gago, Zienkiewicz, and Babuska}]{kelly1983posteriori}
\bibinfo{author}{D.~W. Kelly}, \bibinfo{author}{J.~De~SR~Gago}, \bibinfo{author}{O.~C. Zienkiewicz}, \bibinfo{author}{I.~Babuska},
\newblock \bibinfo{title}{{A posteriori error analysis and adaptive processes in the finite element method: Part I—error analysis}},
\newblock \bibinfo{journal}{Int. J. Numer. Methods Eng.} \bibinfo{volume}{19} (\bibinfo{year}{1983}) \bibinfo{pages}{1593--1619}.
%Type = Article
\bibitem[{Pierce and Giles(2000)}]{pierce2000adjoint}
\bibinfo{author}{N.~A. Pierce}, \bibinfo{author}{M.~B. Giles},
\newblock \bibinfo{title}{{Adjoint recovery of superconvergent functionals from PDE approximations}},
\newblock \bibinfo{journal}{SIAM Rev.} \bibinfo{volume}{42} (\bibinfo{year}{2000}) \bibinfo{pages}{247--264}.
%Type = Article
\bibitem[{Bohn and Feischl(2021)}]{bohn2021recurrent}
\bibinfo{author}{J.~Bohn}, \bibinfo{author}{M.~Feischl},
\newblock \bibinfo{title}{Recurrent neural networks as optimal mesh refinement strategies},
\newblock \bibinfo{journal}{Comput. Math. Appl.} \bibinfo{volume}{97} (\bibinfo{year}{2021}) \bibinfo{pages}{61--76}.
%Type = Inproceedings
\bibitem[{Yang et~al.(2023)Yang, Dzanic, Petersen, Kudo, Mittal, Tomov, Camier, Zhao, Zha, Kolev et~al.}]{yang2023reinforcement}
\bibinfo{author}{J.~Yang}, \bibinfo{author}{T.~Dzanic}, \bibinfo{author}{B.~Petersen}, \bibinfo{author}{J.~Kudo}, \bibinfo{author}{K.~Mittal}, \bibinfo{author}{V.~Tomov}, \bibinfo{author}{J.-S. Camier}, \bibinfo{author}{T.~Zhao}, \bibinfo{author}{H.~Zha}, \bibinfo{author}{T.~Kolev}, et~al.,
\newblock \bibinfo{title}{Reinforcement learning for adaptive mesh refinement},
\newblock in: \bibinfo{booktitle}{International Conference on Artificial Intelligence and Statistics}, \bibinfo{organization}{PMLR}, \bibinfo{year}{2023}, pp. \bibinfo{pages}{5997--6014}.
%Type = Article
\bibitem[{Foucart et~al.(2023)Foucart, Charous, and Lermusiaux}]{foucart2023deep}
\bibinfo{author}{C.~Foucart}, \bibinfo{author}{A.~Charous}, \bibinfo{author}{P.~F. Lermusiaux},
\newblock \bibinfo{title}{Deep reinforcement learning for adaptive mesh refinement},
\newblock \bibinfo{journal}{J. Comput. Phys.} \bibinfo{volume}{491} (\bibinfo{year}{2023}) \bibinfo{pages}{112381}.
%Type = Article
\bibitem[{Raissi et~al.(2019)Raissi, Perdikaris, and Karniadakis}]{raissi2019physics}
\bibinfo{author}{M.~Raissi}, \bibinfo{author}{P.~Perdikaris}, \bibinfo{author}{G.~E. Karniadakis},
\newblock \bibinfo{title}{{Physics-informed neural networks: A deep learning framework for solving forward and inverse problems involving nonlinear partial differential equations}},
\newblock \bibinfo{journal}{J. Comput. Phys.} \bibinfo{volume}{378} (\bibinfo{year}{2019}) \bibinfo{pages}{686--707}.
%Type = Article
\bibitem[{Karniadakis et~al.(2021)Karniadakis, Kevrekidis, Lu, Perdikaris, Wang, and Yang}]{karniadakis2021physics}
\bibinfo{author}{G.~E. Karniadakis}, \bibinfo{author}{I.~G. Kevrekidis}, \bibinfo{author}{L.~Lu}, \bibinfo{author}{P.~Perdikaris}, \bibinfo{author}{S.~Wang}, \bibinfo{author}{L.~Yang},
\newblock \bibinfo{title}{{Physics-informed machine learning}},
\newblock \bibinfo{journal}{Nat. Rev. Phys.} \bibinfo{volume}{3} (\bibinfo{year}{2021}) \bibinfo{pages}{422--440}.
%Type = Book
\bibitem[{Rall(1981)}]{rall1981automatic}
\bibinfo{author}{L.~B. Rall}, \bibinfo{title}{{Automatic differentiation: Techniques and applications}}, \bibinfo{publisher}{Springer}, \bibinfo{year}{1981}.
%Type = Article
\bibitem[{Mao et~al.(2020)Mao, Jagtap, and Karniadakis}]{mao2020physics}
\bibinfo{author}{Z.~Mao}, \bibinfo{author}{A.~D. Jagtap}, \bibinfo{author}{G.~E. Karniadakis},
\newblock \bibinfo{title}{{Physics-informed neural networks for high-speed flows}},
\newblock \bibinfo{journal}{Comput. Methods Appl. Mech. Eng.} \bibinfo{volume}{360} (\bibinfo{year}{2020}) \bibinfo{pages}{112789}.
%Type = Article
\bibitem[{Liang et~al.(2024)Liang, Song, Zhao, and Bian}]{liang2024continuous}
\bibinfo{author}{H.~Liang}, \bibinfo{author}{Z.~Song}, \bibinfo{author}{C.~Zhao}, \bibinfo{author}{X.~Bian},
\newblock \bibinfo{title}{Continuous and discontinuous compressible flows in a converging--diverging channel solved by physics-informed neural networks without exogenous data},
\newblock \bibinfo{journal}{Sci. Rep.} \bibinfo{volume}{14} (\bibinfo{year}{2024}) \bibinfo{pages}{3822}.
%Type = Article
\bibitem[{Meng et~al.(2020)Meng, Li, Zhang, and Karniadakis}]{Meng2020a}
\bibinfo{author}{X.~Meng}, \bibinfo{author}{Z.~Li}, \bibinfo{author}{D.~Zhang}, \bibinfo{author}{G.~E. Karniadakis},
\newblock \bibinfo{title}{{PPINN: Parareal physics-informed neural network for time-dependent PDEs}},
\newblock \bibinfo{journal}{Comput. Methods Appl. Mech. Eng.} \bibinfo{volume}{370} (\bibinfo{year}{2020}) \bibinfo{pages}{113250}.
%Type = Article
\bibitem[{Cai et~al.(2021)Cai, Wang, Wang, Perdikaris, and Karniadakis}]{cai2021physics}
\bibinfo{author}{S.~Cai}, \bibinfo{author}{Z.~Wang}, \bibinfo{author}{S.~Wang}, \bibinfo{author}{P.~Perdikaris}, \bibinfo{author}{G.~E. Karniadakis},
\newblock \bibinfo{title}{{Physics-informed neural networks for heat transfer problems}},
\newblock \bibinfo{journal}{J. Heat Transfer} \bibinfo{volume}{143} (\bibinfo{year}{2021}).
%Type = Article
\bibitem[{Wang and Perdikaris(2021)}]{wang2021deep}
\bibinfo{author}{S.~Wang}, \bibinfo{author}{P.~Perdikaris},
\newblock \bibinfo{title}{{Deep learning of free boundary and Stefan problems}},
\newblock \bibinfo{journal}{J. Comput. Phys.} \bibinfo{volume}{428} (\bibinfo{year}{2021}) \bibinfo{pages}{109914}.
%Type = Article
\bibitem[{Jin et~al.(2021)Jin, Cai, Li, and Karniadakis}]{jin2021nsfnets}
\bibinfo{author}{X.~Jin}, \bibinfo{author}{S.~Cai}, \bibinfo{author}{H.~Li}, \bibinfo{author}{G.~E. Karniadakis},
\newblock \bibinfo{title}{{NSFnets (Navier-Stokes flow nets): Physics-informed neural networks for the incompressible Navier-Stokes equations}},
\newblock \bibinfo{journal}{J. Comput. Phys.} \bibinfo{volume}{426} (\bibinfo{year}{2021}) \bibinfo{pages}{109951}.
%Type = Article
\bibitem[{Qiu et~al.(2022)Qiu, Huang, Xiao, Wang, Zhang, Yue, Zeng, and Wang}]{qiu2022physics}
\bibinfo{author}{R.~Qiu}, \bibinfo{author}{R.~Huang}, \bibinfo{author}{Y.~Xiao}, \bibinfo{author}{J.~Wang}, \bibinfo{author}{Z.~Zhang}, \bibinfo{author}{J.~Yue}, \bibinfo{author}{Z.~Zeng}, \bibinfo{author}{Y.~Wang},
\newblock \bibinfo{title}{Physics-informed neural networks for phase-field method in two-phase flow},
\newblock \bibinfo{journal}{Phys. Fluids} \bibinfo{volume}{34} (\bibinfo{year}{2022}).
%Type = Article
\bibitem[{Zhu et~al.(2024)Zhu, Kong, Deng, and Bian}]{zhu2024physics}
\bibinfo{author}{Y.~Zhu}, \bibinfo{author}{W.~Kong}, \bibinfo{author}{J.~Deng}, \bibinfo{author}{X.~Bian},
\newblock \bibinfo{title}{Physics-informed neural networks for incompressible flows with moving boundaries},
\newblock \bibinfo{journal}{Phys. Fluids} \bibinfo{volume}{36} (\bibinfo{year}{2024}).
%Type = Article
\bibitem[{Lu et~al.(2021)Lu, Meng, Mao, and Karniadakis}]{lu2021deepxde}
\bibinfo{author}{L.~Lu}, \bibinfo{author}{X.~Meng}, \bibinfo{author}{Z.~Mao}, \bibinfo{author}{G.~E. Karniadakis},
\newblock \bibinfo{title}{{DeepXDE: A deep learning library for solving differential equations}},
\newblock \bibinfo{journal}{SIAM Rev.} \bibinfo{volume}{63} (\bibinfo{year}{2021}) \bibinfo{pages}{208--228}.
%Type = Article
\bibitem[{Wu et~al.(2023)Wu, Zhu, Tan, Kartha, and Lu}]{wu2023comprehensive}
\bibinfo{author}{C.~Wu}, \bibinfo{author}{M.~Zhu}, \bibinfo{author}{Q.~Tan}, \bibinfo{author}{Y.~Kartha}, \bibinfo{author}{L.~Lu},
\newblock \bibinfo{title}{A comprehensive study of non-adaptive and residual-based adaptive sampling for physics-informed neural networks},
\newblock \bibinfo{journal}{Comput. Methods Appl. Mech. Eng.} \bibinfo{volume}{403} (\bibinfo{year}{2023}) \bibinfo{pages}{115671}.
%Type = Article
\bibitem[{Chew(1989)}]{chew1989constrained}
\bibinfo{author}{L.~P. Chew},
\newblock \bibinfo{title}{{Constrained Delaunay Triangulations}},
\newblock \bibinfo{journal}{Algorithmica} \bibinfo{volume}{4} (\bibinfo{year}{1989}) \bibinfo{pages}{97--108}.
%Type = Inproceedings
\bibitem[{Shewchuk(1996)}]{shewchuk1996triangle}
\bibinfo{author}{J.~R. Shewchuk},
\newblock \bibinfo{title}{{Triangle: Engineering a 2D quality mesh generator and Delaunay triangulator}},
\newblock in: \bibinfo{booktitle}{Workshop on applied computational geometry}, \bibinfo{organization}{Springer}, \bibinfo{year}{1996}, pp. \bibinfo{pages}{203--222}.
%Type = Article
\bibitem[{Shewchuk(2002)}]{shewchuk2002delaunay}
\bibinfo{author}{J.~R. Shewchuk},
\newblock \bibinfo{title}{Delaunay refinement algorithms for triangular mesh generation},
\newblock \bibinfo{journal}{Comput. Geom.} \bibinfo{volume}{22} (\bibinfo{year}{2002}) \bibinfo{pages}{21--74}.
%Type = Article
\bibitem[{Guo and Babu{\v{s}}ka(1986)}]{guo1986hp}
\bibinfo{author}{B.~Guo}, \bibinfo{author}{I.~Babu{\v{s}}ka},
\newblock \bibinfo{title}{{The hp version of the finite element method: Part 1: The basic approximation results}},
\newblock \bibinfo{journal}{Comput. Mech.} \bibinfo{volume}{1} (\bibinfo{year}{1986}) \bibinfo{pages}{21--41}.
%Type = Article
\bibitem[{Chorin(1968)}]{chorin1968numerical}
\bibinfo{author}{A.~J. Chorin},
\newblock \bibinfo{title}{{Numerical solution of the Navier-Stokes equations}},
\newblock \bibinfo{journal}{Math. Comput.} \bibinfo{volume}{22} (\bibinfo{year}{1968}) \bibinfo{pages}{745--762}.
%Type = Article
\bibitem[{Rhie and Chow(1983)}]{rhie1983numerical}
\bibinfo{author}{C.~M. Rhie}, \bibinfo{author}{W.-L. Chow},
\newblock \bibinfo{title}{Numerical study of the turbulent flow past an airfoil with trailing edge separation},
\newblock \bibinfo{journal}{AIAA J.} \bibinfo{volume}{21} (\bibinfo{year}{1983}) \bibinfo{pages}{1525--1532}.
%Type = Article
\bibitem[{Kingma and Ba(2014)}]{kingma2014adam}
\bibinfo{author}{D.~P. Kingma}, \bibinfo{author}{J.~Ba},
\newblock \bibinfo{title}{{Adam: A method for stochastic optimization}},
\newblock \bibinfo{journal}{arXiv:1412.6980}  (\bibinfo{year}{2014}).
%Type = Article
\bibitem[{Liu and Nocedal(1989)}]{liu1989limited}
\bibinfo{author}{D.~C. Liu}, \bibinfo{author}{J.~Nocedal},
\newblock \bibinfo{title}{{On the limited memory BFGS method for large scale optimization}},
\newblock \bibinfo{journal}{Math. Program.} \bibinfo{volume}{45} (\bibinfo{year}{1989}) \bibinfo{pages}{503--528}.
%Type = Article
\bibitem[{LeCun et~al.(2015)LeCun, Bengio, and Hinton}]{lecun2015deep}
\bibinfo{author}{Y.~LeCun}, \bibinfo{author}{Y.~Bengio}, \bibinfo{author}{G.~Hinton},
\newblock \bibinfo{title}{{Deep learning}},
\newblock \bibinfo{journal}{Nature} \bibinfo{volume}{521} (\bibinfo{year}{2015}) \bibinfo{pages}{436--444}.
%Type = Article
\bibitem[{Abadi et~al.(2016)Abadi, Agarwal, Barham, Brevdo, Chen, Citro, Corrado, Davis, Dean, Devin et~al.}]{abadi2016tensorflow}
\bibinfo{author}{M.~Abadi}, \bibinfo{author}{A.~Agarwal}, \bibinfo{author}{P.~Barham}, \bibinfo{author}{E.~Brevdo}, \bibinfo{author}{Z.~Chen}, \bibinfo{author}{C.~Citro}, \bibinfo{author}{G.~S. Corrado}, \bibinfo{author}{A.~Davis}, \bibinfo{author}{J.~Dean}, \bibinfo{author}{M.~Devin}, et~al.,
\newblock \bibinfo{title}{{Tensorflow: Large-scale machine learning on heterogeneous distributed systems}},
\newblock \bibinfo{journal}{arXiv:1603.04467}  (\bibinfo{year}{2016}).
%Type = Book
\bibitem[{Dolej{\v{s}}{\'\i} and May(2022)}]{dolejvsi2022anisotropic}
\bibinfo{author}{V.~Dolej{\v{s}}{\'\i}}, \bibinfo{author}{G.~May}, \bibinfo{title}{{Anisotropic hp-Mesh Adaptation Methods}}, \bibinfo{publisher}{Springer}, \bibinfo{year}{2022}.
%Type = Article
\bibitem[{Vassberg and Jameson(2010)}]{vassberg2010pursuit}
\bibinfo{author}{J.~C. Vassberg}, \bibinfo{author}{A.~Jameson},
\newblock \bibinfo{title}{{In pursuit of grid convergence for two-dimensional Euler solutions}},
\newblock \bibinfo{journal}{J. Aircr.} \bibinfo{volume}{47} (\bibinfo{year}{2010}) \bibinfo{pages}{1152--1166}.

\end{thebibliography}

%% else use the following coding to input the bibitems directly in the
%% TeX file.

% \begin{thebibliography}{00}

% %% \bibitem[Author(year)]{label}
% %% Text of bibliographic item

% \bibitem[ ()]{}

% \end{thebibliography}
\end{document}